\documentstyle[11pt,aaspp4]{article}
\lefthead{PARRAVANO, HOLLENBACH, AND McKEE}
\righthead{Time dependent FUV Field}

\begin{document}

\title{Time Dependence of the Ultraviolet Radiation Field in
the Local Interstellar Medium}

\author{Antonio Parravano\altaffilmark{1,2}, 
David J. Hollenbach\altaffilmark{2},  and
Christopher F. McKee\altaffilmark{3}}

\altaffiltext{1}{Universidad de Los Andes, Centro De Astrof\'{\i}sica
Te\'orica, M\'erida 5101a, Venezuela}

\altaffiltext{2}{NASA Ames Research Center, MS 245-3, Moffett Field,
 CA 94035}

\altaffiltext{3}{Physics Department and Astronomy Department
University of California at Berkeley, Berkeley, CA 94720} 

\def\beq{\begin{equation}}
\def\eeq{\end{equation}}
\def\avg #1{\langle #1\rangle}
\def\caln{{\cal N}}
\def\calr{{\cal R}}
\def\Nsh{${\cal N}_{*h}$}
\def\Nsm{$\caln_*(m)$}
\def\ssh{$\varsigma _{*h}$}
\def\ssm{$\varsigma _{*}(m)$}
\def\ns{$\varsigma _{*t}$}
\def\dns{$\dot \varsigma _{*t}$}
\def\dNsm{$\dot\caln_*(m)$}
\def\dssh{$\dot \varsigma _{*h}$}
\def\dNsh{$\dot {\cal N}_{*h}$}
\def\dNshT{$\dot {\cal N}_{*h,T}$}
\def\emuh{\mu_h}
\def\eNsm{\caln_*(m)}
\def\eNsh{\caln_{*h}}
\def\eNshtyp{\caln_{*h,\, \rm typ}}
\def\eNshtypmax{\caln_{*h,\, \rm typ\; max}}
\def\eNshl{{\caln_{*h,l}}}
\def\eNshu{{\caln_{*h,u}}}
\def\essh{\varsigma _{*h}}
\def\essm{\varsigma _{*}(m)}
\def\edNsm{\dot\caln_*(m)}
\def\eNa{{\cal N}_a}
\def\edsa{\dot \varsigma _{a}}
\def\edss{\dot \varsigma _{*}}
\def\edssh{\dot \varsigma _{*h}}
\def\edsshtyp{\dot \varsigma _{*h,\, \rm typ}}
\def\edssm{\dot \varsigma _{*}(m)}
\def\edNsh{\dot {\cal N}_{*h}}
\def\edNshT{\dot {\cal N}_{*h,T}}
\def\edNs{\dot {\cal N}_{*}}
\def\eavtion{\avg{t_{\rm ion}}}
\def\dens{\dot \varsigma _{*t}}
\def\emsun{M_\odot}
\def\msun{$M_\odot$}
\def\etsim{{t_{\rm sim}}}
\def\vecr{{\bf r}}
\def\Nsh{${\cal N}_{*h}$}
\def\nsh{$\sigma _{*h}$}
\def\ns{$\sigma _*$}
\def\dns{$\dot \sigma_*$}
\def\dnsh{$\dot \sigma _{*h}$}
\def\dNsh{$\dot {\cal N}_{*h}$}
\def\densh{\dot \sigma _{*h}}
\def\deNsh{\dot {\cal N}_{*h}}
\def\dens{\dot \sigma _{*t}}
\def\Nh{{\cal N}_{*h}}
\def\Nu{{\cal N}_{*h,u}}
\def\Nl{{\cal N}_{*h,l}}
\def\Non{{\cal N}_{*h,1}}
\def\Noncr{{\cal N}_{*h,1cr}}
\def\Ntw{{\cal N}_{*h,2}}
\def\Ntwcr{{\cal N}_{*h,2cr}}
\def\Nth{{\cal N}_{*h,3}}
\def\d{n_{ac}(\Nh)}
\def\dd{n_a(\Nh)}
\def\ddd{{\dot n}_{ac}(\Nh)}
\def\dddd{{\dot n}_{a}(\Nh)}
\def\sd{\varsigma _a(\Nh)}
\def\sdd{{\dot {\varsigma}} _a(\Nh)}
\def\s0{{\dot {\varsigma}} _{a0}}
\def\Ut{U_{\rm band}}
\def\Us{U_s}
\def\Ub{U_{\rm back}(\Us)}
\def\Um{<\Ut >}
\def\Umed{U_{\rm med}}
\def\Umeds{U_{{\rm med},s}}
\def\Uz{U_0}
\def\Uo{U_1}
\def\Uoa{U_{1a}}
\def\Utw{U_2}
\def\Uth{U_3}
\def\del{\Delta _{\rm band}}
\def\H{\sqrt {\pi}H_*}
\def\ph{\Phi _s(\Us)}
\def\Ps{P_s(>\Ut)}
\def\Pss{P_s(>\Us)}
\def\Pt{P_t(>\Ut)}
\def\ps{p_s(\Ut)}
\def\pss{p_s(\Us)}
\def\rn{r_{\Nh}(\Us)}
\def\rco{r_1}
\def\rcoa{r_{1a}}
\def\rct{r_2}
\def\Lf{L_{\rm band}(\Nh)}
\def\L0{L_{0,\rm band}}
\def\Lfuv{L_{0,\rm FUV}}
\def\lm{\ell _{\rm mfp}(>\Ut)}
\def\ta{t_U(>\Ut )}
\def\t0{t_{\rm band}}

\begin{abstract}
Far Ultraviolet (FUV, 6 eV$ < h\nu < $13.6 eV) radiation has been
recognized as the main source of heating of the neutral interstellar
gas, and, as a consequence, it 
determines whether the thermal balance of the neutral gas results in
cold ($T\sim 50 - 100 K$) clouds (CNM), warm ($T \sim 10^4 K$) clouds (WNM),
or a combination of the two.  High FUV fields convert the neutral gas
to WNM, while low fields result in CNM.  
The knowledge of how these fractions depend on the
FUV sources (i.e. the star formation rate, the IMF, and the size
distribution of associations) is a basic step in building any detailed
model of the large scale behavior of the ISM and the mutual relation
between the ISM and the star formation rate in a galaxy.

The sources of FUV
radiation are the short-lived massive stars that generally originate in
associations that form in Giant Molecular Clouds  present
in the galactic disk. 
Using McKee \& Williams' (1997) distribution of birthrates for
OB associations in the Galaxy, we determine
the expected behavior of the time-dependent FUV field for 
random positions in the local ISM.
The FUV field is calculated in two bands ($912-1100$~\AA\ and $912-2070$~\AA)
and at the wavelength 1400~\AA.
In terms of $U_{-17}\equiv U/(10^{-17}$~erg cm$^{-3}$ \AA$^{-1}$),
where $U$ is the energy density of the radiation field in some
band, we find (mean, median) values at the
solar circle of $U_{-17}=$(15.7, 7.4) and (14.2, 7.2) for
the [912-1100~\AA] and [912-2070~\AA] bands, respectively.
At $1400~\AA$ we find (mean, median) values of $U_{-17}=$(14.4, 7.5).
Our median value for the [912-2070~\AA] band is $G_0=1.6$ times
Habing's (1968) value for the radiation field at the solar circle
in this band,  and quite close to 
Draine's (1976) value, $G_0=1.7$.
Both the latter values are based on observations of sources
of FUV radiation in the solar neighborhood, so all three
values are close to observed values. 
Due to attenuation by dust, only associations 
within about 500 pc 
contribute significantly
to the energy density at a given point.  Large angle scattering
produces a diffuse field that is about 10\% of the field produced
by the sum of 
direct and small angle ($<5^o$) scattering from
discrete sources (the associations), as observed.
At a point exposed to the median radiation field,  the brightest
association typically produces about 20\% of the total energy density.
At a point exposed to an above average
radiation field, the brightest association
produces most of the energy density. Therefore, the FUV field
is asymmetric at a given point, and the asymmetry
grows for higher fields.

The FUV field fluctuates with a variety of
amplitudes, the larger ones being less frequent. 
The mean field is about twice the median field because of
these fluctuations, or spikes, in the radiation field. These
spikes, which last
$\sim 30$ Myr, are caused by the infrequent birth of  nearby 
associations.  
For spikes that are significantly
higher than the mean field, the time interval between spikes is 
$\sim 2 U_{-15}^{3/2}$ Gyr.  
We also
model  shorter duration spikes caused by runaway OB stars.  
The presence of a fluctuating heating rate created by the 
fluctuating FUV field
converts CNM to WNM and vice versa.  

\end{abstract}


\section{Introduction}

Neutral interstellar gas is the dominant mass component of the
interstellar medium (ISM), and the atomic gas exists as two phases in rough 
pressure equilibrium: the warm neutral medium (WNM) with hydrogen
densities at the solar circle of 
$n\sim 0.3$ cm$^{-3}$ and $T\simeq 8000$ K, and the
cold neutral medium (CNM) with $n\sim 40$ cm$^{-3}$ and $T\sim 70$ K
(Wolfire et al. 1995, 2002). With sufficient shielding column density, 
$N > 10^{20-21}$
cm$^{-2}$, the CNM clouds harbor molecular interiors, and for
$N > 10^{22}$ cm$^{-2}$ they become gravitationally bound and
stars may form in their molecular interiors (e.g., McKee 1989).  Most of the
star formation in the Galaxy occurs in Giant Molecular Clouds (GMCs)
with large masses $M\sim 10^{5-6}$ M$_\odot$ and columns $N\sim 10^{22}$
cm$^{-2}$.

In spiral galaxies FUV radiation from OB stars heats, photodissociates, 
and controls the state
of the neutral interstellar gas.  High FUV fields heat CNM to WNM; low
FUV fields allow WNM to cool to CNM (Wolfire et al. 1995, 2002). If, as seems
likely, the formation rate of stars in galaxies depends on the mass
fraction of CNM gas, then a possible feedback process occurs between 
stars and the ISM that may control the global rate of star formation
in a galaxy 
(Parravano 1988, 1989, 1991; Corbelli \& Salpeter 1995).  
High star formation rates lead to large numbers of OB stars
that create high FUV fluxes that destroy the sources of star formation.
Low rates of star formation, on the other hand, may lead to enhanced
growth rates of star-forming clouds and thereby to 
enhanced star formation rates.  In this paper we focus on the
calculation of the fluctuating FUV field in the local region of
the Milky Way Galaxy, deferring to future papers the study of
the fractional amounts of CNM and WNM produced by the FUV field, and
the possible regulation of star formation rates in galaxies.

	OB stars are generally born in associations that form inside 
GMCs present in the galactic disk.
The  FUV field at a
given point in a galaxy then depends
largely on the distribution of stellar masses in relatively nearby
associations
and on the birth-rate and size distribution of these associations.
The FUV luminosity  of an association is an increasing function
of its size, but the birth-rate of associations decreases with size.
This fact, coupled with the fact that the probability of a young association
existing within a distance $r$ of  a point decreases rapidly with decreasing
$r$, leads to a fluctuating FUV field, with
the larger fluctuations being less frequent. 
The abundant low luminosity or distant associations help to build
an FUV background, over which is added bursts of FUV flux
coming from  relatively
rare events (the birth of nearby large associations) that can significantly
increase the FUV field
for about $\sim 30$ to 50 Myr.  This period corresponds
to the time that an association 
produces stars ($\sim 20$ Myr, generally in a mode of several bursts) plus 
the time that early B  stars, which dominate the FUV production,  
remain alive after the end of the star formation process in the GMC.
The {\it time} variation of the FUV field at any given point in a galaxy
maps directly onto the {\it spatial} variation of the FUV field in the
galaxy at any instant of time: those regions very close to luminous
associations will have exceptionally high FUV fluxes, whereas large
volumes of space will have much lower FUV fluxes.

	The presence of a time and space-dependent FUV field 
 has a major impact on the structure of the interstellar medium.
In a subsequent paper we calculate the response of 
interstellar gas to this fluctuating FUV field,
estimate the average fractions of the gas in the CNM and WNM phases,
 and determine the typical time that the gas remains in each phase. The
knowledge of these quantities and how they depend on the gas
properties (i.e., mean density and composition) and on the
FUV-sources (i.e., star formation rate SFR, initial mass function IMF, 
and size distribution of associations) is
a basic step in building any detailed model of the large scale behavior
of the ISM and the mutual relation between the ISM and the SFR.

	FUV photons heat the neutral interstellar gas primarily by
photoejecting  energetic ($\sim 1$ eV) electrons into the gas
from interstellar dust grains 
(Watson, 1972; Jura, 1976; de Jong, 1977; Draine, 1978)
and large stable molecules like polycyclic aromatic hydrocarbons
(PAHs) (d'Hendecourt and L\'eger, 1987; 
Lepp and Dalgarno, 1988; Verstraete et al., 1990; Bakes and Tielens, 1994).
The relevant wavelength range for this ``grain photoelectric heating''
process is determined
by the frequency dependence of
the absorption cross sections and the photoelectric
yields of grains and PAHs. Since we are interested in the 
neutral ISM, the Lyman threshold (912~\AA) is the shortest wavelength
of the relevant band. The long wavelength limit is somewhat arbitrary. 
It depends
on the wavelength at which both the absorption cross section
and the yield has fallen significantly for grains as well as for PAHs
(Draine 1978; Bakes \& Tielens 1994, and references therein).
In general, the band between 6 eV$\le h\nu \le 13.6$ eV 
(i.e. $[912-2070$~\AA]) has been considered as the primary band for
grain photoelectric heating, and a mean value of the flux
in the band is often used  as a parameter to calculate the thermochemical 
equilibrium of the neutral gas. 
However, for a given total flux in the band,
a change in its energy distribution can have a large effect 
on the photodissociation rates and even the heating rates. 
In particular, the main processes of $\rm{H}_2$ and CO destruction 
is  photodissociation 
by radiation in the band $[912-1100$~\AA]. 
We therefore follow  two primary bands:\\
a) {\bf{The band $[912-2070$~\AA], hereafter the ``FUV band"}}. 
The mean energy density in this band is related to the
parameter $G_0$ that is often used to parametrize the photoelectric 
dust heating rate (Tielens \& Hollenbach 1985). 
Let $U_{\rm band}\equiv U({\rm band})/\Delta_{\rm band}$
be the average radiation energy density
per \AA\ in the wavelength band $\Delta_{\rm band}$.
The energy density of FUV radiation is generally expressed in
units of the Habing (1968) typical energy density at the solar circle
averaged  between
6 eV$\le h\nu \le 13.6$  eV
(i.e. $[912-2070$~\AA]),
\beq
G_0\equiv\frac{U_{\rm FUV}}{U_{\rm FUV}^{\rm H}}=\frac{U_{\rm FUV}}
	{4.6\times 10^{-17}~{\rm erg\,  cm^{-3}\,  \AA^{-1}}}.
\eeq
If the radiation field is one-dimensional, it is convenient
to express the Habing field as a flux, $cU_{\rm FUV}^{\rm
H}\Delta_{\rm FUV}=
1.6 \times 10^{-3}~{\rm erg\,  cm^{-2}\,  s^{-1}}.$ 

b) {\bf{ The band $[912-1100$~\AA], hereafter the `` H$_2$ band''}}. 
The flux 
in this band determines the dissociation rate of $\rm{H}_2$ and CO.  It 
is also useful to consider separately the contribution of this band to the 
photoelectic heating. Even though the $\rm{H}_2$ band covers only
about 10\% of 
the whole FUV wavelength band, both the absorption cross section and the yield 
of dust are significantly higher in the H$_2$ band, so that it
contributes disproportionately to the gas heating.  On the other hand, the 
extinction in the $\rm{H}_2$ band is about 
twice the extinction in the FUV band, which reduces the importance of
the H$_2$ band heating. 

In addition to these two rather broad bands, simulations in a 1~\AA\ band
around 1400~\AA\ are performed in order to compare
the averages of the time dependent FUV model with observations. 
Many observations have been made of the interstellar field at
1400~\AA, although the observations are averages over bands
considerably broader than 1 \AA.

In our model, we include both direct and scattered FUV, and therefore, at 
a given time the local value of the FUV interstellar radiation field
(hereafter FUV-ISRF) can be reasonably determined
from the location of the FUV sources and the distribution and properties
of the absorbing and scattering dust particles. 
In order to construct a time dependent model for FUV interstellar radiation
we adopt:
a) an initial mass function for  stars in associations 
(\S 2.1);
b) a stellar-mass dependence of the FUV and H$_2$ band luminosities, and 
the main sequence
lifetime for stars of various masses (\S 2.2);
c) the dependence of the 
birth-rate of  associations on the
number of stars in the association, the history of the star
formation rate in an association, and the distribution in the galactic
plane of the association birthplaces from McKee \& Williams (1997;
hereafter MW97)
(\S 3); and
d) a vertical distribution of associations and dust (\S 4).
We also present in \S 4 model simulations of the time dependence of the 
FUV-ISRF, together with the corresponding  asymptotic values. 
The typical
FUV-ISRF is computed as the median value in the time-sequence. We compare
in \S 5 the mean and median values derived with the time-dependent model
with previous estimations of the local field. 
In \S 6 we quantify the effect of runaway OB stars on the 
detailed time evolution of the FUV-ISRF.
Finally, we summarize the results in \S 7.

\section{Stellar Properties}

\subsection{Stellar IMF and PDMF}

	Let $\caln_*(>m)$ be the number of stars with masses greater
than $m\,M_\odot$, and let $d\eNsm\equiv -d\caln_*(>m)$ be the number
of stars with masses between $m$ and $m+dm$.
We express the present day mass function (PDMF)
as $d\eNsm/d\ln m$, which
is simply the number of stars in a logarithmic mass
interval.  We denote the PDMF per unit area of Galactic disk by 
\beq
\essm\equiv \frac{d^2\eNsm}{dA\; d\ln m}.
\eeq
This quantity is related to the quantity $\phi_{\rm ms}$ defined by Miller \&
Scalo (1979; hereafter MS79) 
by $\essm=\phi_{\rm ms}\log e=0.434\phi_{\rm ms}$.

	The stellar birthrate as a function of mass---the IMF---is
given by $d\edNsm/d\ln m$; note that $\edNsm$ includes only stellar
birth, not stellar death.  The IMF per unit area of Galactic disk is
$\edssm$, which
is related to MS79's $\xi(\log m)$ by
$\int\dot\varsigma(m,t)dt=0.434\xi(\log m)$. Sometimes the IMF is described
by the fraction of stars above mass $m$ [e.g., $F_n(>m)$--- MS79], but
this depends on the uncertain and possibly variable
number of low-mass stars in the IMF.
Since we are interested in
the stars that produce the FUV radiation, we shall generally
characterize the IMF in terms of quantities related to the high-mass
end of the IMF.

	 For massive stars, the IMF can be approximated as
a cut-off power law that extends up to a mass $m_u$ (MW97),
\beq
\frac{d\edNsm}{d\ln m}=\dot\caln_{*u}\left(\frac{m_u}{m}\right)^
	{\Gamma}~~~~~(m\leq m_u),
\label{eq:dNsm}
\eeq
where $\dot\caln_{*u}$ is approximately equal to the birthrate of
stars with masses between $0.5m_u$ and $m_u$ (for $\Gamma=1$ this
relation is exact). 
Young star clusters have PDMFs that are of this form also, although
the parameters $m_u$ and $\Gamma$ that enter
this distribution may differ from those in the IMF due to stellar mass
loss and, eventually, to supernovae.
The birthrate of stars more massive than $m$ is
\beq
\dot\caln_*(>m)=\frac{\dot\caln_{*u}}{\Gamma}
	\left[\left(\frac{m_u}{m}\right)^\Gamma -1\right].
\label{eq:nsgm}
\eeq
In most associations, there are very few stars with masses
near $m_u$, so it is convenient to express the IMF in terms
of \dNsh, the birthrate of high-mass stars (those with masses $m\geq m_h$),
\beq
\frac{d\edNsm}{d\ln m}=\edNsh\left(\frac{\Gamma}{\phi_h}\right)
	\left(\frac{m_h}{m}\right)^\Gamma,
\label{eq:dNsm2}
\eeq
where $\phi_h\equiv 1-(m_h/m_u)^\Gamma$ is a numerical parameter
that is of order unity; indeed, for the fiducial values
of the parameters that we adopt below, $\phi_h=0.974$.

	The value of $m_h$ is somewhat arbitrary; we adopt 
$m_h=8$ since most ionizing photons are produced
by stars above this mass and since this is the minimum mass for
core-collapse supernovae.  
The value of $\Gamma$ is also somewhat uncertain.
Van Buren (1983) carried out 
a careful study of the high stellar mass IMF of the Galaxy, and 
concluded that $\Gamma=1.03$. 
A fit to the high-mass end of
Scalo's (1986) IMF gives $\Gamma\simeq 1.5$.
In a study of OB associations in the northern Milky Way,
Massey, Johnson, \& DeGioia-Eastwood (1995) found
a mean value $\Gamma=1.1\pm 0.1$.  
In the star cluster R136 located in the LMC,
Massey and Hunter (1998) found $\Gamma\simeq 1.35$.
While it remains controversial whether the IMF is a universal function
(Scalo 1998), there is no evidence that the variations seen in
the IMF are anything but statistical in nature.  We
adopt a single power-law IMF with
$\Gamma = 1.35$ for stars more massive than 8 $M_\odot$.
As described below and in Appendix
A, the probabilistic distribution used here allows for fluctuations around
this power law in a given association.

	The existence of an upper limit to the mass function
is also controversial; for example, Massey and Hunter (1998) suggest
that the apparent cutoff is due to the sparcity of very massive
stars.  This issue was addressed in a general way by MW97.  
If the distribution
of a quantity $x$ can be approximated by a cutoff-power law
of the form in equation (\ref{eq:nsgm}),
\beq
\caln(>x)=\frac{\caln_u}{\Gamma}\left[\left(\frac{x_u}{x}\right)^\Gamma
	-1\right],
\eeq
then the expected number of objects with $x>x_u$ in the absence
of a physical cutoff is $\caln_u/\Gamma$.  If this number is
large, then the cutoff must be physically significant, whereas
if it is $\la 1$ then it is not.  In R136, Massey and Hunter (1998) find
29 stars more massive than 75 $M_\odot$, but none more massive
than $155 M_\odot$ (based on the stellar data of Vacca, Garmany,
\& Shull 1996).  Setting $m_u=155$ in this case, we find
$\caln_u/\Gamma=17.4\gg 1$.  Hence, we conclude that the upper mass
limit in R136 is physically significant.  In the Galaxy, it
is not clear that $m_u$ is as large as 155. In their survey
of OB associations in the northern Milky Way, Massey et al. (1995)
found only one star more massive than 120 $M_\odot$, whereas
many more would have been expected if the IMF continued on
to higher masses.  We shall adopt $m_u=120$ here.

\subsection{Stellar Sources of FUV Radiation}

As they evolve,  stars change their radii and effective temperatures and
therefore their FUV luminosities.  However, we are interested 
in the FUV radiation
field from a large population of stars of different ages and in 
different associations, and therefore
the use of a mean luminosity during the  main sequence 
(MS) lifetime, $t_{\rm ms}$, is an adequate approximation. 
During the large but infrequent fluctuations of the FUV field that occur 
when a single
nearby association dominates the radiation field, 
this approximation is not as good as in
the quiescent  period when associations of different ages contribute to the
FUV field.
However, even in a single association, several generations of stars 
can contribute to the total luminosity of the association, and
the use of time-averaged 
FUV luminosities  provides
an adequate description of the total FUV luminosity of the association. 
Therefore, in our model the time variation
of the total FUV luminosity of an association is due to the 
birth and death of its 
massive stars, and not to the luminosity changes in individual stars during
their MS lifetime.

To compute the mean main-sequence luminosities we have used the
set of stellar evolutionary tracks generally referred as the Padova tracks 
(Bruzual 1999, and references therein), 
with the stellar flux in the $\rm{H}_2$ and FUV bands kindly provided
by G. Bruzual and J. Mateu (based on the library of synthetic 
stellar spectra compiled by Lejeune et al. 1997ab).
Hereafter, we denote as $L_{\rm FUV}$ and $L_{H2}$ the mean main-sequence 
luminosities 
in the FUV band, $[912-2070$~\AA], and in the $\rm{H}_2$ band, 
$[912-1100$~\AA], respectively. 
For stars with solar metallicity, Figures 1a and 1b  show respectively
the resulting mean 
luminosities in the band $[1100-2070$~\AA] (i.e. $L_{\rm FUV}-L_{H2}$) and in 
the $\rm{H}_2$ band (i.e. $L_{H2}$) as a function of the zero age main
sequence stellar mass $m$ in units of $M_\odot$. 
In  Figures 1a and 1b the triangles correspond to the calculated mean MS 
luminosities using the stellar evolutionary tracks, and the lines are the 
eight-segment power law used to approximate the mass dependence of the
luminosity in the simulations.
Table 1 shows the adopted power laws for $L_{\rm FUV}-L_{H2}$, $L_{H2}$,
$s_{49}$ (the ionizing photon luminosity, see below), and the main
sequence lifetime $t_{\rm ms}(m)$.
The power law approximations for the luminosities are good to within 10\%,
and those for the main sequence lifetime are good to within 5\%.
 
	When a massive star is young and embedded or near its natal cloud, 
the natal cloud absorbs a fraction
of the emitted radiation. 
Due to their shorter
lifetimes, more massive stars are more affected by this obscuration.
We assume that the fraction of radiation locally absorbed goes linearly from
$3/4$ to zero in a time $t_{\rm obsc}$. In the nearest region of massive star
formation, the Orion Molecular Cloud, the Trapezium generation of massive
stars and the other Blaauw subgroups of massive stars 
apparently broke out of the 
obscuring cloud material in few million years.
A more exact analysis can be done by quantitatively considering the
breakout of OB stars from their clouds, as was done in the models of
Whitworth (1979) and Williams \& McKee (1997).  The latter models
indicate $t_{\rm obsc}\simeq 4$ Myr, the number that we adopt.  Because
B stars are dominant contributors to the interstellar FUV field (see
below), and because they live considerably longer than  $t_{\rm obsc}$, the
exact value of $t_{\rm obsc}$ does not significantly affect our results,
i.e., most of the FUV radiation is not absorbed locally in the natal clouds
of the OB stars. 

In order to show the time averaged contribution of the different 
mass ranges to the FUV radiation field, 
Figures 1c and 1d  show the energy output in each band 
for a mass distribution of stars following a Present Day Mass Function (PDMF)
of MS stars. 
The PDMF $\essm$ 
has been constructed assuming a constant star formation
rate (appropriate for the solar circle--see \S 2.4 below),
a single power-law for the IMF (\S 2.1), and weighting the stars by
their main-sequence lifetimes. Rocha-Pinto et al. (2000) have 
presented evidence that the Galactic star formation rate changes
on time scales of order 400 Myr, which corresponds to the
main-sequence lifetime of a 2.9 \msun\ star.  Almost all the FUV
radiation is produced by stars more massive than this, so such a
time variation in the Galactic star formation rate would not affect
our results.
We note that the results in Figures 1c and 1d are
useful only to make time-averaged estimates.
To follow the evolution of the FUV-ISRF when the SFR is time and position 
dependent, it is necessary to consider that obscuration acts 
during the period $t_{\rm obsc}$, starting at the moment when each generation
of stars are formed.

In the FUV band 
the corresponding mean emissivity per unit disk area 
is $\sim 13.4 \, {L}_\odot/\rm{pc}^2$
(i.e., the area below the heavy line curve in Fig. 1c).
After correcting for obscuration, the emissivity is 
$\sim 10.9 \, {L}_\odot/\rm{pc}^2$ 
(i.e. the area below the thin line curve in Fig. 1c).
Therefore, $\sim 2.5 \, {L}_\odot/\rm{pc}^2$ is 
absorbed by the natal clouds, representing $\sim 19\%$
of the total mean emissivity per unit disk area.{\footnote
{This calculation has interesting implications. If the diffuse ISM
is optically thick to FUV photons, then 81\% of the IR luminosity
from galactic dust heated by FUV photons will arise from the diffuse gas.
Therefore, when comparing IR observations of galaxies made with
rather large beams, encompassing many GMCs and many FUV optical depths
in the galactic plane, with PDR models (e.g., Malhotra et al. 2001,
Kaufman et al. 1999) one will need to separate the IR continuum
and [CII] and [OI] fine structure line emission into two components:
one from the natal GMCs and one from the diffuse ISM.}}
The distribution of emissivity with stellar mass has a maximum at
$\sim 6 \, {M}_\odot$ (i.e., B5), but only $\sim 16 \%$ of 
the total emission is produced
by stars with masses below this maximum.
Half of the FUV radiation is produced by
stars with $m>18$, but half of the {\it escaping} 
FUV radiation is produced by stars 
with $m>13$ 
(i.e., earlier than B1).
Stars with masses less than $3 \, {M}_\odot$ only contribute  $\sim 3\%$.
Therefore, the SFR history in the past $\sim 350 \, \rm{Myr}$ 
[i.e. $t_{\rm ms}(3 \, M_\odot)]$ is sufficient to simulate the present FUV 
radiation field.

     The characteristics of the FUV-emitting stars can be
described in terms of their mean lifetime and luminosity.
Ignoring absorption by the natal cloud, 
we calculate the luminosity-weighted lifetime as 
\beq
\avg{t_{\rm FUV}}=\frac{\int L_{\rm FUV}(m) t_{\rm ms}d\edNs}
{\int L_{\rm FUV}(m) d\edNs} ,
\label{eq:tfuv}
\eeq
which is 7.8 Myr for the stellar parameters we 
have adopted. 
By comparison, the corresponding lifetime for ionizing radiation is
4.0 Myr (\S 2.3).
The mean FUV luminosity per high-mass star in the IMF is 
\beq
\avg{L_{\rm FUV}}_h\equiv\frac{1}{\edNsh}\int L_{\rm FUV}(m) d\edNs
        =\frac{\Gamma}{\phi_h}\int L_{\rm FUV}(m)
        \left(\frac{m_h}{m}\right)^\Gamma d\ln m.
\eeq
The mean FUV emissivity per unit disk area is related to the
high-mass star formation rate as
\beq
\Sigma_{\rm FUV}=\edssh \avg{L_{\rm FUV}}_h \avg{t_{\rm FUV}}.
\eeq
For the stellar parameters we have adopted 
$\avg{L_{\rm FUV}}_{h}=5.8 \times 10^4 L_\odot$,
and using the high-mass star formation rate derived in \S 2.4,
$\Sigma_{\rm FUV}=13.4 \, {L}_\odot/\rm{pc}^2$ in accordance
with the value derived
from the PDMF above. Since $\Sigma_{\rm FUV}=10.9 \, {L}_\odot/\rm{pc}^2$
after correcting for obscuration, the obscuration-corrected mean
luminosity is 
$\avg{L_{\rm FUV}}_{h}=4.7 \times 10^4 L_\odot$.

In the   $\rm{H}_2$ band,
the radiation field is dominated by more massive stars.
The mean emissivity per unit disk area 
is $\sim 3.2 \, L_\odot/\rm{pc}^2$.
After correcting for obscuration, the emissivity is 
$\sim 2.5 \, L_\odot/\rm{pc}^2$. 
The maximum contribution
per unit mass occurs at $m=9$ (i.e., B2 stars).
After correcting for obscuration, about $85\%$ of the total emission in the 
band is produced by stars with 
masses $m>9$, and stars with masses less 
than $5 \, M_\odot$ only contribute  $\sim 2\%$ to
the total emission (lifetime $\sim 100 \, \rm{Myr}$). 
Half of the total escaping emission 
is produced by stars with masses over $\sim 21 \, M_\odot$
(lifetime of $\sim 9 \, \rm{Myr}$). 
Since the extinction in the $\rm{H}_2$ band is larger
than the extinction in the FUV band, the radiation field 
in the $\rm{H}_2$ band is more dominated by nearby associations. Therefore, the
amplitudes of the fluctuations in the $\rm{H}_2$ band
are expected to be larger than in the FUV band, where more sources
contribute to the field.
For the stellar parameters we have adopted and
ignoring absorption by the natal cloud, 
the luminosity-weighted lifetime is 
$\avg{t_{\rm H_2}}=6.3$~Myr and the
mean H$_2$ luminosity per high-mass star in the IMF is 
$\avg{L_{\rm H_2}}_{h}=1.8 \times 10^4 L_\odot$.
Since $\Sigma_{\rm H_2}=2.5 \, {L}_\odot/\rm{pc}^2$
after correcting for obscuration, the obscuration-corrected mean
luminosity is
$\avg{ L_{\rm FUV} }_{h} =1.0 \times 10^4 L_\odot$.

Most of the conclusions in the preceding two paragraphs apply only  
to a distribution of 
FUV sources that is approximately constant
in both space and time. 
In fact, the FUV sources are located in stellar associations
that are spatially and temporally very inhomogeneous.
As a result, the local SFR, and therefore the FUV and H$_2$ band emissivities,
are below their mean values most of the time (see \S 4.3). In addition, 
massive stars are scarce and therefore distant, so that
they are subject to relatively large
interstellar extinction.
All of these effects will be included in our time dependent
model.  In any case, the results in Figures 1c and 1d show that it is
sufficient to consider only the contribution
of stars with masses larger than $2$ or $3\, M_\odot$ to compute the
time dependence of the FUV radiation field.

\subsection{Stellar Sources of Ionizing Radiation}

Ionizing radiation from hot stars is considered in the present model
because we are calculating the FUV-ISRF in the neutral gas
and therefore it is necessary to verify that the sampling point is
outside of HII regions. In addition, we use the observations of the
total ionizing photon production rate in the Galaxy to constrain the birth rate
of associations (MW97; see \S 4).

   Let $s(m)$ be the time-averaged main-sequence ionizing photon luminosity 
(in photons s$^{-1}$) of a star of initial mass $m$.  
Values of $s(m)$ were kindly provided by Vacca (private communication) 
 and are plotted in Figure 2a as 
$s_{49}=s/(10^{49}$ photons s$^{-1}$). 
An analytic fit to these data is given in Table 1.
In Vacca, Garmany, \& Shull (1996) the
ionizing luminosities are given as function of the spectroscopic 
mass and the evolutionary mass, but here the 
zero age main sequence mass is used because the IMF (\S 2.1) refers 
to this initial mass.

	The contribution of the different mass ranges 
to the ionizing photon production is shown in Figure 2b.
Again a mass distribution of stars following a PDMF was assumed.
For the adopted distribution (constant SFR and an IMF with power law 
$\Gamma=1.35$, see \S 2.1), stars of about
$30 \, M_{\odot}$ make the maximum contribution. Half of the ionizing
photons are produced by stars above $\sim 55 \, M_\odot$ 
(lifetime $\sim 3.8$ Myr),
and $\sim 95\%$ are produced by stars above $\sim 20 \, M_\odot$.

	The characteristics of the ionizing luminosity from a group
of stars following an IMF that is a cutoff power law can be
described by two quantities (McKee 1989).  The first is the mean
ionizing photon luminosity per high-mass star,
\beq
\avg{s}_h\equiv\frac{1}{\edNsh}\int s(m) d\edNs
	=\frac{\Gamma}{\phi_h}\int s(m)
	\left(\frac{m_h}{m}\right)^\Gamma d\ln m,
\eeq
where the second equation follows from equation (\ref{eq:dNsm2}).
The second quantity is the mean lifetime of an ionizing star,
\beq
\avg{t_{\rm ion}}=\frac{\int s(m) t_{\rm ms}d\edNs}{\int s(m) d\edNs}.
\label{eq:tion}
\eeq
If each star produces $Q(m)=s(m) t_{\rm ms}$ ionizing photons during its
lifetime, then these equations imply that
in a steady state the expected value of
the ionizing luminosity from
a group of stars is
\beq
S=\int Q(m) d\edNs=\edNsh\avg{s}_h\avg{t_{\rm ion}}.
\eeq
For the stellar parameters we have adopted,
$\avg{s}_{h,49}=0.42$ and $\avg{t_{\rm ion}}=4.0\times 10^6$ yr, so that
\beq
\edNsh=6.0\times 10^{-7}\phi_S S_{49},
\label{eq:sfn}
\eeq
where \dNsh\ is measured in yr$^{-1}$ and
$\phi_S$ allows for deviations from our adopted values of
$\avg{s}_h$ and $\avg{t_{\rm ion}}$.  
If $m_u$ differs from our adopted value of 120,
$\phi_S$ changes.  For example, $\phi_S=1.32$ for $m_u=80$ and
$\phi_S=0.88$ for $m_u=150$; a power-law approximation
over this range gives $\phi_S\simeq (120/m_u)^{0.65}$.
By comparison MW97,
who adopted $\Gamma=1.5$ and used a different set of stellar
lifetimes but the same value of $m_u$, found 
$\avg{s}_{h,49}=0.34$ and $\avg{t_{\rm ion}}=3.7\times 10^6$ yr, 
corresponding to $\phi_S=1.32$.

\subsection{Star Formation in the Galaxy}

	What is the relation between the star formation rate
locally and that in the Galaxy as a whole?
Let $\edssh(R)$ be the birthrate of
massive stars ($m>m_h$) per unit area at Galactocentric
radius $R$.  
MW97 found that the spatial distribution of massive star formation
in the Galaxy (omitting the Galactic Center)
can be approximated as $\edssh(R)\propto \exp(-R/H_R)$ for
11 kpc $>R>$ 3 kpc, with $H_R=3.5$ kpc.
The total rate of massive star formation in the Galaxy
is then
\begin{eqnarray}
\edNshT &=& \edssh\int_{3~{\rm kpc}}^{11~{\rm kpc}}
	2\pi \exp[(R_0-R)/H_R] R^2dR\\
	&\equiv &
	\edssh A_{\rm eff},
\label{eq:aeff}
\end{eqnarray}
where $R_0=8.5$~kpc is the distance to the Galactic Center
and $\edssh$ without an argument is the value of the massive
star formation rate at the solar circle. [This definition of
$A_{\rm eff}$ differs by
a factor $\exp(R_0/H_R)$ from that in MW97].
Numerically, the effective area for
massive star formation in the Galaxy is $A_{\rm eff}=530 \, \rm{kpc}^2$.
It should be borne in mind that star formation is observed
to occur beyond 11 kpc, but there are no giant radio HII regions,
and therefore few if any large associations, there.

	Radio surveys of the Galaxy provide an estimate for the total ionizing
photon production rate, $S_T=2.6\times 10^{53}$ photons s$^{-1}$,
good to a factor of about 1.5 (MW97).  We use $S_T$ to set \dssh,
since equations (\ref{eq:sfn}) and (\ref{eq:aeff}) yield
\begin{equation}
\label{ssh}
\edssh=6.0\times 10^{-7}\phi_S\left( \frac{S_{T,49}}{A_{\rm eff}}
		\right) \rightarrow 29.4\phi_S ~~~ \rm{kpc}^{-2} \, 
	\rm{Myr}^{-1} ,
\end{equation}
where we have expressed the result in convenient units.
This result is compared with other estimates in the literature
in Figure 3. We note that the value
$\edssh=38 \, \rm{kpc}^{-2} \, \rm{Myr}^{-1}$ derived in MW97 was obtained
by assuming $\Gamma = 1.5$ and adopting a different set of stellar
lifetimes.  
 
	In the Galaxy as a whole, the birthrate of massive stars
is $6.0\times 10^{-7} \phi_S 
S_{49,T}=0.0156\phi_S$~yr$^{-1}$.  Correspondingly,
the time between core-collapse supernovae in the Galaxy
is $t_{SN} \sim (64/\phi_S)$ yr. 

	The resulting PDMF for $m>m_h$ agrees reasonably well with the
one derived by MS79 from observed data. Our PDMF is systematically higher
than the one derived by Scalo (1986), 
although it is close to the top of his error bars.
This trend can be explained by the fact that our PDMF corresponds to the
mean star formation rate at the solar circle
as determined from the Galactic distribution of emission of ionizing
photons, whereas the observed local PDMF likely
corresponds to the median star formation rate (see \S 4.3). Also, 
massive stars embedded or close to molecular clouds might be missing
in the observed PDMF. 

\section{Associations}

The sources of FUV radiation are the short-lived massive stars that 
generally form
in associations. Therefore, these associations can be considered as
point sources with an FUV luminosity that decreases with time as their massive 
members
die. These associations tend to expand (typical velocities are 
$\sim 1-2$~km s$^{-1}$),
but this effect can be neglected because 
the associations remain small during the time that most of the FUV
radiation is produced---recall that 
a large fraction of 
the FUV radiation that
escapes from the natal cloud escapes during the first 16 Myr (\S 2.2).
The fact that the main contributors to the FUV field are grouped has a
tremendous effect on the temporal behavior of the radiation field. 
For example, the occurrence
of a close association can raise the local radiation field
by orders of magnitude until its massive star population dies. Therefore,
we must model 
the birth-rate and size-distribution of associations in order to determine
the time and space dependence of the interstellar
FUV radiation field. We adopt the MW97 model, 
which explicitly gives the
birthrate of associations as a function of the number of stars  formed in
the association.
The birth-rate distribution in MW97 is derived from observations of HII
regions, and reproduces  the total ionizing luminosity,
the local ionizing luminosity,
and the rate of core collapse supernovae observed in the Galaxy.
Moreover, the observed OB associations in the
solar neighborhood (i.e., those that
are young and large enough to be identified) are 
consistent with the  MW97 model, as we discuss below.
The MW97 model is based on observations of the effects produced by 
Lyman continuum photons radiated by
a small fraction of the stellar population (i.e. the very hot stars).
A larger fraction of the star population contributes to the
FUV radiation field; therefore an additional test to the MW97 model will be
provided here.

\subsection{Evolution of the Luminosity of Associations}

The typical period of time that an association forms stars is
of the order of $15-20 \, \rm{Myr}$ (Blaauw 1964, 1991; Heiles 1990). 
The rate at which
stars form in an association is not continuous but occurs in bursts during
the 15-20 Myr interval. We idealize this situation
by assuming that each association produces 
five bursts, or
generations, each of which has the same number of stars 
 with a stellar mass distribution given by equation (\ref{eq:dNsm}). 
We characterize the size of the association by $\eNsh$, the number of
massive stars that form over the 5 generations.
The stars in each generation are assumed to 
form simultaneously.
As in MW97, 
we assume that these five generations occur with a period
of $t_g=3.7 \, \rm{Myr}$.
At any given time after the birth of an association, the 
luminosities ${\cal L}_{\rm FUV}$ and ${\cal L}_{\rm{H}_2}$ of the 
entire association are the
sum of the luminosities of the stars still living.
However, as discussed in \S 2.2, not all of this UV luminosity can escape
the parent cloud of the association. We estimated in \S 2.2 that for each
generation  the
fraction of the radiation that escapes, $f_{\rm esc}$, goes 
linearly from $1/4$ to one
in a time $t_{\rm obsc}\sim 4$~Myr.
In this case, the effective
luminosity of an association of age $t$ that will eventually form 
a total of $\eNsh$ massive stars is
\begin{equation}
\label{Lasso}
{\cal L}_{\rm FUV, H_2}(\eNsh,t) = \sum_{j=1}^{\caln_g}
 \left\{ f_{\rm esc}[t-(j-1) \,t_g]
\sum_{i\geq i_{\rm live}(j,t)} L_{\rm FUV, H_2}(m_i) \right\}
\ \ .
\end{equation}
Here the first summation is over the $\caln_g \leq 5$ generations or bursts 
that occurred during the association age $t$
[i.e., those for which $t-(j-1) \, t_g > 0$].
The second summation is over all the stars 
born in generation $j$
and that are still on the main sequence at time $t$. 
(The stars in each generation are labeled
in order of decreasing mass, and $i_{\rm live}(j,t)$
labels the most massive star from generation $j$ still on
the main sequence at time $t$.)
In practice, the second 
summation can be stopped at $m< 2 M_{\odot}$ due to the 
negligible FUV luminosities of such stars.

Figure 4 shows the
evolution of the FUV luminosity ${\cal L}_{\rm FUV}(\eNsh,t)$ 
for associations
characterized by generations with  $\eNsh/5=
200$, 20, and 2 high-mass stars (i.e., the associations ultimately
form 1000, 100, and 10 high-mass stars,
respectively).
The decreases in ${\cal L}_{\rm FUV}$ are caused by the deaths of the
most massive stars born in the early generations.
The first generation is born at age $t=0$.
The initial increase is due to the gradual dispersion of the dust
from the natal cloud around the first generation. The final 
decrease is due to the death of OB stars
after the last generation has formed. 
The right vertical axis gives the contribution
to the energy density $U_{\rm FUV}$ 
at a point located $100$ pc away assuming
that there is no intervening extinction 
[i.e., $U_{\rm FUV} = (1/4 \pi r^2 c) \, {\cal L}_{\rm
FUV}/(2070~$\AA$-912$~\AA)].
Figure 4 shows that the maximum luminosity of an association
is attained at about the time the fifth generation dissipates the absorbing
material around it (i.e., about 18 Myr after the first generation
begins to form). At this time, the mean FUV luminosity per high-mass
star formed (in the five generations) is 
$L_{{\rm max},h} \sim 1.6\times 10^4 {L}_\odot$. Note that $L_{{\rm max},h}$
is only about 28\% of the value $\avg{L_{\rm FUV}}_h$ calculated in \S
2.2 since many of the most luminous stars in the first four
generations have died by this point.
The contributions from generations 1 to 5 
to $L_{{\rm max},h}$ are respectively
5.4\%, 7.9\%, 12.4\% 21.6\% and 52.7\%. 
At 20 Myr, the mean FUV luminosity per high-mass
star formed drops to $\sim 1.0\times 10^4 {L}_\odot$.

\subsection{Size Distribution of Associations}
 
	Using compilations of radio HII regions in the Galaxy and
optical studies of HII regions in nearby galaxies, MW97 showed that
the luminosity distribution of giant radio HII regions in the Galaxy
can be fit by a truncated power law of the form
${{\cal N}}_a(>S)={{\cal N}}_{au}[(S_u/S)-1]$, 
where $S$ is the initial ionizing
photon luminosity, ${\cal N}_a(>S)$ is the number of associations
with luminosity at least $S$, $S_u$ is the upper limit of
the distribution, and ${\cal N}_{au}$ is the number of the most luminous
associations, with  luminosities between $0.5 \, S_u$ and $S_u$.

	  The luminosity distribution of OB associations does not
continue as a power law in $S$ to low luminosities because when
the luminosity is dominated by a single star, the distribution is
determined by the IMF.  MW97 assumed that the underlying dependence
on the number of high-mass stars in the association did not change at
this point, however, so that $\caln_a(>\eNsh)=\caln_{au}[(\eNshu/\eNsh)-1]$.
Here $\eNsh$ is the number of high-mass stars in all the generations
of the association, and $\eNshu$ is the number of high-mass stars in
an association in which each generation has an ionizing luminosity
$S_u$  (MW97 argued that the different generations of an association had
comparable ionizing luminosities).
If the lifetime of the association is $t_a$,
the birthrate of associations in which at least $\eNsh$ 
high-mass stars form is then
\begin{equation}
\dot{{\cal N}}_a(>\eNsh)=
\frac{{\cal N}_{au}}{t_a} \left( 
\frac{\eNshu}{\eNsh} - 1 \right).
\label{eq:dna}
\end{equation}
MW97 assumed that associations had 5 generations, each lasting
3.7 Myr, so that $t_a=18.5$ Myr; we shall adopt this value here.
For the Galaxy, MW97 estimate an upper bound 
$S_u=490 \times 10^{49} \, \rm{photons} \, \rm{s}^{-1}$,
which corresponds to $S_u/\avg{s}_h=1180$ massive stars per generation,
or $\eNshu=5900$ massive stars over the life of the association.
This is somewhat smaller than the value $\eNshu=7200$ estimated by MW97
because of the flatter IMF we have adopted here.
The lower limit of the distribution is 
${\cal N}_{*h,l}={\cal N}_{*h,u} \exp[-S_T/({\cal N}_{au} \, S_u)]$,
which is obtained by requiring that the integrated ionizing photon
emission by the entire distribution of associations match the observed
value.
With $S_T=2.6 \times 10^{53} \, \rm{photons} \, \rm{s}^{-1}$ 
and ${\cal N}_{au}=6.1$ (MW97), we find
${\cal N}_{*h,l} \simeq 1.0$.
  
The birthrate of associations per unit disk area is
$\edsa(>\eNsh)=\dot\caln_a(>\eNsh)/A_{\rm eff}.$
At the solar circle the effective area is $A_{\rm eff}\sim 530$ kpc$^2$,
so that the birthrate
of associations per unit disk area is
\begin{equation}
\label{rate_asso}
\edsa(>\eNsh)=
6.2\times 10^{-4} \,
\left(\frac{5900}{\eNsh} - 1 \right)
\ \ \ \, \rm{kpc}^{-2} \, \rm{Myr}^{-1}\ \ \ .
\end{equation}
 
\subsection{The Typical Star Formation Rate is Below Average}

Equation (\ref{rate_asso}) reveals that, on average,
one association (of any size,
${\cal N}_{*h,l}<\eNsh<{\cal N}_{*h,u}$) forms in one $\rm{kpc}^2$ 
each $\sim 0.3 \rm{Myr}$ at the solar circle.
The number of massive stars formed
in a given area $A$ and over a given time interval $\Delta t$
will depend on 
the size $\eNsh$ and number  ${\cal N}_a$ of  associations formed.
For small areas (e.g., $\sim 1 \, \rm{kpc}^2$) large fluctuations are 
expected even if $\Delta t$ is tens of millions of years. 
We shall show that as 
a result, the typical rate of massive star formation $\edsshtyp$
can be significantly
less than the average rate in the Galaxy for small values
of $A\Delta t/(A_{\rm eff}t_a)$.
Since the FUV-ISRF is determined by the average rate of massive
star formation in
a small fraction of the disk over a relatively short time, this
means that the typical value of the radiation field is significantly
less than the average value.

     The number of associations 
${\cal N}_a (>\eNsh)$ with at least $\eNsh$ high-mass stars
that are expected 
to form during a period of time $\Delta t$ inside an area $A$ is
\begin{equation}
\label{NaAt}
{\cal N}_a (>\eNsh,\Delta t , A)=
\frac{ 
\dot{{\cal N}}_a(>\eNsh)
\, A \, \Delta t}
{A_{\rm eff}} \,
\ \ \ .
\end{equation}
Conversely, the largest $\eNa$ associations
formed in an area $A$ and time $\Delta t$, 
can be expected to have at least 
\beq
\label{NhNa}
\eNsh(\eNa)=\frac {{\cal N}_{*h,u}}{
1+ \left(\frac {\displaystyle A_{\rm eff} t_a}
{\displaystyle {\cal N}_{au} \, A \, \Delta t}\right)\eNa
}
\eeq
high-mass stars, where we have used equation (\ref{eq:dna}) to
evaluate $\dot\eNa$.
For example, consider a large sample of sets of associations formed
in $A\Delta t$; the median size of the
largest association in each set will be $\eNsh(\eNa=\frac 12)$ (i.e.,
it is expected that half of the sets of associations formed would have 
an association with $\eNsh$ larger than this value).  
As a numerical example, $A=1$ kpc$^2$ and $\Delta t=25$ Myr,
give $\eNsh(\eNa=\frac 12)=178$, 
which is substantially less than $\eNshu=5900$.

This analysis carries over into comparing the mean and median rates
of star formation in a given area of the Galaxy.
The important point is that, averaged over a finite region of
space and time, the star formation rate is most likely a relatively low
value, but that if one waits long enough or chooses a large enough area,
a very massive association will form in this region that raises the
{\it mean} rate above the {\it median} or typical rate. 
For an area $A$ and time interval $\Delta t$, 
the typical star formation rate can be estimated as the
rate of formation in all but the largest association plus the rate
of formation in the typical largest association,
\beq
\edssh\simeq\frac{1}{A_{\rm eff}}\int_{\eNshl}^{\eNsh(1)}\eNsh d\dot\eNa
+\frac{1}{A_{\rm eff}}\eNsh({\scriptstyle\frac 12})
\dot\eNa[\eNsh(1)].
\eeq
The ratio of this typical rate to the mean rate can be evaluated with
the aid of equation (\ref{eq:dna}),
\beq
\label{sfr2mean}
\frac{\dot\varsigma_{*h,\rm typ}}{\avg{\edssh}}\simeq
\frac {
\ln\left[\frac{\displaystyle\eNsh(1)}{\displaystyle\caln_{*h,l}}\right]
+2\left[1-\frac{\displaystyle \eNsh({\scriptstyle \frac 12})}
{\displaystyle\eNshu}\right]}
{\displaystyle\ln\left(\frac{{\cal N}_{*h,u}}{{\cal N}_{*h,l}}\right)}
\ \ \ .
\end{equation}

For example, if the typical star formation rate is measured by
observing the O and B stars, then we are only probing
a time $\Delta t \sim 25$ Myr, and the typical to mean ratio within 1 kpc is 
$\sim 0.86$.
We show in \S 4 that most of the typical FUV-ISRF originates
from associations closer than $\sim 400$ pc and younger than $\sim 30$
Myr; in this case, the ratio of typical to mean is 0.69.
Moreover, when averaging over very long periods of time, 
about half of the local 
mean FUV-ISRF is produced by very infrequent large associations 
within less than 
$H_*\sim 85$ pc, the observed scale height of OB 
stars.  The ratio of the typical to mean star formation within $H_*$ during
a period of 30 Myr is $\sim 0.34$. 
In other words, for very long periods
of time there are no large associations within $H_*$
and the FUV field is relatively low.  Then, 
sporadically, a large association forms within $H_*$ and raises
the field so much that, even though its duration is brief, it pushes
the mean significantly above the median. 
Properly accounting for this effect, the 
typical FUV-ISRF is about half the mean value!

As a sidenote to this section, it is interesting to compare the
predictions of the MW97 model with observations of nearby OB associations.
In total, about 7 associations younger than $25$ Myr
are observed within 1 kpc from the Sun. Equation (\ref{NaAt}) 
predicts that the 7 associations  will each produce
more than $\sim 40$ high-mass stars.
The completeness of the observed sample is unknown, but
the observations appear consistent with the model. 
MW97 came to the same conclusion by a slightly different approach:
They calculated that there should be 7.3
associations within a kpc of the Sun that currently
have at least 30 high-mass stars (i.e., excluding some 
young associations that have yet to reach their full complement
of high-mass stars) and are younger than 25 Myr.
In summary, the MW97 model for the  OB associations is in good agreement with
the observed associations in the Solar neighborhood.   However, 
the observational bias is such that these
associations are large and
young. A significant contribution to the FUV-ISRF
comes from smaller and/or older associations that 
are less readily detected. 
We shall use the MW97 model to describe these associations as well.

\section{Time dependent FUV-ISRF}

To calculate the evolution of the FUV-ISRF at a given point in the Galaxy,
we assume that
the birthrate of associations per unit volume has a Gaussian distribution
with respect to the  height $z$ above the galactic plane. We take
the scale height of the birthrate to be 
$H_* = 85$ pc, the observed scale height of molecular gas and OB stars 
(Ferri\`ere 1998, and references therein). 
Since the surface number density of young associations is small, the
above distributions must be regarded as probabilistic.
In the plane the distribution is assumed to be uniform (random) since
almost all the FUV-ISRF originates within $\sim$1 kpc, the scale length 
for dust attenuation of FUV, and therefore radial
gradients in the birthrates are negligible.
Large scale patterns of star formation, such as spiral arms,  are present
in the Galaxy, but it is not necessary to include the corresponding
spatial variation in the star formation rate since
the FUV field at a given point is mainly determined by 
the population of massive
stars in a rather small region around it (within $\sim $ 0.4 kpc).
Our adopted star formation rate is based on the global model of MW97
and is therefore an average over arm and interarm regions of the
Galaxy.  Since the fluctuations in the ISRF we calculate are purely
statistical, they are a lower limit to the fluctuations that occur as
a point moves between arm and interarm regions.

Figure 5 shows the projection on the plane of the 
positions of associations formed in a period
of 500 Myr (left panel) and 30 Myr (right panel).
The birthrate corresponds to the average value at the solar circle.
Note that although there are more small associations than large associations,
there are equal total numbers of stars in small and large associations.
For a point at the center, most of the FUV radiation comes from
within the circle. The FUV flux varies as function of
position (or time), depending on whether the given point is close
to large associations.

The distribution of attenuating dust with respect to $z$ is assumed
to follow the distribution of HI as given by Dickey \& Lockman (1990), 
i.e., 
\begin{equation} 
\label{nneut}
n_{\rm H} \, (z) =  0.566 \, [ \, 0.69 \, e^{-(z/127\,\rm pc)^2} +
 0.189 \, e^{-(z/318\,\rm pc)^2} +
 0.113 \, e^{-|z|/403\,\rm pc}\, ]\,\rm{cm}^{-3}  \, \,  .
\end{equation}  
Note that $n_{\rm H}$ is the space-averaged number of H atoms in a large
volume that might contain a significant amount of hot gas
(McKee \& Ostriker 1977).
The molecular gas is not considered due to its small filling factor.
The dust optical depth to a given association is then
assumed to be proportional to the column density of H atoms, $N$(HI), along the
line of sight. We adopt the physical properties of dust proposed by
Witt and Gordon (1999) to estimate the average absorption and scattering
by dust grains in the $912-1100$~\AA band and in the $1100-2070$~\AA band.
The scattering and absorption optical depths are given by
$\tau_{s,a}=\kappa _{s,a} N$(HI), where the values of
$\kappa_{s,a}$ are given in Table B1.
The assumption of a constant extinction per unit column density is
justified in statistical studies of the
expected radiation field, but this approximation is not
useful in calculating the contribution of a particular source because the
extinction per unit column density is known to depend
on the particular line of sight.
Scattering removes photons from the direct line of sight to the
associations, but adds photons from other directions.
The effect of scattering is considered in Appendix B.

The contribution of an association labeled "$i$" to the radiation density 
at a distance $r$ can be approximated as
\begin{equation}
\label{uasso}
u_{\rm band} (i) =
\frac{1}{4 \pi r^2 c} \, \frac{{\cal L}_{\rm band}(i)}{\Delta_{\rm band}}  
\, e^{-\tau_{\rm band}(i)} \ \ \ ,
\end{equation}
where $c$ is the light speed, and
${\cal L}_{\rm band} (i)$ is the effective luminosity (eq. \ref{Lasso}) 
of the association labeled $i$, which has
$\eNsh$ high-mass stars and whose age is $t$. 
The term $e^{-\tau_{\rm band}}$ accounts for both absorption and scattering
(see Appendix B).

At a given time $t_{\rm sim}$ of the simulation, 
the total energy density in each band $U_{\rm band}(t_{\rm sim})$ 
is the sum of the contributions $u_{\rm band} (i)$ from all 
the associations formed between $t=0$ and $t=t_{\rm sim}$.
Due to the $1/r^2$ term in equation (\ref{uasso}),
the resulting radiation density $U_{\rm band}$ at a
given point is very sensitive to the spatial distribution of the 
associations; in particular,
a nearby luminous source can dominate the radiation field. The effect
of extinction is very important for distant sources.  Without extinction
the contribution of FUV sources in a disk grows logarithmically with distance;
with extinction the total FUV field comes from sources within a few optical
depths. 
In practice it is sufficient to consider only associations within about 1 kpc
for conditions at the solar circle of the Galaxy.
Additionally, old associations (with ages exceeding the lifetime of 
stars with masses $\sim2-3$ 
$M_\odot$) make a negligible contribution to $U_{\rm FUV}$.
These two facts are useful in reducing the number of associations that
must be tracked in simulations to obtain an accurate accounting of 
the time dependence of the radiation field at a given point.

\subsection{The Mean FUV-ISRF}

Before presenting the results from simulations, it is useful to estimate
the mean FUV energy density assuming a smooth emissivity per unit volume 
corresponding to the PDMF and a smooth spatial distribution; hereafter the 
``continuous method". This approach is necessary because
the ``discrete method" used in the simulation (assuming that FUV sources 
are highly discrete in space and time) is useful to get the median 
value of the radiation 
field, but is very inefficient in calculating
the mean value because of the large integration times needed to obtain a good
sample of rare events. However, the ``continuous method" does not show the
interesting time dependence  of the FUV-ISRF. 
Knowledge of the ratio between
the median and the mean value of the radiation field is important because
low resolution
observations (for example, in external galaxies) that average over
huge spatial volumes can provide an estimate
of the mean FUV-ISRF, but the interstellar gas is subject typically to the
median FUV-ISRF.

The mean energy density $<U_{\rm band}>$ in the plane can be
calculated by adding
the diluted extincted and scattered radiation from a smooth distribution of 
stars following a PDMF. 
Let
\beq
n_*(m,{\bf r})\equiv\frac{d^2\caln_*(m,{\bf r})}{dV d\ln m}
\eeq
be the PDMF per unit volume at a point ${\bf r}$.  
The mean energy density of radiation at the origin is then
\begin{equation}
\label{meanU}
\avg{U_{\rm band}}=
\frac{1}{c \Delta_{\rm band}}
\int
\frac{e^{-\tau_{\rm band}({\bf r})}}
{4\pi r^2} \, dV\,
\int_{m_{l}}^{m_{u}} n_*(m,{\bf r}) L_{\rm band}(m)\, d\ln m
\ \ \ ,
\end{equation}
where  $L_{\rm band}(m)$ is the mean luminosity of a star of
mass $m$ after subtracting the
fraction lost in the parent cloud.  As shown in Appendix B,
we approximate the optical depth, including both absorption and 
scattering, as $\tau_{\rm band}=\tau_a+\tau_s(1-{\cal R})$,
where ${\cal R}$ is 
a factor that is somewhat less than unity for scattering
that is strongly peaked in the forward direction.

     Since the stars that dominate the FUV flux are relatively
nearby, we can ignore the Galactic radial gradient in the density of
stars and
express the volume density of stars in terms of the surface density
as 
\beq
n_*(m,{\bf r})=\essm \, D(z)
\eeq
where $z$ is the distance normal to the plane and, 
for a Gaussian distribution,
\begin{equation} 
D(z)=\frac{e^{-(z/H_*)^2}}{\sqrt{\pi} H_*}\ .
\end{equation}   
Recall that for a constant birthrate, the PDMF for stars younger
than the age of the disk is related to the
star formation rate by $\essm=\edssm t_{\rm ms}(m)$.
Equation (\ref{meanU}) can then be rewritten as
\begin{equation}
\label{<Uband>}
\avg{U_{\rm band}}=
\frac{4\pi{\cal {J}}_{\rm band} \,  \Sigma_{L_{\rm band}}}
{c\Delta_{\rm band}}
\ \ \ ,
\end{equation}
where
\begin{equation}
\label{<Slband>}
\Sigma_{L_{\rm band}}=
\int_{m_{l}}^{m_{u}} \essm \, L_{\rm band}(m) \, d\ln m
\end{equation}
is the mean rate of
emission of radiation (after subtracting the absorbed emission in 
the
natal clouds) per unit disk area in the band, and
\begin{equation} 
\label{calJ}
4\pi {\cal{J}}_{\rm band}=
\int\frac{
e^{-\tau_{\rm band}(\vecr)} \, D(z)}{4\pi r^2} \, dV
\ \ \  
\end{equation}
is $4\pi$ times the mean intensity due to unit
FUV emissivity in the disk.

	At the solar circle $\Sigma_{L_{\rm H2}} 
\sim 2.5 L_\odot/\rm{pc}^2$ and
$\Sigma_{L_{\rm FUV}} \sim 10.9 L_\odot/\rm{pc}^2$. 
Numerical integration of equation (\ref{calJ}), taking into account multiple
scattering (Appendix B), gives respectively 
0.88 and 1.19 for the value of $4\pi \cal{J}_{\rm band}$ in the 
H$_2$ and FUV$-$H$_2$ bands. From
equation (\ref{<Uband>}) one obtains 
$\avg{U_{\rm H2}}=15.7 \times 10^{-17} \rm{erg} \, 
\rm{cm}^{-3} \, {\rm \AA} ^{-1}$, 
$\avg{U_{\rm FUV-H2}}=13.8 \times 10^{-17} \rm{erg} \, \rm{cm}^{-3} \,
{\rm \AA} ^{-1}$, and
$\avg{U_{\rm FUV}}=14.2 \times 10^{-17} \rm{erg} \, 
\rm{cm}^{-3} \, {\rm \AA} ^{-1}$.
Additionally, 
$\avg{U_{1400{\rm \AA}}}=
14.4 \times 10^{-17} \rm{erg} \, \rm{cm}^{-3} \, {\rm \AA} ^{-1}$.
Hereafter, we denote these mean values as $\avg{U_{\rm
band}}_{\infty}$ since they correspond to
the mean energy density for an infinite period of time.

    To clarify how different regions contribute to the mean radiation field,
consider a simple model consisting of two regions: the first
is a sphere of radius
$H_*$ centered at a point in the midplane, and the second is a zero 
thickness disk that covers the midplane but excludes a radius $H_*$
around the point in question.
The volume number density per unit mass of FUV sources in the sphere is 
$n_*(m) = \essm/2 H_*$, and the surface density in the disk is 
$\essm$. The UV extinction is assumed to be $e^{-k\, r}$, where
$k=[\kappa _a +\kappa _s(1-{\cal R})]n_{\rm H} 
\simeq 7\times 10^{-22}$ cm$^{-1}$
(see Appendix B).
Then,
\begin{equation}
\avg{U_{\rm band}}\simeq
\frac{1}{c \Delta_{\rm band}}
\,
\,\biggl [
\frac{1}{2 \, H_*}
\int_{0}^{H_*}
e^{-k\, r} dr \,
+
\frac{1}{2}
\int_{H_*}^{\infty}
\frac{e^{-k\, r}}{r} dr
\,\biggr ]
\,
\int_{m_{l}}^{m_{u}} \essm \, L_{\rm band}(m) \, d\ln m
\ \ \ .
\end{equation}
The two terms in the bracket correspond to $4\pi{\cal{J}}$ 
in this approximation. 
Since $k \, H_* \ll 1$, the contribution of
the sources in the sphere to the function  $4\pi{\cal{J}}$ is $ \sim 1/2$. This
term represents sources out to a distance of order the scale height $H_*$ of
the disk.
Beyond $H_*$, the contribution of the sources in the disk 
is $\frac{1}{2} \, E_1(k \, H_*)$,
where $E_1$ is the first exponential integral. Since 
$k \, H_* \simeq 0.18$ and
$E_1(0.18) \simeq 1.3$,
the second term is about $0.65$.  In other words, in calculating
the mean field, we find that roughly half the radiation comes from
sources closer than $\sim H_*=85$ pc, and half the radiation arises
from more distant sources in the disk.
However, as shown
in equation (\ref{sfr2mean}), the typical
number of FUV sources inside of 85 pc
is much less than the mean value. Therefore, most of the time, the FUV 
radiation field is expected to be well below its mean value.
We also see that beyond $r>k^{-1}\sim 0.46$ kpc the 
contribution to the FUV field
rapidly falls off due to attenuation of the distant FUV flux by dust. 

	In order to quantify the simulation time necessary to get a good
estimate of $\avg{U}$,
we calculate the accumulated {\it typical} FUV radiation field 
during a period of time $t_{\rm sim}$,
\begin{equation}
\label{eq:utsim}
\avg{U_{\rm band}}_{t_{\rm sim}}=
\frac{1}
{c\Delta_{\rm band}}
\int \frac{e^{-\tau_{\rm band}(\vecr)} \, D(z)}{4\pi r^2}\, dV
\int \varsigma_{*,\rm typ}(m,A=\pi r^2,\etsim) L_{\rm band}(m) \, d\ln m
\ \ \ . 
\end{equation}
Here the typical PDMF is
\begin{eqnarray}
\varsigma_{*\,\rm typ}(m,A,\etsim) 
		& = & \int_0^{{\rm min}(\etsim,\, t_{\rm ms})}
		\dot\varsigma_{*\rm typ}(m,A,\etsim) \, dt\\
		& = & \left\{\frac {
		\ln\left[\frac{\displaystyle\eNsh(1)}
		{\displaystyle\caln_{*h,l}}\right]
		+2\left[1-\frac{\displaystyle \eNsh({\scriptstyle \frac 12})}
		{\displaystyle\eNshu}\right]}
		{\displaystyle
		\ln\left(\frac{{\cal N}_{*h,u}}{{\cal
		N}_{*h,l}}\right)}
		\right\}
		\edssm{\rm min}(\etsim,\, t_{\rm ms})\ \ \ ,
\end{eqnarray}
where the typical maximum value of $\eNsh$ in the area $A$ during
time interval $\Delta t=\etsim$ is given by equation (\ref{NhNa}).
This equation applies only if $A\etsim$ is large enough that
$\eNshtypmax>\caln_{*h,l}$; for smaller values of $A\etsim$,
$\varsigma_{*\,\rm typ}(m,A,\etsim) =0$.

Note that in computing $\avg{U_{\rm band}}_\etsim$ 
for a given spatial point, the area
of simulation $A_{\rm sim}$ must be chosen so that 
$\tau_{\rm band}\gg 1$ at the outer
boundary $r_{\rm max}$ (see eq. (\ref{eq:utsim})). Otherwise, the simulation
area would not include significant contributors to the UV flux at the
central point.  This is equivalent to $r_{\rm max}\gg k^{-1}$, which for
our standard values of $n_{\rm H}$ and $\kappa_{a,s}$ results in $r_{\rm max}
\gg 0.46$ kpc.  Assuming a square simulation area, this
corresponds to $A_{\rm sim} \gg
0.85$ kpc$^2$.  
Given that $A_{\rm sim}\simeq 5$ kpc$^2$,
the simulation area is no longer a factor in $\avg{U}$; only the 
simulation time enters. 
Figure 6 shows the ratio $\avg{U_{\rm band}}_{t_{\rm sim}}/
\avg{U_{\rm band}}_{\infty}$ as
function of the time $t_{\rm sim}$. 
Extremely long times are required
to get a good estimate of $\avg{U_{\rm band}}$ from the simulations. 

\subsection{Simulation in the Galactic Plane at the Solar Circle}
 
Over a sufficiently large area $A_{\rm sim}$ to include all possible
contributions to the FUV field not severely attenuated by dust, 
the associations form 
following the truncated birthrate law given by equation (\ref{rate_asso}).
We assume a constant interval of time $\Delta t_{a,\, \rm form} $ between the 
formation of
associations in the simulation area $A_{\rm sim}$. Equation 
(\ref{rate_asso}) gives this interval as 
\begin{equation}
\label{Delta_t}
\Delta t_{a,\, \rm form}= \frac{ A_{\rm eff}t_a}
{{\cal N}_{au} \, \left( \frac{\displaystyle{\cal N}_{*h,u}}
{\displaystyle{\cal N}_{*h,l}} -1 \right) \, A_{\rm sim}}
\ \ \ .
\end{equation}
Substituting ${\cal N}_{au}=6.1$, $t_a=18.5$ Myr,
${\cal N}_{*h,u}\, =\, 5900$ and 
${\cal N}_{*h,l}\, =\, 1$ (see \S 3.2), we obtain
$\Delta t_a \, =514\ (A_{\rm eff}/A_{\rm sim})$ yrs.  At the solar circle
$A_{\rm eff}=530$ kpc$^2$ and we simulate a disk area $A_{\rm sim}=5$ kpc$^2$,
so that an association forms in $A_{\rm sim}$ every $5.5\times 10^4$ years.

   In the simulation, the position of each association is 
taken to be completely 
random in the plane $(x,y)$,
but in the vertical direction $(z)$ the random distribution is weighted by a
Gaussian profile of scale height $H_* = 85$ pc.
The masses of the stars in each generation of each association are chosen  
using equations (A1) and (A2).
The adopted gas distribution from which we derive the dust
distribution is given by equation (\ref{nneut}), and the 
methods for including dust absorption and scattering are given in Appendix B.
To avoid boundary effects, the evolution of $U_{\rm band}$ is studied only in 
the central part of $A_{\rm sim}$.

	Figure 7 shows the time evolution of $U_{\rm FUV}$ at 
a point located in the 
center of simulation area and in the galactic plane.
The radiation field energy density $U_{-17}$ is expressed in units of 
$10^{-17} \rm{erg} \, \rm{cm}^{-3} \, {\rm \AA} ^{-1}$. 
We note that Habing (1968)
found that the local interstellar radiation field is approximately
$U_{\rm FUV}^{\rm H}=4.6 \times 10^{-17} \rm{erg} \, \rm{cm}^{-3} 
\, {\rm \AA} ^{-1}$. 
The first association in the simulation volume forms at $t_{\rm sim}=0$.
Although difficult to see in Figure 7, it takes  
$\sim$ 40 Myr to build up the FUV-ISRF
once star formation commences.
After this time, a ``steady state'' is achieved with
large time fluctuations of $U_{-17}$.  
 In addition, large spatial
gradients are derived
when $U_{-17}$ is evaluated at different points in the plane at a given 
instant of time.
Note that although the probability is small, a massive star can form 
in a small association (see Appendix A). These small associations 
are abundant, and many of the HII regions encompassing the point are produced
by these unusually massive stars in  small associations.

The differences in the behavior of the radiation field
in the H$_2$ band [$912-1100$~\AA] and in the FUV$-$H$_2$ band 
[$1100-2070$~\AA]
are shown in Figures 8a,b. 
Note the high concentration of points around the median value,
which corresponds to $U_{\rm{H2}} \sim U_{\rm{FUV-H2}}$.
The ordered sequence of points outside the crowded region in Figure 8b 
corresponds to periods of time when the radiation field is completely
dominated by one association. In Figure 8b, 
two larger excursions to high values
of $U$ have been connected by lines in order to show how the field evolves
as generations form and stars die. 
The large fluctuation that starts at
$t_{\rm sim}=1875$ Myr is produced by an association at 190 pc with
2830 high-mass stars. 
The fluctuation that starts at $t_{\rm sim}=3975$ Myr
is produced by an association at 32 pc with 440 high-mass stars.
The main trend seen in Figure 8b is that, as $U_{\rm{FUV}}$ increases, the
ratio $U_{\rm{H_2}}$ to $U_{\rm{FUV-H_2}}$ increases. This expected
correlation arises because high $U_{\rm{FUV}}$ is often associated with
massive (and therefore hot) nearby stars.   

 Figure 8c shows the fraction of $U_{\rm band}$ 
in the [$1100-2070$\AA] band that comes to the observer as 
scattered light (see Appendix B), as function of $U_{-17}$.
The median value of $U_{\rm scatt}/U_{\rm tot}$ is $\sim 0.18$. 
We discuss this median value in comparison with observational data in
\S 5.

Figure 8d shows for the FUV band the fraction of total field $U_{-17}$ 
that comes to the observer from the source that provides the larger contribution.
For values of $U_{-17}$ around the median ($\sim 7$) the dominant source
provides in average 20\% of the energy density $U_{-17}$, but the values range
from 10\% to 50\%. For large values of $U_{-17}$ the radiation field
is totally dominated by only one source. The distribution of points in 
Figure 8d indicate the degree of anisotropy of the field as function of $U_{-17}$.

In order to calculate the mean and median values of the FUV radiation 
field in the plane, a very long simulation was performed with
$t_{\rm sim} \sim 57$ Gyr.
Excluding the first 400 Myrs,
Figure 9a shows the corresponding distribution of 
the ``fraction of occurrence''
of $U_{\rm band}$ values. The fraction of occurence is defined as
$-dP(>U)/d\, ln U$, where $P(>U)$
is the fraction of the simulation time that 
the point has an FUV energy density exceeding $U$. The fraction of occurrence
can be thought of 
approximately as the probability of lying within a factor
of 1.6
of $U_{\rm band}$.
The arrows indicate the mean and median values of the distributions.
The two arrows for each mean value indicate the value including (the higher)
or omitting (the lower)
the time steps when the point is inside an HII region.
The difference with and without HII regions is very small for the 
median because a given point is in an HII region
only about 2\% of the time. 
The difference for the mean is larger because $U_{\rm band}$ is
very high when the point is in a HII region.  
The expected mean value of $U_{\rm band}$ (including HII regions)
from equation \ref{eq:utsim} and Figure 6 
at 57 Gyr of simulation is $\sim 0.9$ 
of the theoretical asymptotic value calculated in \S 4.1. 
Therefore, even when 57 Gyr of simulation are considered, the simulation does 
not provide an accurate estimate of the mean values because
extremely rare events with large
associations very close to the point in question
tend to be missing. 
On the other hand, the mean value of $U_{\rm band}$ excluding HII regions
is closer to its asymptotic value
since regions of space very near the association are excluded by the
requirement that the point be outside the HII region.
Note that the energy density in the FUV band is never lower than 
$U_{\rm FUV,\,min}=
2.4  \, \times \, 10^{-17} \rm{erg} \, \rm{cm}^{-3} \, {\rm \AA} ^{-1}$;
in the $H_2$ band the minimum value is 
$1.5  \, \times \, 10^{-17} \rm{erg} \, \rm{cm}^{-3} \, {\rm \AA} ^{-1}$. 
These minimum values correspond to
the periods of time when no young associations are present in the central
region of the simulation volume.

Figure 9b shows the probability $P(>U)$ that, at a given instant in the 
simulation, the FUV energy density is greater than $U$.
As stated above, the energy density in the FUV band is never lower than
$U_{\rm FUV,\,min}$
in the simulation, and therefore $P(>U_{\rm FUV,\,min})=1$.
The curves labeled 1 Gyr, 4 Gyr, and 56 Gyr, correspond to
the probabilities $P(>U)$ calculated from simulations extending
over time intervals $\Delta t =$ 1 Gyr, 4 Gyr, and 56 Gyr,
respectively.
Note that the peak energy density decreases as $\Delta t$
drops. Even for a simulation time of 56 Gyr, events
with $U_{-17}>4000$ are missing. For 
$U_{-17} > 10$,
the probability $P(>U)$  
can be fitted by the power law $P(>U) \simeq 10  \, U_{-17}^{-3/2}$.
For $U_{-17}>300$ a better fit is $P(>U) \simeq 15  \, U_{-17}^{-3/2}$.
This dependence arises from the fact that 
the volume of the sphere around a source of luminosity ${\cal L}$ 
where the energy density exceeds $U$ is proportional to $({\cal L}/U)^{3/2}$.
The volume filling factor $\phi(>U)$ of these spheres therefore scales as
$U^{-3/2}$. 
For $\phi << 1$, we have $P(>U)=\phi(>U)$, so that in this case
$P(>U) \propto U^{-3/2}$. This relation
is not exact because, for example, extinction is neglected, 
and the overlapping of spheres from different sources is not considered. 
However, the fit becomes better at high values of
$U$ since these effects become unimportant
there, since the volume around the sources 
is then small and consequently overlapping and extinction can be neglected.
Figure 9b also shows the typical time $t_U$ (right vertical axis)
between two events that
raise the energy density to $>U$ at a fixed point in space.
It can be shown (Hollenbach et al. 2002)
that $t_U = t_{\rm FUV} / \phi(>U)$, where $t_{\rm FUV}$
is the typical time that a source (an association) emits strongly 
in the FUV band (see Fig. 4).  
The power law $t_U = t_{\rm FUV} / P(>U) = 2 \, U_{-17}^{3/2}$ Myr assumes
$t_{\rm FUV} = 20$ Myr and is good for 
$U_{-17} > 10$.
For $U_{-17}>300$ a better fit is $t_U = 1.3  \, U_{-17}^{3/2}$ Myr.
From this relation one can see that for $t_{\rm sim}= 57$ Gyr only the 
events with $U_{-17} < 1000$ are well sampled.
 
For the FUV band, Figure 10 shows the evolution 
of the mean $\avg{U}$ and the median $U_{\rm med}$ 
radiation fields
(including HII regions) normalized to the theoretical value 
$\avg{U}_{_{\infty}}$ derived in the previous section. 
The calculation of these averages is over the period of time $t_{\rm sim}$
starting at time zero when associations begin to form in the simulation area.
The theoretical ratio $\avg{U}_{t_{\rm sim}}/{\avg{U}_{\infty}}$ 
(plotted in Fig. 6) is also shown in Figure 10 for comparison. 
Note that the median initially grows for $\sim 200$ Myr 
and then stabilizes, showing
that the simulation is very efficient for calculating the 
median value of the field.
The amplitude of fluctuations tends to decrease with time since the 
averages are made over the entire
history of the simulation.
The  large fluctuation in 
$\avg{U_{\rm sim}}/\avg{U}_{{\infty}}$ at $\sim 100$ Myr is due to an 
association of $\sim 300$ high-mass stars
(the most massive star
with a mass of $86 \, M_{\odot}$) born at $t_{\rm sim} \sim 60$ Myr 
at a distance
of $\sim 120$ pc (with $z \sim -80$ pc) from the central point. 
At $t_{\rm sim} \sim 75$ Myr, 
$U \sim 25 \, \times \, 10^{-17} \rm{erg} \, \rm{cm}^{-3} \, 
{\rm\AA} ^{-1}$,
mainly due to the FUV flux from this association.
The second large fluctuation at $\sim 600$ Myr is due to an
association of $\sim 41$ 
high-mass stars born at $t_{\rm sim} \sim 540$ Myr at a distance
of $\sim 30$ pc (with $z \sim +23$ pc) 
from the central point. 
This association is
relatively small, but its  most massive star has a mass of 
$114 \, M_{\odot}$. 
At $t_{\rm sim} \sim 560$ Myr, 
$U \sim 80 \, \times \, 10^{-17} \rm{erg} \, \rm{cm}^{-3} \, 
{\rm\AA} ^{-1}$,
mainly due to the FUV flux from this star.
The large fluctuation at $\sim 21$ Gyr is due to a very rare event: an
association of 3120 high-mass stars at a distance
of $\sim 35$ pc (with $z \sim -9$ pc) 
that raises the FUV radiation field to a peak value of $U_{-17} \sim 3200$.
The radiation field has
$U_{-17}>1000$ for over 20 Myr, enough to raise 
$\avg{U_{\rm sim}}$ over $\avg{U}_{{\infty}}$ even when the average is
made over 21 Gyr.
  
As stated in \S  3.3, the star formation rate is expected to fluctuate
during the simulation. Figure 11 shows the evolution
of \ssh , the surface (number) density of high-mass stars in the simulation.
The thick line
corresponds to the average surface density over $A_{\rm sim}$ 
(i.e. 5 $\rm{kpc}^2$), whereas the
thin line corresponds to the average within 500 pc from the central point.
As expected, the fluctuations are larger for the  surface density 
within 500 pc.
For a simulation interval between 0.1 to 57 Gyr, the mean value of 
\ssh \  over $A_{\rm sim}$ is $\sim 540 \ \rm{kpc}^{-2}$, and within 500 pc
is $\sim 530 \ \rm{kpc}^{-2}$. Due to the large simulation time, 
these mean values 
of \ssh \ are similar and correspond to a SN II every 62 yrs in the whole 
galaxy (we take $A_{\rm eff}=530 \ \rm{kpc}^2$).

\section{The Local FUV Radiation Field: Observations and Theory}

\subsection{The Diffuse FUV Radiation Field}

Observationally, it has been established that diffuse radiation
coming from regions where no sources are observed (i.e. far from
the regions of recent star formation) is a minor component of the FUV-ISRF.  
This diffuse component arises primarily from large angle scattering
by dust grains of FUV photons from Galactic sources, with a smaller
contribution from extragalactic sources.
Measurements of the diffuse FUV
radiation field from these regions place an upper limit of
$200-400 \, \rm{photons}\, \rm{s}^{-1}\, \rm{sr}^{-1}\, {\rm \AA}^{-1}$
in the [$912-1200$~\AA] band 
(Murthy et al. 1991; Bowyer 1991; Henry et al. 1980).
This upper limit corresponds to a radiation density
$U_{\rm diffuse} \sim 1.7 - 3.4 \times 10 ^{-18}  
\rm{erg} \, \rm{cm}^{-3}\,$\AA\
and represents less than $10\%$ of the estimates of local FUV energy density.
By comparing a map resulting from summing  the fluxes from
individual FUV bright stars (Gondhalekar, et al. 1980)
to a map constructed from large angle detections, Henry (1991)
demonstrated that the FUV radiation field is dominated by the
direct (or forward scattered) flux from relatively nearby FUV-producing stars.
In the previous section we presented model results that showed
that the median value of $U_{\rm scatt}/U_{\rm tot}$ is $\sim 0.18$, which
is higher than the above estimates of 
$U_{\rm diffuse}/U_{\rm tot}\leq 0.1$.  
This difference arises because the ``diffuse field'' $U_{\rm diffuse}$ 
comes from 
regions in the sky far from strong stellar sources, whereas
the scattered field
$U_{\rm scatt}$ refers to all scattered flux, including that
scattered through small angles. In fact, we find (Appendix B) that
about 1/3 of the once-scattered radiation from a source comes from lines
of sight within a cone of five degrees from the observer to the source.
Taking this into account, we conclude that
our estimate of the scattered flux is
consistent with the observed diffuse field. 

\subsection{The Observed FUV Radiation Field in the Solar Vicinity}

Neglecting diffuse radiation, the spectral energy density $U_\lambda$ at 
wavelength $\lambda$ in the immediate solar vicinity 
can be directly measured by adding the observed flux 
$F_{\lambda}(i)$ of all the FUV stellar sources $i$.
By including all the data in the S2/68 Survey (more than 50,000 stars), 
Gondhalekar et al. (1980) estimated the flux at several wavelengths in the
FUV band. At $1400$\AA\ the estimated energy density is about
$5 \times 10^{-17} \rm{erg} \, \rm{cm}^{-3} \, {\rm\AA} ^{-1}$.
Habing (1968), like Gondhalekar, added the fluxes from individual stars to 
estimate $U_\lambda$ in the immediate solar neighborhood.
The estimated value for $U_{\rm 1400~\AA}$ was 
$4.2 \times 10^{-17} \rm{erg} \, \rm{cm}^{-3} \, {\rm \AA} ^{-1}$  
when the direct light from all the bright stars B9 and earlier were included. 
For the extreme case of albedo 0.9, the Habing's value with diffuse light
included was $U_{\rm 1400~\AA} = 
7.5 \times 10^{-17} \rm{erg} \, \rm{cm}^{-3} \, {\rm \AA} ^{-1}$.
However, this value depended on the assumptions 
made to infer the FUV flux for each object from its visual 
magnitude $m_V$ and its color excess $E(B-V)$.    
Jura (1974)  recomputed $U_\lambda$ near 1000~\AA\ and obtained a value
$\sim 7 \times 10^{-17} \rm{erg} \, \rm{cm}^{-3} \, {\rm\AA} ^{-1}$ 
by using updated calculations of stellar atmospheres and including all 
stars B5 and earlier listed in the {\it{Catalogue of Bright Stars}}. 
Utilizing a different technique, Henry et al. (1980)
have directly measured the local $U_\lambda$ by taking large beam ( $12^o \, 
\times 12^o$) observations over about one third of the sky and
extrapolating to obtain the unobserved portion.  Henry et al. found  
that the average energy density is 
$5.8 \times 10^{-17} \rm{erg} \, \rm{cm}^{-3} \, {\rm\AA} ^{-1}$ 
in the band [$1180-1680$~\AA], and  $5.3 \times 10^{-17} \rm{erg} 
\, \rm{cm}^{-3} \, {\rm \AA} ^{-1}$  at 1400~\AA.  Draine (1978)
compiled a number of the observations and models of that epoch,
and did a fit to some of them (primarily Henry et al (1977),
Witt and Johnson (1973), Belyaev et al (1971), and Jura (1974)) to
obtain what has become another standard (besides Habing) field,
sometimes called the ``Draine field''.  Averaged over the FUV band,
it is approximately 1.7 times higher than the Habing field, i.e.,
the Draine field has $G_0=1.7$.
A summary of these estimates is given in Table 2.

The asymmetry of nearby FUV point sources results in a non-isotropic
local FUV field.   Most, if not all,  OB stars are born in
associations,  and there are few associations that are young and close
enough to produce a significant contribution to the FUV field.
A sample of the nearby OB associations reveal  about 7
associations within a kiloparsec of the Sun (see \S 3.3). The distribution
of their distances above and below the galactic plane is
quite symmetric, but their distribution in the plane is totally asymmetric.
When the observed FUV field is plotted in Gould coordinates
(Galactic Coordinates
tipped by about $19^o$), a clear excess toward the active section of the Gould
Belt is apparent (Henry et al. 1980).
Even when our simulations do not consider correlation between
the positions where coeval associations born, the results in Fig 8d
shows that because FUV sources form in groups, FUV field is in 
general non-isotropic, in particular when the radiation field is high.

\subsection{The Mean and Median FUV Radiation Field at the Solar Circle}

Complementing these more or less direct determinations of $U_\lambda$ 
in the immediate solar neighborhood are the 
estimates, such as ours,
 obtained by modeling a spatial distribution of the relevant 
spectral type stars and the dust. 
The advantages of modeling are that it  
corrects for the incompleteness in the catalogs, it is useful to quantify 
the variations from different distributions, and it generalizes to
other locations than the immediate solar neighborhood. 
The  disadvantage is the 
necessity to introduce simplifying assumptions such 
as smooth dust distributions 
and prototype stellar luminosities. However, the ``direct" procedure is only 
useful to give a snapshot of the local FUV-ISRF at the current time, 
whereas the theoretical models 
permit the study of the temporal and spatial behavior of the UV field
in the interstellar medium, which is the main subject of the 
present paper. 

In the previous section, the mean FUV field was calculated assuming the model 
ingredients, but, for the sake of comparison,
we reproduce here one of the classical calculations of Habing. 
The ``continuous approximation" in the Habing (1968) 
model includes the contribution 
of 10 associations and 577 field stars between B0 and B5. 
The FUV emission
of these sources is assumed to be uniformly distributed in the plane 
(the field stars in 1 $\rm{kpc}^2$ and the associations within 1 kpc).
The distribution is proportional to
$\exp[-(z/120\ {\rm pc})^2]$ in the vertical direction.
Habing assumed a homogeneous plane-parallel layer of dust of thickness 150 pc
producing an extinction of 3.5 mag kpc$^{-1}$.
By integrating over this continuous medium (there are neither stars nor 
associations, but an equivalent emitting medium), he obtained 
what we term the {\it observed-source mean} 
radiation energy density
at $1400\AA$ of 
$\avg{U_{1400\AA}}=3.4 \times 10^{-17} \rm{erg} \, \rm{cm}^{-3} \, \AA ^{-1}$.
(This value assumes that the dust has zero albedo, so there is
no scattered light. For an albedo of 0.9, Habing obtain 
$\avg{U_{1400\AA}}=5.6 \times 10^{-17} \rm{erg} \, \rm{cm}^{-3} \, \AA ^{-1}$).
The distinction between Habing's observed-source mean and our mean is
that ours is a mean based on the star formation rate
in the entire Galaxy, whereas Habing's is based on the observed FUV sources
in the solar vicinity.
The surface emissivity $\Sigma_{L_{\rm FUV}}$ in Habing's  model
is about 1/3 of the value in our model; this is plausible,
since we expect his value to be
closer to our typical value than to our mean (\S 3.3).  Because his dust
opacity is somewhat higher than ours,
his observed-source mean radiation density is smaller yet,
about 1/4 of the
value of $\avg{U_{\rm FUV}}$ calculated in \S 4.1. 

	The observed-source mean FUV energy density estimated by Habing 
(1968) and the mean FUV energy density calculated in this paper are summarized
in Table 3, while the typical values are summarized in 
Table 4. Habing's typical value is different from ours: his
estimate is obtained 
by excluding the space in the neighborhood of 
associations (about 10\% of the total volume) and
taking the average over the remaining space. 
As discussed in \S 4, we estimate the typical FUV radiation by calculating
median value of the FUV field at a point in a large simulation of the model.
Compared to the band-averaged FUV ``Habing field'', our model results
give $G_0=(1.6,3.1)$ for the (median, mean) radiation field at the solar
circle.
Table 4 also gives the value 
below and above which the FUV energy density exists
 20\% of the simulation time.
A more complete characterization of the FUV radiation field is
given by the probability distribution in 
Figure 9. This representation obscures the temporal correlations
of the radiation field, however, so it is also useful to use
the FUV field history shown in 
Figures 7 and 8a to characterize the typical behavior of the 
FUV field. 

\section{Effect of Runaway OB Stars and Supernova}

\subsection{Runaway OB Stars}

"Runaway" OB stars, as defined by Blaauw (1961),
are OB stars with peculiar space velocities
that exceed 30 km s$^{-1}$ and that are directed away from known clusters and
associations. About 10-30\% of O stars and 5-10\% of B stars
are observed to be runaways.
These runaway stars have large space velocities
(30-200 km s$^{-1}$), and most of them appear to be single
(Gies 1987; Stone 1991; Hoogerwerf et al. 2000a,b).
The most viable scenarios for the formation of runaway stars
are: (i) a supernova explosion in a massive binary (Zwicky 1957;
Blaauw 1961), in which the primary becomes a compact object and
the secondary moves away with a velocity comparable to its
pre-explosion orbital velocity; and (ii) dynamical ejection from
a dense stellar cluster (Poveda et al. 1967; Gies \& Bolton 1986),
in which the most efficient interaction is the encounter of two
hard binary systems. Recently, milli-arcsecond astrometry provided
by Hipparcos and radio observations have provided specific evidence for
both scenarios (Hoogerwerf et al. 2000a,b), but the relative
importance of the two mechanisms remains unknown.

Since the OB stars produce the FUV radiation field,
and a non-negligible fraction of them are runaways, we
must estimate their effect on the evolution of the FUV-ISRF. We
assume that about 20\% of O stars and
7\% of B stars in each generation are ejected in random directions
with a velocity weighted by the probability distribution
\begin{equation}
p(v) \propto v \, \exp \left[-\left(\frac{v-v_c}{\sigma}\right)^2\right] 
\end{equation}
with $v_c=\sigma=$ 30 km s$^{-1}$, appropriate for runaways.
We consider only stars with masses larger than
8 $\rm{M}_\odot$; i.e. O and early B stars.
Our primary assumption is that
OB-star runaways (regardless of the ejection mechanism) are
ejected at the moment they form.
In the dynamical ejection scenario, the runaways are ejected
very early in the cluster evolution, so that this assumption is well justified.
In the binary-supernova
scenario, the creation of the runaway is delayed until the explosion
of the more massive component of the binary system. Therefore, the
instantaneous ejection assumption overestimates the size of the sphere of
influence of the SN-ejected runaways
and provides an upper limit on the effect of runaways.

To estimate the effect of runaway stars on the FUV radiation field,
we compute the difference $\Delta U_{-17,\rm run}$ of
the radiation energy density produced by the stars with masses
over 8 $\rm{M}_\odot$ in two cases,
with and without runaways.
Figure 12 shows the effects of including runaways
in the simulation performed in
\S 4.2; that is, the simulation parameters are exactly those used
to generate the time evolution of the FUV-ISRF in Figures 7 and 8,
so that in both simulations the associations are formed with the
same size, and at the same place and time.
Additionally, we have assumed periodic boundary conditions;
runaway stars that leave the simulation area from one boundary
are reinserted at the opposite boundary.

The heavy line in Figure 12 shows $\Delta U_{-17,\rm run}(t_{\rm sim})$ for
the same time interval
in Figure 8a. The dotted line shows the total energy density $U_{\rm FUV}$
from the simulation in \S 4.2 (i.e. considering stars of all masses and
assuming that they all remain at their birth places).
On average, at a given moment of time, there are about
200 runaway stars in the simulation area ($\sim 40$ O runaways and
$\sim 160$ early B runaways in 5 $\rm{kpc}^2$), but
very few of them make a passage close to the observer and
create a spike in the radiation field.  Most of these spikes
are produced by nearby associations;
in the plot, only one significant spike is produced by
a runaway coming from a distant (844 pc) association. 
For FUV spikes characterized by peak strengths $U_{-17}$, the
duration of the spike is of order $\Delta t_U \sim 10^6 L_5^{1/2}v_7^{-1}
U_{-17}^{-1/2}$ years, where $L_5=L_{FUV}/10^5\;L_\odot$ is
the stellar luminosity and $v_7=v/100$ km s$^{-1}$ is the runaway speed.
Spikes that dominate the FUV field occur roughly once every $\sim 100$
Myr. The
negative values of $\Delta U_{-17,\rm run}$
are also produced by nearby associations, since on average
their runaway stars are at greater distances from the point in
question than is the natal association.

The median value of $\Delta U_{-17,\rm run}$ in the simulation
is 0.016 (this slightly positive value is due to the fact that
it is assumed that the parent cloud does not absorb radiation from 
the runaways).
The ratio of the fluence in the spikes to that in the rest of the
radiation field is $\sim 0.025$. Therefore, the conclusions drawn in
previous sections (assuming that all the stars remain at their place
of birth) do not change significantly if runaways are considered.
However, very sporadically, the FUV-ISRF is dominated by the radiation from
a runaway.
These dominant spikes tend to occur when $U_{\rm FUV}$ is high
(i.e., there are nearby associations).

\subsection{Supernovae}

Supernovae produce very large FUV luminosities for
very short periods of time.
Core-collapse supernovae produce
UV radiation by two separate phenomena. One is the FUV radiation
burst due to shock breakout, and the second is the FUV emission
associated with the visible supernova.
Scalo et al. (2001) estimate that each process
typically produces about $10^{47}$ erg in UV photons,
over time intervals of less than an hour and $2-3$ months, respectively.
The luminosities in these phases are very large
compared with the mean FUV luminosity per main sequence
high-mass star
(i.e. $\avg{L_{\rm FUV}}_{h}=5.8 \times 10^4 L_\odot$, see \S 2.2)
However, the luminosity-weighted lifetime
of main sequence high-mass stars (i.e. $\avg{t_{\rm FUV}}=7.8 \times 10^6$ yr)
is much longer than the short periods where SN are actively emitting FUV
photons, and as a result
the total FUV energy emitted per main sequence high-mass star
($\avg{L_{\rm FUV}}_{h} \times \avg{t_{\rm FUV}}\sim 6 \times 10^{52}$ erg)
is about five orders of magnitude larger than the FUV fluence
per SN ($\sim$ few $10^{47}$ erg); indeed, the emitted FUV energy
is about 60 times larger than the typical {\it total} energy
released by a SN ($\sim10^{51}$ erg).
Therefore, SN FUV spikes make a negligible contribution
to the median and mean values of the radiation field
calculated in previous sections.

However, as pointed out by Scalo et al. (2001), these FUV spikes may be
relevant for atmospheric chemistry and for inducing
mutations in living creatures. Scalo et al. (2001) calculate the
frequency of events exceeding the solar flux in the $2000 - 3000 \AA$
at 1 AU and conclude that such events have occurred several times during
the history of the Earth. Scalo et al. (2001) have considered an homogeneous
distribution of SN. If as in the present work,
massive stars are assumed to form in associations,
high $U$ events tend to occur in groups because high mass
members of a given association explode within a relatively short period
of time and at similar distances.
Then, the time intervals between events (in different groups)
exceeding a given flux increase
by a factor $\edssh/\edsa(>1)=ln \, \eNshu=8.7$. 
In addition, we note that
the spectrum of the shock breakout is dominated by extreme ultraviolet and
soft X-ray emission (Matzner \& McKee 1999), so only a small fraction of
the emission is in the FUV band.  When this energetic radiation strikes
the Earth's atmosphere, much of it will be reprocessed to lower energies.
The analysis of this process and of the effects on atmospheric chemistry
is beyond the scope of the present work.

\section{Conclusions}

The presence of a time and space-dependent FUV field may have a major
impact on the structure of the neutral interstellar medium. 
In particular, the main process of $\rm{H}_2$ and CO destruction
is  photodissociation by radiation in the H$_2$ band (912-1100 \AA), 
and the primary heating process of the neutral interstellar gas is
grain photoelectric heating by radiation in the FUV band (912-2070 \AA).
The FUV controls the amount of interstellar material in the cold
(CNM) and warm (WNM) neutral phases.

In this paper, we have presented a study of the 
temporal behavior of the radiation field in these bands, as well as at 
1400~\AA.
To simulate the evolution of the radiation field at the solar circle
of the Milky Way, we have adopted: 
a) a power-law initial mass function with an upper mass
cutoff for high-mass stars in associations (\S 2.1);  
b) a stellar-mass dependence of the FUV and H$_2$
band luminosities, and the main sequence lifetime for stars of
various masses (\S 2.2); 
c) a stellar-mass dependence of the ionizing 
luminosity from high-mass stars (\S 2.3), which, coupled with 
radio observations
of the total ionizing luminosity in the Galaxy, determines the rate of
high-mass star formation (\S2.4); 
d) a history of the star formation rate in an association (\S 3.1);
e) a dependence of the birth-rate of
associations on the number of high-mass
stars ${\cal N}_{*h}$ in the association (\S 3.2); and 
f) a vertical distribution of associations and dust (\S 4).
Our calculations are based on the distribution of
OB associations in the Galaxy as determined by
MW97 and are completely independent of observations of
the local interstellar radiation field.

The most important conclusions are the following:

1) We have computed the typical (or median) value of the FUV energy
density at the solar circle in various bands and at 1400~\AA.
Our computed value of the median FUV energy density at 1400 \AA\
at the solar circle
($\sim 7.5 \times 10^{-17} \, \rm{erg} \, \rm{cm}^{-3}$ \AA$^{-1}$) 
compares reasonably well with the currently measured 
local value at 1400 \AA\  
($5.3 \times 10^{-17} \, \rm{erg} \, \rm{cm}^{-3}$ \AA$^{-1}$; 
Henry et al. 1980) and 
with Habing's and Draine's recommended values 
for the typical local interstellar
medium ($\sim 5 \times 10^{-17}$ and $\sim 9.8 \times 10^{-17}\, 
\rm{erg} \, \rm{cm}^{-3}$ \AA$^{-1}$, respectively).
The observationally estimated local value at 1400 \AA\ is 
near the
most probable FUV energy density values in the simulation (Fig. 9).  
Therefore, the local neighborhood has neither an exceptional high
or low number of OB associations. 
We find band-averaged median energy densities of 
(7.4, 7.1, 7.2)$\times 10^{-17}$
erg cm$^{-3}$ \AA$^{-1}$ for the ([912-1100 \AA], [1100-2070 \AA], 
[912-2070 \AA]) bands, respectively.  
Our value for the FUV band
is 1.6 times the value recommended by Habing ($G_0=1.6$), 
which is close to Draine's (1978) suggested value of $G_0=1.7$.

2) MW97 proposed a distribution for the birth rate of associations 
in terms of the number of high-mass stars in the association,
$d\dot\eNa/d\ln\eNsh\propto 1/\eNsh$,
based on observations of giant radio HII regions. 
Here we have shown that this 
distribution produces a median value of the
solar circle FUV-ISRF in agreement with the currently measured local value.
This is consistent with the extension of the
MW97 distribution to small associations, as they proposed.

3) We have included scattering and absorption by interstellar dust.
The absorption by dust limits the region that contributes to the
FUV field at a given point to distances less than about 500 pc. The
scattering produces a ``diffuse'' field
that is observed to be less than 10\% of the total FUV field.  
Our models are
consistent with these observations.  Thus, most
of the FUV energy density  at a given point at the solar circle comes
from the sum of discrete sources.

4) We have computed the average (or mean) of the FUV energy
density in various bands and at 1400 \AA. The mean value is nearly
twice the median value, due to infrequent spikes in the radiation
field.  The mean energy density at 1400 \AA \ is 
14.4 erg cm$^{-3}$ \AA$^{-1}$.
The band-averaged mean energy densities are $U_{-17}=(15.7, 13.8,
14.2)$ for the ([912-1100 \AA], 
[1100-2070 \AA], [912-2070 \AA]) bands, respectively. 

5) The fluctuations in the H$_2$ band [912-1100 \AA] are larger than 
those in the FUV band, 
and the ratio of the mean to median  energy density in this band
is larger than that in the FUV band. This is both because the extinction in 
the H$_2$ band is larger than in the FUV band and because the
sources contributing to the H$_2$ band are rarer than the 
those contributing to the FUV band.

6) The FUV field undergoes substantial fluctuations, with
the larger ones being less frequent. 
Figures 7 and 8 show that at the solar circle
the energy density of the radiation $U$ fluctuates above
the median value by a factor
2 to 3 every $100-200 $ Myr.
For $U>>U_{\rm med}$, the time interval $t_U$ between episodes 
with fields greater than $U$
is roughly given by $t_U \simeq 2 \, U_{-17}^{3/2}$ Myr or
$2 \, U_{-15}^{3/2}$ Gyr (Figure 9).
These infrequent epochs of high fields last of order $30$ Myr, the lifetime
of the nearby OB association that creates the high field.
 As we shall show in a subsequent paper, these
fluctuations affect the relative fractions of cold neutral medium
and warm neutral medium, and thereby create a fluctuating state
of the ISM as a function of space and time.

7) A point at the solar circle is inside the H~II region of a nearby
OB association about 2\% of the time, based on the assumption
that the size of the H~II region is that of a Str\"omgren sphere
in a medium with a density of $n=1$ cm$^{-3}$.

8) In general, a significant fraction of the FUV radiation
field comes from a single source.
For values of $U_{-17}$ around the median ($\sim 7$) the dominant source
provides on average 20\% of the energy density, but the values range
from 10\% to 50\%. For large values of $U_{-17}$, the radiation field
is totally dominated by only one source. Therefore, the FUV field is 
asymmetric at a given point, the asymmetry growing for higher fields.
Very low values of the radiation field are very infrequent---indeed,
values below 1/3
of the median FUV field are totally absent from our
simulations.

9) OB runaways produce short-lived spikes in 
the radiation field. At a given moment of time at the solar circle, 
we estimate that there are about 40 runaway stars per  $\rm{kpc}^2$ 
($\sim 8$ O runaways and $\sim 30$ early B runaways), but very few of 
them make a passage close to the observer and create a spike in the 
radiation field.
However, very sporadically (on the order of once every $\sim 100$ Myr, see
Figure 12), the
FUV-ISRF is briefly dominated by the radiation from
a runaway. These dominant spikes tend to occur when $U_{\rm FUV}$ is high
(i.e., there are nearby associations). However, the duration of the spikes
is short (about one Myr) and the total fluence in the spikes is negligible
in comparison with the total fluence from the stars in the associations.

10) Supernovae raise the radiation field significantly above the median value
every $\sim 5\times 10^4$ years. However, their duration
is so brief that the total fluence in the spikes is negligible
in comparison with the total fluence from the stars in the associations.

We acknowledge support from the NASA Astrophysical Theory Program in RTOP
344-04-10-02, which funds The Center for Star Formation Studies, a consortium of
researchers at NASA Ames, University of California at Berkeley, and 
University of California at Santa Cruz. 
AP was supported as a Senior Associate
for part of this research by the National Research Council and by
the Universidad de Los Andes by CDCHT project C-1098-01-05-B. 
The research of CFM is supported in part by NSF grant AST-0098365.  
We thank Juan Mateu and Gustavo Bruzual for providing
theoretical models of evolutionary spectra from high mass stars. 

\eject

%--------------- TABLE 1 ------------------------------
\begin{tabular}{c}
\hspace{3.7cm}TABLE 1\\
\hspace{3.7cm}Stellar Properties\\
\end{tabular}\\
\begin{tabular}{lclrcl}
\hline
\hline
 Quantity & Units & {\hspace{0.2cm} Approximation} & & & 
{\hspace{-1.55cm} Mass Range} \\ \hline
MS Life Time & Myr & $7.65\times 10^3 \ \ \ \ m^{-2.80}$ & 1.2 &  -  & 3 
$M_\odot$ \\
 & & $4.73\times 10^3 \ \ \ \ m^{-2.36}$ & 3 & - & 6  \\
 & & $2.76\times 10^3 \ \ \ \ m^{-2.06}$ & 6 & - & 9  \\
 & & $1.59\times 10^3 \ \ \ \ m^{-1.81}$ & 9 & - & 12  \\
 & & $7.60\times 10^2 \ \ \ \ m^{-1.57}$ + 2.30 & 12 & - & 120  \\
\hline
$L_{\rm H2} $ & $L_\odot$ & $1.98 \times 10^{-14} \ m^{26.6}$ & 1.8 & - & 3  \\
 & & $2.86 \times 10^{-8} \ \ m^{13.7}$ & 3 & - & 4  \\
 & & $1.35 \times 10^{-4} \ \ m^{7.61}$ & 4 & - & 6  \\
 & & $1.10 \times 10^{-2} \ \ m^{5.13}$ & 6 & - & 9  \\
 & & $1.07 \times 10^{-1} \ \ m^{4.09}$ & 9 & - & 12  \\
 & & $5.47 \times 10^{-1} \ \ m^{3.43}$ & 12 & - & 15  \\
 & & $9.07 \times 10^{0} \ \ \ \ m^{2.39}$ & 15 & - & 30  \\
 & & $9.91 \times 10^{1} \ \ \ \ m^{1.69}$ & 30 & - & 120  \\
\hline
$L_{\rm FUV}-L_{H2} $ & $L_\odot$ & $2.77\times 10^{-4} \ \ m^{11.8}$ & 1.8 
& - & 2  \\
 & & $1.88\times 10^{-3} \ \ m^{9.03}$ & 2  &-  & 2.5  \\
 & & $1.19\times 10^{-2} \ \ m^{7.03}$ & 2.5 & - & 3  \\
 & & $1.47\times 10^{-1} \ \ m^{4.76}$ & 3 & - & 6  \\
 & & $8.22\times 10^{-1} \ \ m^{3.78}$ & 6 & - & 9  \\
 & & $2.29\times 10^{0}  \ \ \ \ m^{3.31}$ & 9 & - & 12  \\
 & & $2.70 \times 10^{1} \ \ \ \ m^{2.32}$  & 12 & - & 30  \\
 & & $3.99 \times 10^{2} \ \ \ \ m^{1.54}$ & 30 & - & 120  \\
\hline
${s_{49}} $ & $10^{49} \rm {phot/s}$ & $2.23\times 10^{-15} \ m^{11.5}$ & 5 
& - & 7 \\
 & & $3.69 \times 10^{-13} \ m^{8.87}$ & 7 & - & 12  \\
 & & $4.80 \times 10^{-11} \ m^{7.85}$ & 12 & - & 20  \\
 & & $3.12 \times 10^{-8} \ \ m^{4.91}$ & 20 & - & 30  \\
 & & $2.80 \times 10^{-5} \ \ m^{2.91}$ & 30 & - & 40  \\
 & & $3.49 \times 10^{-4} \ \ m^{2.23}$ & 40 & - & 60  \\
 & & $2.39 \times 10^{-3} \ \ m^{1.76}$ & 60 & - & 120  \\
\hline
\end{tabular}

\eject

%--------------- TABLE 2 ------------------------------
\begin{tabular}{c}
\hspace{4cm}TABLE 2\\
\hspace{4cm}The Local Energy Density \\
\end{tabular}\\
\begin{tabular}{llll}
\hline
\hline
Authors & Method & Wavelength &
\hspace{0.6cm}$U \, [10^{-17} \rm{erg} \, \rm{cm}^{-3} \, \AA ^{-1}]$ \\ \hline
Habing 1968 & A & 1000\AA &\hspace{2cm} 3.3 \\
 &  & 1400\AA &\hspace{2cm} 4.2 \\
 &  & 2200\AA &\hspace{2cm} 2.2 \\
Jura 1974 & B & 1000\AA &\hspace{2cm} 7.0 \\
Gondhalekar et al. 1980 & C & 1000\AA &\hspace{2cm} 4.9 \\
 &  & 1400\AA &\hspace{2cm} 5.0 \\
 &  & 2200\AA &\hspace{2cm} 2.5 \\
Henry et al. 1980 & D & 1400\AA &\hspace{2cm} 5.3 \\
 &  & 1180-1680\AA &\hspace{2cm} 5.8 \\
Draine 1978 & E & 1000\AA &\hspace{2cm} 6.7 \\
 &  & 1400\AA &\hspace{2cm} 9.8 \\
 &  & 2200\AA &\hspace{2cm} 3.7 \\
\hline
\end{tabular}
\\
\begin{center}
{\bf {Notes to Table 2}}
\end{center}
\small{

\begin{itemize}
\item{
{Method A.-} The FUV flux from bright stars B9 and earlier was inferred from
its
visual magnitude $m_V$ and its color excess $E(B-V)$.
}
\item{
{Method B.-} Similar to method A but using updated stellar atmospheres model
and including
only bright stars B5 and earlier.
}
\item{
{Method C.-} Similar to method B but including more than 50,000 stars in the
S2/68 Survey.
}
\item{
{Method D.-} Extrapolation from direct measurements of the spectrum in
regions of
$12^o \, \times 12^o$ over about one-third of the sky.
}
\item{
{Method E.-} Values from the analytic expresion proposed by Draine (1978)
for the ``Standard UV background". His formula is based on estimates
of the typical FUV energy density at the solar circle and in the
solar neighborhood (Habing (1968); Gondhalekar and Wilson (1975);
Witt and Johnson (1973) and Jura (1974)) and on local measurements
(Hayakawa, et al. (1969); Belyaev et al. (1971) and Henry et al. (1977)).}
\end{itemize}
}

\eject

%--------------- TABLE 3 ------------------------------
\begin{tabular}{c}
\hspace{4cm}TABLE 3\\
\hspace{4cm}The Mean Energy Density at the Solar Circle\\
\end{tabular}\\
\begin{tabular}{llll}
\hline
\hline
Authors & Method & Wavelength &
$\hspace{0.2cm}\langle U\rangle \, 
[10^{-17} \rm{erg} \, \rm{cm}^{-3} \, \AA ^{-1}]$ \\
\hline
Habing 1968 & A & 1000\AA &\hspace{1.1cm} 3.9 - 6.7\\
 &  & 1400\AA &\hspace{1.1cm} 3.4 - 5.6\\
 &  & 2200\AA &\hspace{1.1cm} 1.4 - 2.3\\
This paper & B & H2: [912-1100\AA] &\hspace{1.5cm} 15.7\\
 &  & FUV-H2: [1100-2070\AA] &\hspace{1.5cm} 13.8 \\
 &  & FUV: [912-2070\AA] &\hspace{1.5cm} 14.2\\
 &  & 1400\AA &\hspace{1.5cm} 14.4 \\
This paper & C & H2: [912-1100\AA] &\hspace{1.1cm} 15.9 - 11.4\\
 &  & FUV-H2: [1100-2070\AA] &\hspace{1.1cm} 12.1 - 9.8\\
 &  & FUV: [912-2070\AA] &\hspace{1.1cm} 12.7 - 10.0\\
 &  & 1400\AA &\hspace{1.1cm} 12.5 - 10.3\\
\hline
\end{tabular}
\\
\begin{center}
{\bf {Notes to Table 3}}
\end{center}
\small{
\begin{itemize}
\item{
{Method A.-} ``Continuous method" (see \S 5.3).  Contribution
of 10 associations and 577 field stars between B0 and B5.
Left: direct star light; right: diffuse light included.
Habing 1968, Table 6 and moderate extinction.
}
\item{
{Method B.-} As method A but using the PDMF (see \S 4.1).
Diffuse light included (see \S 6, Appendix B).
}
\item{
{Method C.-} From simulation. Left (right) values include
(exclude) HII regions. Diffuse light included (see Sec. 6, Appendix B).
The quoted values are the averages over 5 points 280 pc apart in a simulation
of 57 Gyr. These mean values vary from point to point in one simulation
and from one simulation to another (see text and Fig. 10).
The variations of mean values excluding HII regions are much smaller.
A point is inside an HII region about 2\% of the simulation time.
}
\end{itemize}
}

\eject

%--------------- TABLE 4 ------------------------------
\begin{tabular}{c}
\hspace{3cm}TABLE 4\\
\hspace{3cm}The Typical Energy Density at the Solar Circle\\
\end{tabular}\\
\begin{tabular}{llll}
\hline
\hline
Authors & Method & Wavelength &
$\hspace{0.6cm}U \, [10^{-17} \rm{erg} \, \rm{cm}^{-3} \, \AA ^{-1}]$ \\ \hline
Habing 1968 & A & 1000\AA &\hspace{1cm} $2.5 < 4.0 < 14.0$\\
 &  & 1400\AA &\hspace{1cm} $2.5 < 5.0 < 15.0$\\
 &  & 2200\AA &\hspace{1cm} $1.5 < 3.0 < 13.0$\\
This paper & median & H2: [912-1100\AA] &\hspace{2cm} 7.4 \\
 &  & FUV-H2: [1100-2070\AA] &\hspace{2cm} 7.1\\
 &  & FUV: [912-2070\AA] &\hspace{2cm} 7.2\\
 &  & 1400\AA &\hspace{2cm} 7.5 \\
This paper & 20\% below & H2: [912-1100\AA] &\hspace{2cm} 4.6 \\
 &  & FUV-H2: [1100-2070\AA] &\hspace{2cm} 5.0\\
 &  & FUV: [912-2070\AA] &\hspace{2cm} 5.0\\
 &  & 1400\AA &\hspace{2cm} 5.4 \\
This paper & 20\% above & H2: [912-1100\AA] &\hspace{1.8cm} 14.8 \\
 &  & FUV-H2: [1100-2070\AA] &\hspace{1.8cm} 12.0\\
 &  & FUV: [912-2070\AA] &\hspace{1.8cm} 12.4\\
 &  & 1400\AA &\hspace{1.8cm} 12.5 \\
\hline
\end{tabular}
\\
\begin{center}
{\bf {Notes to Table 4}}
\end{center}
\small{
\begin{itemize}
\item{
{Method A.-} Habing's 1968 recomended values for  typical interstellar space
near the solar circle (his Table 8). 
The three values given
for each wavelength are the lower
limit, the typical value, and the upper limit.
}
\item{
{Median.-} Median value of the field in a point in a simulation 
with $t_{\rm sim}= 57$ Gyr. The first Gyr is discarded. See \S
4.2 and Figures 9 and 10.
}
\item{
{20\% above.-} The value below which the energy density exists
20\% of the simulation time.
}
\item{
{20\% over.-} The value above which the energy density exists
20\% of the simulation time.
}
\end{itemize}
}

\eject

%--------------- TABLE B1 ------------------------------
\begin{tabular}{c}
\hspace{0.5cm}TABLE B1\\
\hspace{0.5cm}Physical Properties of the Dust$^{\rm a}$\\
\end{tabular}\\
\begin{tabular}{llll}
\hline
\hline
 Wavelength & $\kappa_a$ & $\kappa_s$ & 
$g$ \\ \hline
912-1100\AA & $1.9 \, 10^{-21}$ & $8.95 \, 10^{-22}$ & $0.8$ \\ 
1100-2070\AA & $8.0 \, 10^{-22}$ & $7.5 \, 10^{-22}$ & $0.75$ \\ 
1400\AA & $7.03 \, 10^{-22}$ & $7.81 \, 10^{-22}$ & $0.756$ \\ 
\hline
\end{tabular}
\\
\small{
$^{\rm a}$Units of $\kappa$ are cm$^2$ (H atom)$^{-1}$
}
\eject

\appendix
\section{MASS DISTRIBUTION IN ASSOCIATIONS}

The IMF in equation (\ref{eq:dNsm})
is the probability distribution for the stellar masses in
an association.  Each realization of an association will
have fluctuations about this distribution, with 
the fluctuations being larger for smaller associations.
In order to make a realization for a single generation
of an association  that is
consistent with the IMF, we set 
the masses of stars with $m_i\ge 2 M_\odot$ to
\begin{eqnarray}
m_i=[
\mu_{i}^{-\Gamma}-x_{i}(\mu_{i}^{-\Gamma}-\mu_{i-1}^{-\Gamma})
]^{-1/\Gamma}
\ \ \ (i=1,..,{\cal N}_*)
\end{eqnarray}
where
\begin{eqnarray}
\mu_i=\frac{m_{u}}{\displaystyle
\left[ \frac{i \phi_h}{{\cal N}_{*h}/5} \left(\frac{
m_u}{m_h}\right)^{\Gamma}+1 \right] ^{1/\Gamma}}
\end{eqnarray}
and $\mu_0=m_u$.
We have assumed that there are 5 generations in an association,
so that the generation has $\eNsh/5$ stars.
The mass $m_1$ of the most massive star in each generation is chosen
at random by setting $x_{1}$ equal to a random number between $0$ and $1$.
The masses of the less
massive stars are set assuming $x_2=x_3=...=1/2$.
This procedure enables us to recover the correct IMF slope near $m_u$ when
a distribution of associations is considered.
Since an important fraction of $L_{\rm FUV}$ and $L_{H_2}$ is emitted
by the most massive star in the association, this approach also
accounts for the significant differences that can be present in similar
(i.e. same ${\cal N}_{*h}$) associations.

\section{DIRECT AND DIFFUSE RADIATION}

In this appendix we describe the procedure used to calculate
the scattered radiation, allowing for the possibility
that the source and/or the 
observer are out of the galactic plane.
We assume that the  absorbing and dispersing medium follows the HI 
vertical distribution (eq. \ref{nneut}).
The geometrical parameters of the problem are 
the height of the observer above the plane, $z_{o}$, 
the height of the emitter, $z_{e}$, and the projected
separation between the two, $x_{oe}$, as
indicated in Figure B1. 

To calculate the contribution $dF_{\rm scatt,1}$ to the diffuse flux
emitted at point `e', scattered at point `d', and observed at
point `o', we erect a coordinate system based on the line of
sight `oe' from the observer to the source.  Let `p'
denote the projection of `d' onto this axis, $\varpi$ be the
distance from `p' to `d', and $r_{ep}$ the distance from
the source to `p'.
We then have
\begin{eqnarray}
dF_{\rm scatt,1}= \frac{L_{*}}{4 \, \pi \, r_{ed}^2} \, e^{-\tau_{ed}}
 \,  df_{\rm scatt}
 \, 
\frac{\phi_{\rm scatt}
(\alpha,g)}{r_{do}^2} \, e^{-\tau_{do}} \, d\varpi \,  \varpi \,  d{\theta}
\end{eqnarray} 
where
\begin{eqnarray}
df_{\rm scatt} = \kappa_s \,  n_{\rm H}(z_d) \, dr_{ep}
	=\left[\frac{n_{\rm H}(z_d)}{n_{\rm H}(z_p)}\right]
	d\,\tau_s 
\end{eqnarray} 
is the fraction of the incident flux scattered per unit area 
(measured normal to the axis `oe') at `d'.
In equation (B2), $\kappa_s$ is the 
grain scattering cross section per H atom, 
$\tau_s$ is the scattering optical depth along the direct ray `oe',
and $ n_{\rm H}(z_d)$ is the number density of H atoms at height 
$z_d= z_p + \varpi \, cos(\theta) \, cos(\phi)$ (see Fig. B1)
above the galactic plane.
The term $\phi_{\rm scatt}(\alpha,g)$ in equation (B1) is the scattering
phase function; we adopt the Henyey \& Greenstein (1941) 
phase function 
\beq
\phi_{\rm scatt}
(\alpha,g)=\frac{1\,-\,g^2}{4\,\pi (1\,+\,g^2-2\,g\,\cos\,\alpha)^{3/2}},
\eeq
where 
\beq
g\equiv\frac{1}{4\pi}\int \,d\Omega\,\cos\alpha \phi_{\rm
scatt}(\alpha,g)\,d\Omega
\eeq
is the average of the cosine of the scattering angle.
The factors 
$\exp(-\tau_{ed})$ and $\exp(-\tau_{do})$ in equation (B1) 
are the extinctions along the paths
`ed' and `do', respectively, and include both absorption and scattering.

For the $912 - 1100 \AA$ and $1100 - 2070\AA$ bands, and at $1400\AA$, 
the adopted values for $\kappa_a$, $\kappa_s$ and $g$ are given in Table B1, 
and are derived from the data in
Witt \& Gordon (1999) assuming that
$\kappa_a\, =\, 5.34 \times 10^{-22} \, (\tau/\tau_V) \, (1-a)$ cm$^2$ (H 
atom)$^{-1}$ and 
$\kappa_s\, =\, 5.34 \times 10^{-22} \, (\tau/\tau_V) \, a$ cm$^2$ 
(H atom)$^{-1}$. 

The total flux $F_{\rm scatt,1}$ that arrives at `o' after being
scattered once at `d'
is simply the integration of $dF_{\rm scatt,1}$ over the whole volume. 
In practice, 
the integration is done over the revolution volume around the axis `oe' 
that satisfies $\alpha < \pi/2$, because more than $96\%$ of the 
diffuse flux comes from this volume. 
It is illustrative to compare $F_{\rm scatt,1}$ to the direct flux 
\begin{eqnarray}
F_{\rm dir}= \frac{L_{*}}{4 \, \pi \, r_{oe}^2} \, e^{-(\tau_a+\tau_s)} ,
\end{eqnarray} 
where $\tau_a$ is the absorption 
optical depth along the ray `oe'.
Taking advantage of the fact that the scattering is strongly
peaked in the forward direction ($g\ge 0.75$---Table B1),
we normalize to the case of purely forward scattering ($g=1$). In that case,
the probability of a single scattering is $\tau_s\exp(-\tau_s)$
and $F_{\rm scatt, 1}=F_{\rm dir}\tau_s$.
In the general case in which $g\le 1$,
we express the single scattering flux as:
\begin{eqnarray}
F_{\rm scatt,1} =
F_{\rm dir}\tau_s {\cal R}.
\end{eqnarray} 
We have evaluated $\cal R$ numerically in Figure B2.  
Note
that it is close to unity, as expected from the fact that 
the scattering is strongly peaked in the forward direction.

    The procedure described above considers the diffuse radiation arriving
at `o' after only one scattering. To estimate the contribution of multiple
scattering, we again consider the case of purely forward scattering.
In this case, the 
probability that a photon will arrive at `o' after $n$ scatterings
is $\tau_s^n\exp(-\tau_s)/n!$.  We adopt the {\it ansatz} that
the flux that arrives at
`o' after $n$ scatterings when $g<1$ is
\begin{eqnarray}
F_{{\rm scatt},n} \sim  F_{\rm dir}  \, \frac{(\tau_s \, {\cal R})^n}{n!}.
\end{eqnarray} 
The total flux $F_{\rm tot}$ received at `o' (direct plus scattered) 
is then
\begin{eqnarray}
F_{\rm tot} =  F_{\rm dir} e^{\tau_s\cal R}=
\frac{L_{*}}{4 \, \pi \, r_{oe}^2} \, 
e^{-\tau_a \, - \, \tau_s \, (1 \, - \, {\cal R})}.
\end{eqnarray} 
For purely forward scattering, ${\cal R}=1$ and scattering has
no effect on the observed flux, as expected.
In the actual case, $\cal R$ is close to unity and
scattering has only a small effect.  Figure B3 shows
$\exp-\tau_s (1-{\cal R})$ as a function of the total optical
depth for several different cases.

The contribution of the diffuse radiation to the total
radiation field obtained by this method can not be directly compared
with the estimates obtained from observation because
the latter only counts the diffuse radiation coming from regions
relatively far from sources. For example, the diffuse radiation that
originates in lines of sight close to  regions of recent star formation
such as Gould's Belt are excluded.
In \S  5, the median value of $F_{\rm scatt}/F_{\rm tot}$
from the simulations
is about $0.2$ whereas the estimates from observations are somewhat less
than  $0.1$.
However, if the scattered light that originates from lines of sight close
to the sources are excluded from our calculations, 
our results are closer to the observations.  We find that
about $1/3$ of the single-scattered radiation 
has a scattering angle less than $5\deg$; omitting such
radiation from the ``diffuse'' category gives a diffuse
fraction of about 13\%, close to what is observed. 

\eject

\eject

%fig 1a,b,c,d
\figcaption{(a) The mean luminosity in the band $[1100-2070\AA]$
as a function of the stellar mass $m$. The triangles correspond to the 
calculated mean MS
luminosities using  stellar evolutionary tracks (see text) and the lines are the
eight segment power law fit used to approximate the mass dependence of the
luminosity in the simulations. (b) Same as (a) but for the  $\rm{H}_2$ band 
$[912,
1100\AA]$.
(c) The energy output in the FUV band  $[912-2070\AA]$
for a mass distribution of stars following a Present Day Mass Function (PDMF)
of MS stars.
The thin lines correspond to the case when obscuration
is taken into account. (d) Same as (c) but for the  $\rm{H}_2$ band.
}

%fig 2a,b
\figcaption{(a) The mean main sequence ionizing photon 
production rate $s_{49}$ (units 10$^{49}$ photons s$^{-1}$) as a 
function of the initial stellar mass. 
The seven segment power law fit to these data is shown in broken line.
(b) The contribution of the different mass ranges to the ionizing 
photon production rate assuming a Present Day Mass Function of MS stars.
}

%fig 3
\figcaption{Various estimates
of the IMF parameters in the plane
[$\edssh ,-\Gamma$] together with three curves of constant
total ionizing photon production rate 
$S_{T,53}=S_T/(10^{53}$ photons 
s$^{-1})$.
The letters beside the x symbols labels
the estimates of Lequeux 1979 (L), Miller \& Scalo 1979 (MS),
Van Buren 1983 (VB, b, c),
and Scalo 1986 (Sc86).
The 90\% and 99\% confidence contours correspond to Van Buren data.
The cross near the
label P indicates the adopted values of $\edssh $ and $\Gamma$
in the present work.
}

%fig 4
\figcaption{Evolution of the FUV luminosity for associations
that ultimately form 1000, 100, and 10 high-mass stars in five generations.
The contribution by a single association to the energy density $u_{\rm FUV}$
($u_{-17}=
u/10^{-17}$ erg cm$^{-3}$
\AA$^{-1}$) at a point located  
$100$ pc away, assuming that there is no extinction, is indicated
in the right vertical axis. 
}

%fig 5
\figcaption{The projection on the plane of the
positions of associations born in a period
of 500 Myr (left panel) and 30 Myr (right panel).
Large pentagons, large stars, small stars, and dots correspond
respectively to associations with sizes characterized
by ${\cal N}_{*,h}$ stars with $ 1000 < {\cal N}_{*,h} < {\cal N}_{*h,u}$, 
$100 < {\cal N}_{*,h} < 1000$,
$10 < {\cal N}_{*,h} < 100$, and ${\cal N}_{*h,l} < {\cal N}_{*,h} < 10$.
The right panel reflects the OB associations, since the OB stars die
after about 30 Myr.
}

%fig 6
\figcaption{
Expected increase of the mean energy density 
$\avg{U_{\rm FUV}}_{t_{\rm sim}}$
in the simulation time interval $[0,t_{\rm sim}]$ as $t_{\rm sim}$ increases.
Values normalized to the asymptotic
mean value $\avg{U_{\rm FUV}}_{\infty}$.
}

%fig 7
\figcaption{Time evolution of the FUV energy density at a point 
located at the center of simulation area and in the galactic plane.
The bars in between the arrows indicate the periods of time where
the point is inside an HII region; the labels give the characteristics
of the association that produces the HII region: size ${\cal N}_{*,h}$, distance $r$,
ionizing luminosity $s_{49}$ and the mass $m_1$ of the most massive star.
}

%fig 8
\figcaption{(a) Time evolution of the energy density in the
$[1100-2070\AA]$ band $U_{\rm{FUV-H_2}}$
(continuous line) and in the $[912-1100\AA]$ H$_2$ band $U_{\rm{H_2}}$
(dotted line). 
(b) Ratio of the energy density in the  $[912-1100\AA]$ band, $U_{\rm{H_2}}$,
to the  energy density in the $[1100-2070\AA]$ band,
$U_{\rm{FUV-H_2}}$, 
as a function of the 
FUV energy density $U_{\rm{FUV}}$. 
Each data point corresponds to the radiation field 
every 20 time steps (i.e. $20 \, \Delta t_a \sim$ 1 Myr) and 
the time period plotted is from
$t_{\rm sim}=1000$ to $4600$ Myrs (see Fig 7).
(c) The fraction of $U$ that comes to the observer as
scattered light  as function of $U_{-17}$ in the
$1100 \, - \, 2070\AA$ band. The times are the same as in Figure 8b.
(d) The fraction of $U$ in the FUV band that comes to the observer from 
the dominant source as function of $U_{-17}$ in the FUV band.
}

%fig 9
\figcaption{(a) Distribution of the occurrence
of $U_{\rm{H}_2}$ values (dashed)  and of $U_{\rm{FUV}}$ 
values (non-dashed). The arrows
indicate the mean and median values of the distributions
(see text \S 4.2).
(b) Probability $P(>U)$ that, at a given instant in the
simulation, the FUV energy density is greater than $U$.
The curves labeled 1 Gyr, 4 Gyr, and 56 Gyr, correspond respectively to
the probabilities $P(>U)$ calculated considering only the evolution
of $U$ in the time intervals $\Delta t =$ 1 Gyrs, 4 Gyr, and 56 Gyr.
The negative slope dashed line shows a power law fit to $P(>U)$ 
for large values of $U$.
The positive slope dashed line shows the corresponding power law
for the typical time $t_U$  (right vertical axis) between two events that
raise the energy density to $>U$ at a fixed point in space.
}

%fig 10
\figcaption{Comparison of the evolution of mean energy density $<U>_{t_{\rm 
sim}}$
in a single simulation of time interval $t_{\rm sim}$ and the
expected relation (see Figure 6) of a large number of simulations. 
The corresponding evolution of the median value for a single simulation
is also shown.
Values normalized to the asymptotic
mean value $\avg{U_{\rm FUV}}_{\infty}$.
See text (\S 4.2).
}

%fig 11
\figcaption{Evolution of the surface density of SN progenitors \ssh
\ in the simulation. The continuous line
corresponds to the average surface density over $A_{\rm sim}= 5 \, \rm{kpc}^2$ 
and the 
dotted line corresponds to the average within 500 pc of the central point.
}

%fig 12
\figcaption{ $\Delta U_{-17,\rm run}(t_{\rm sim})$ (heavy line)
and the total energy density $U_{\rm FUV}$
from the simulation (dotted line). The labels at the top give the properties
of the runaway stars producing the most prominent FUV spikes
(i.e. the stellar mass $m$, the velocity of ejection $v_{\rm ejec}$,
the distance of closest
approach  $r_{\rm min}$ to the point in question, and
the generation in which the runaway was produced in the association).
The labels at the bottom give the properties
of the association where the runaway was born
(i.e. the size of the association ${\cal N}_{*h}$ and its distance
to the observer $r_{\rm form}$).
}

\figcaption{ Fig. B1.- Geometrical parameters of the scattering problem. 
}

\figcaption{ Fig. B2.- 
$\cal R$ as a function of the distance $r_{oe}$ for $z_e$ and $z_o$
at the mid plane (continuous line), $z_o=z_e=100$ pc (long dash line),
and $z_o=100$ pc and $z_e=-100$ pc (short dash line).   
}

\figcaption{ Fig. B3.- 
Ratio of total flux to the flux for pure absobtion 
$\exp-\tau_s (1-{\cal R})$ as a function of the total optical
depth for the three cases in Fig. B2.
}

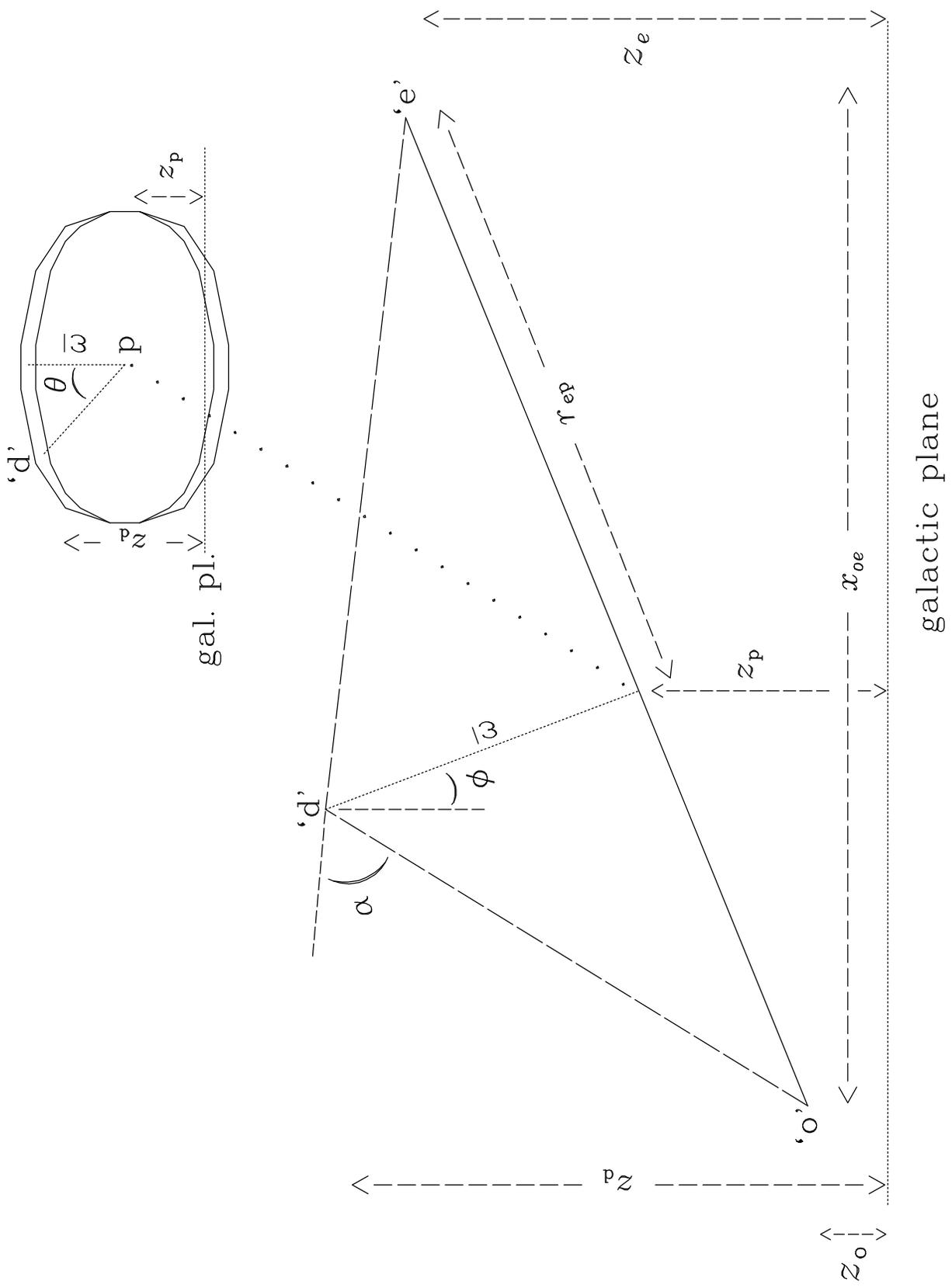
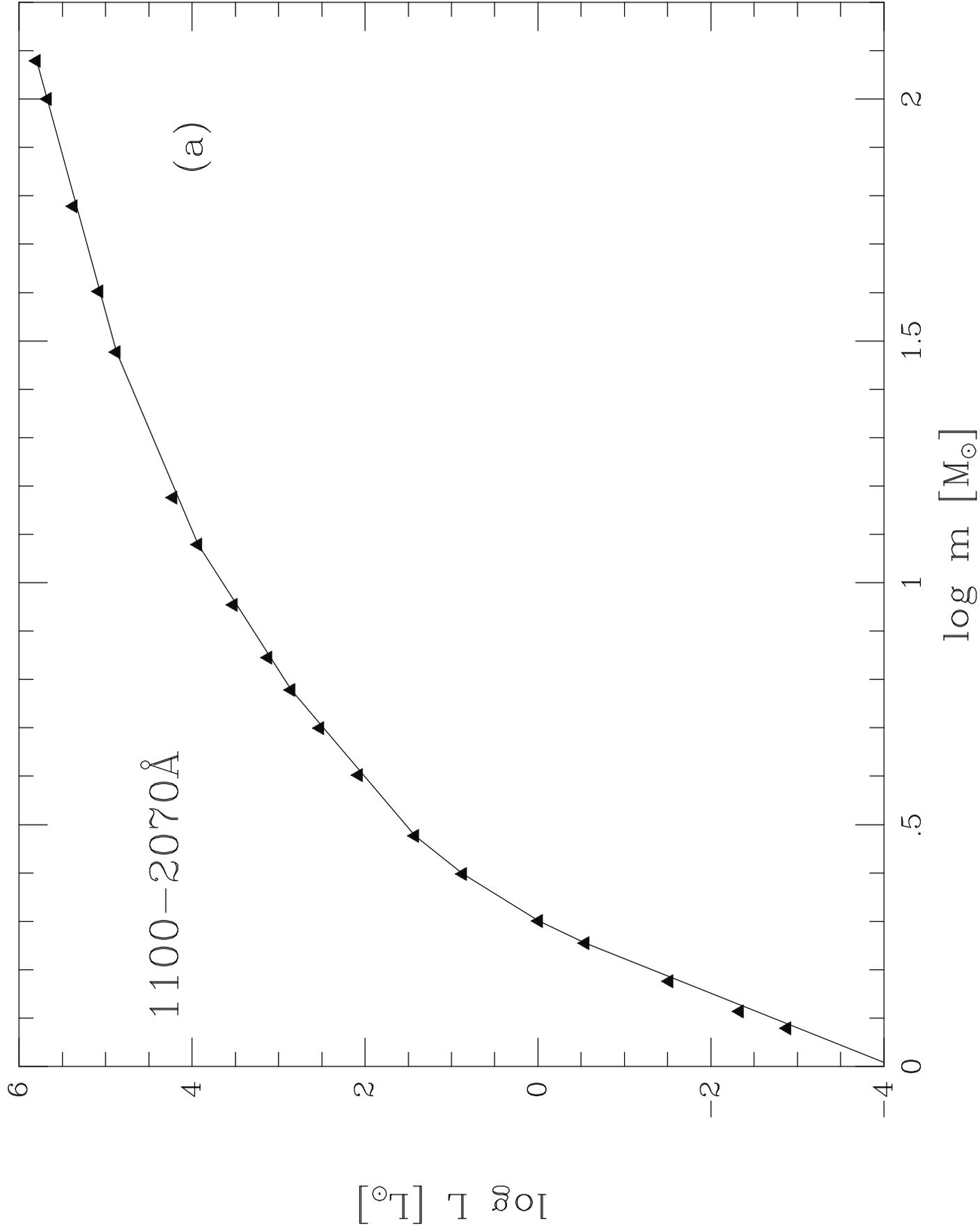
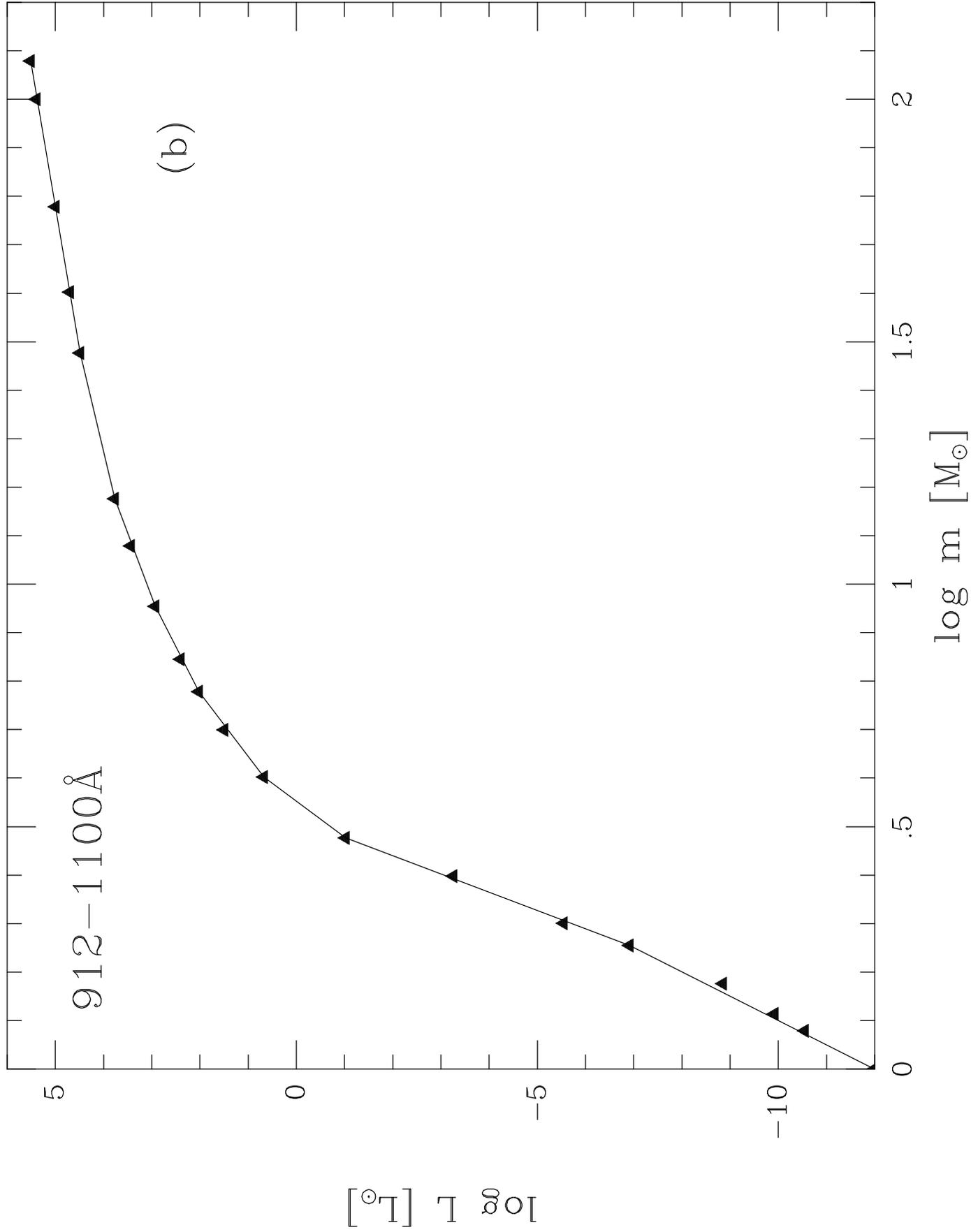
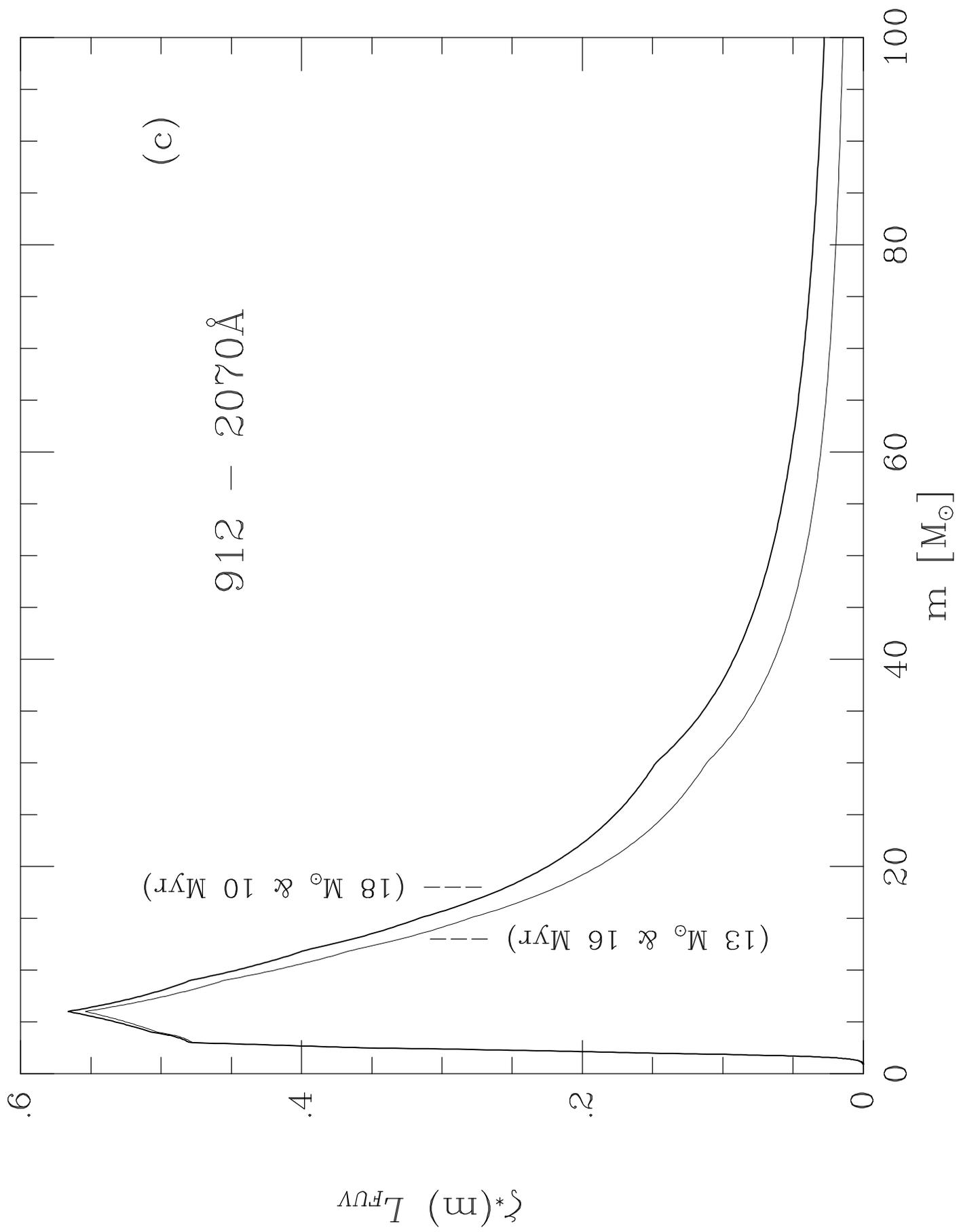
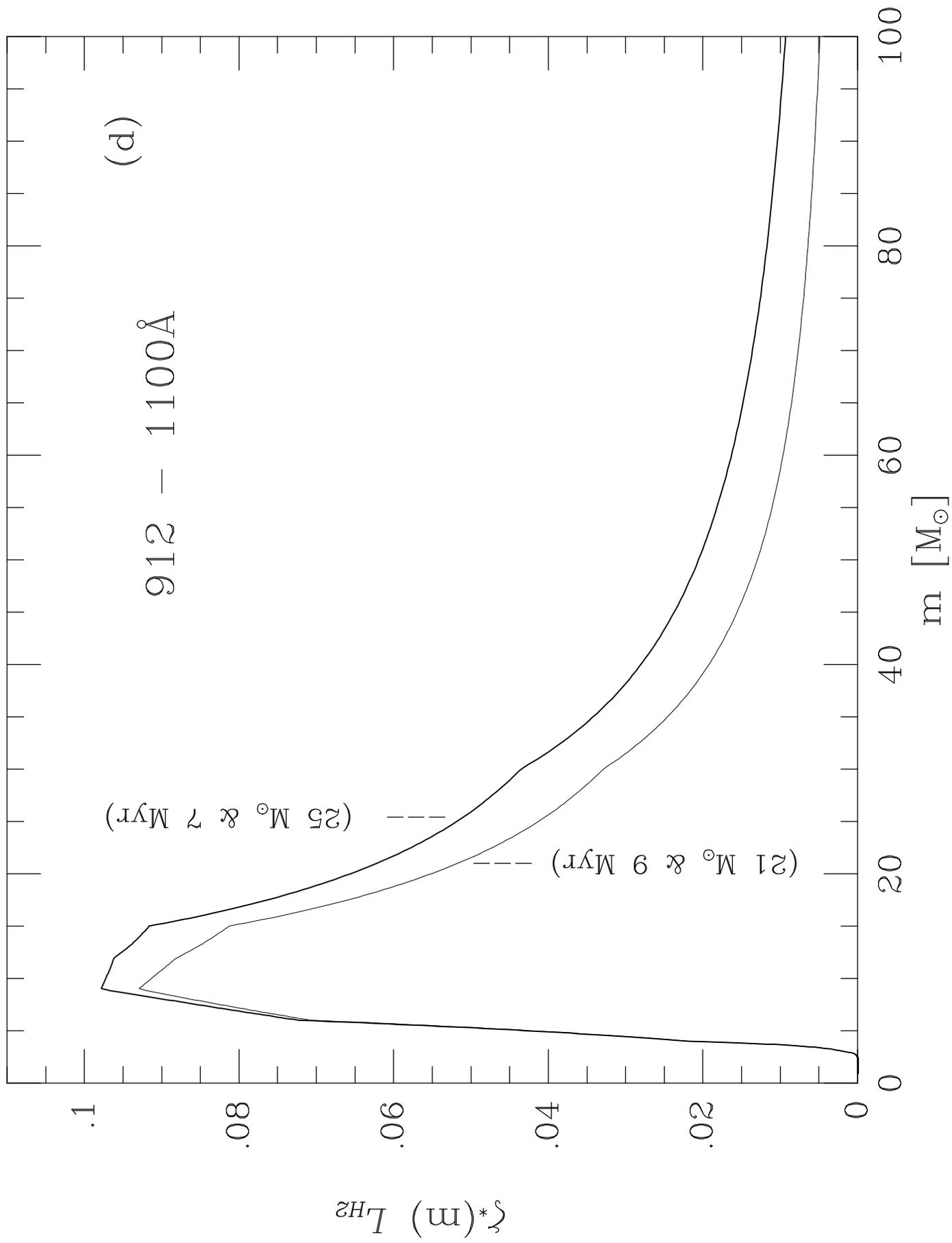
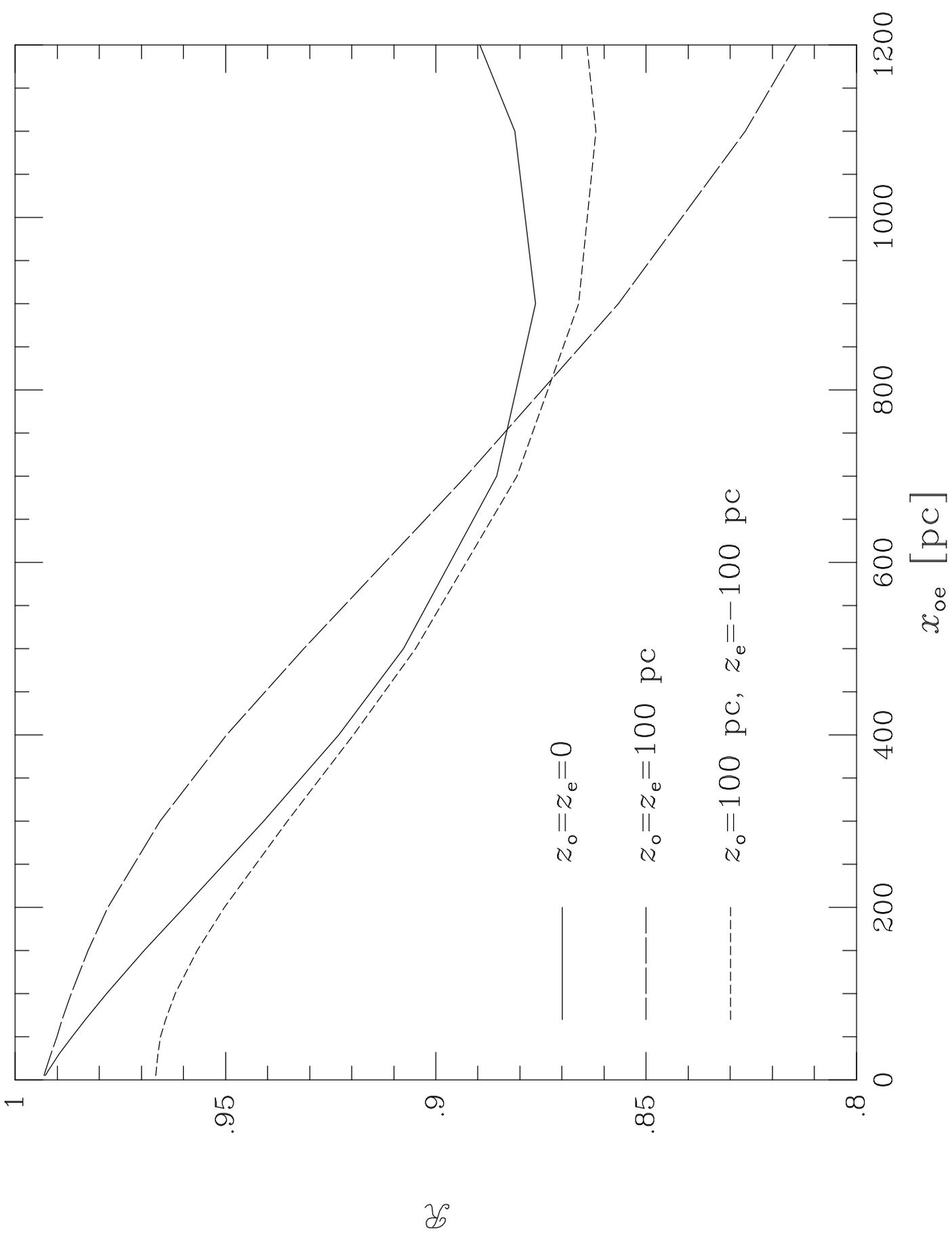
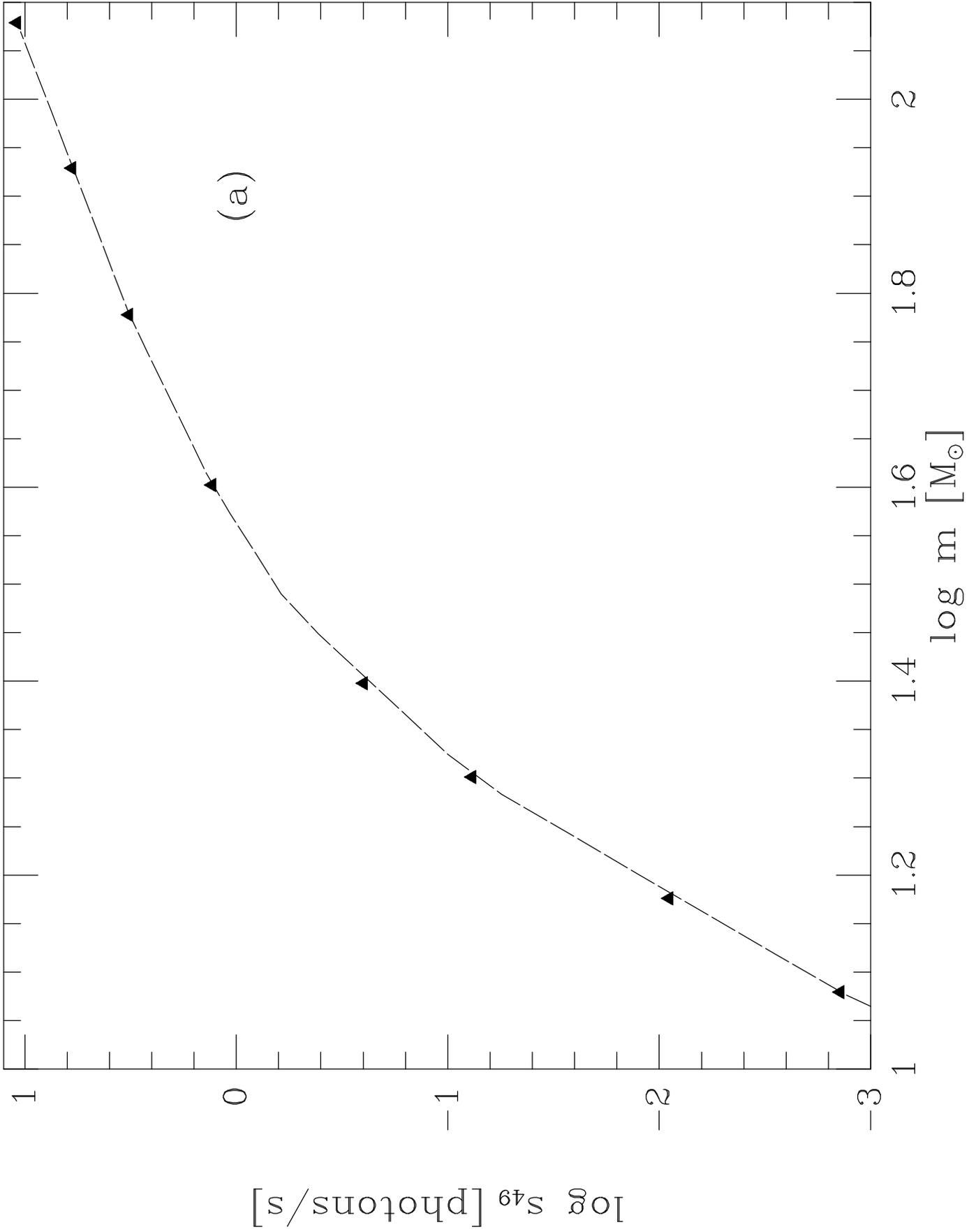
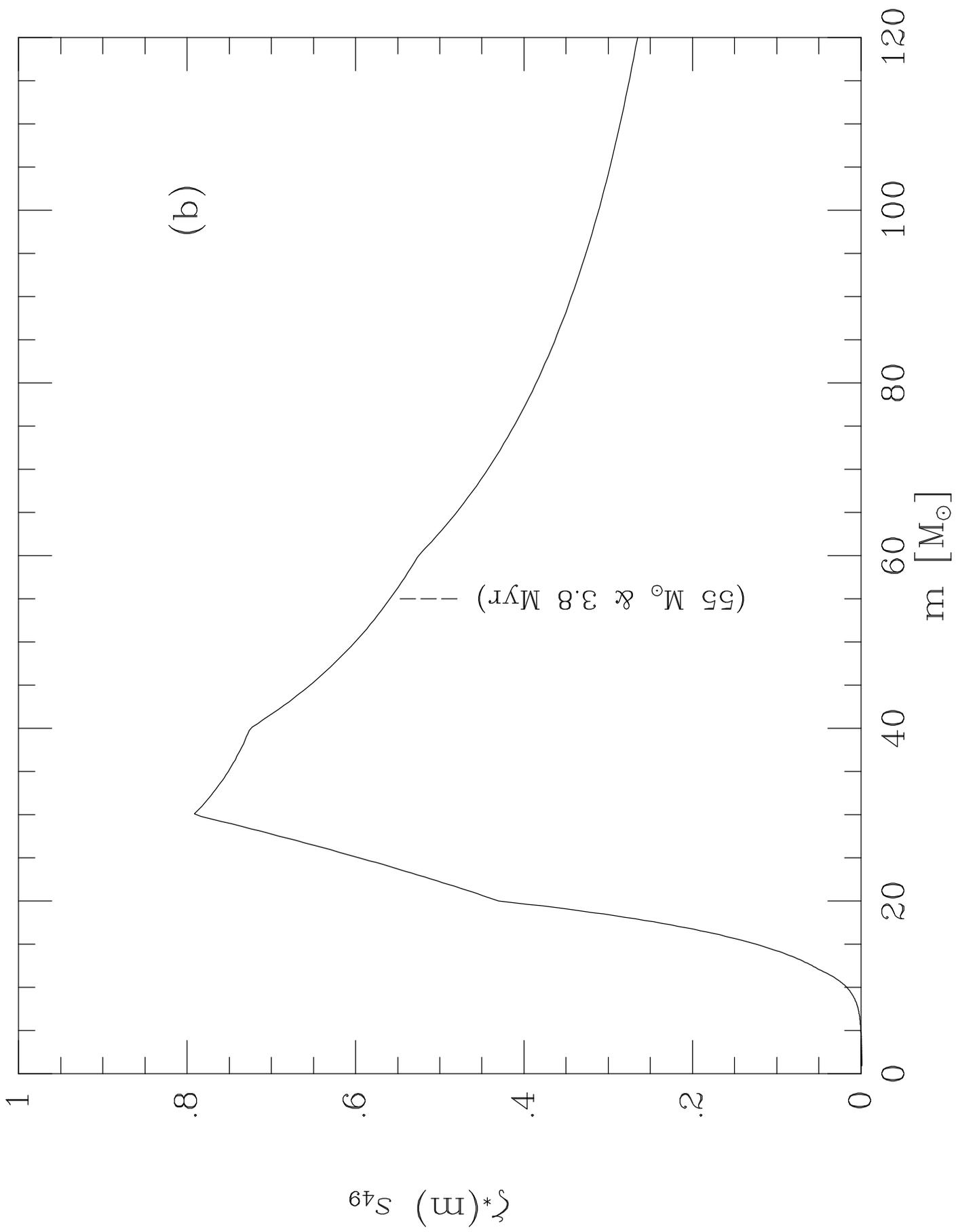
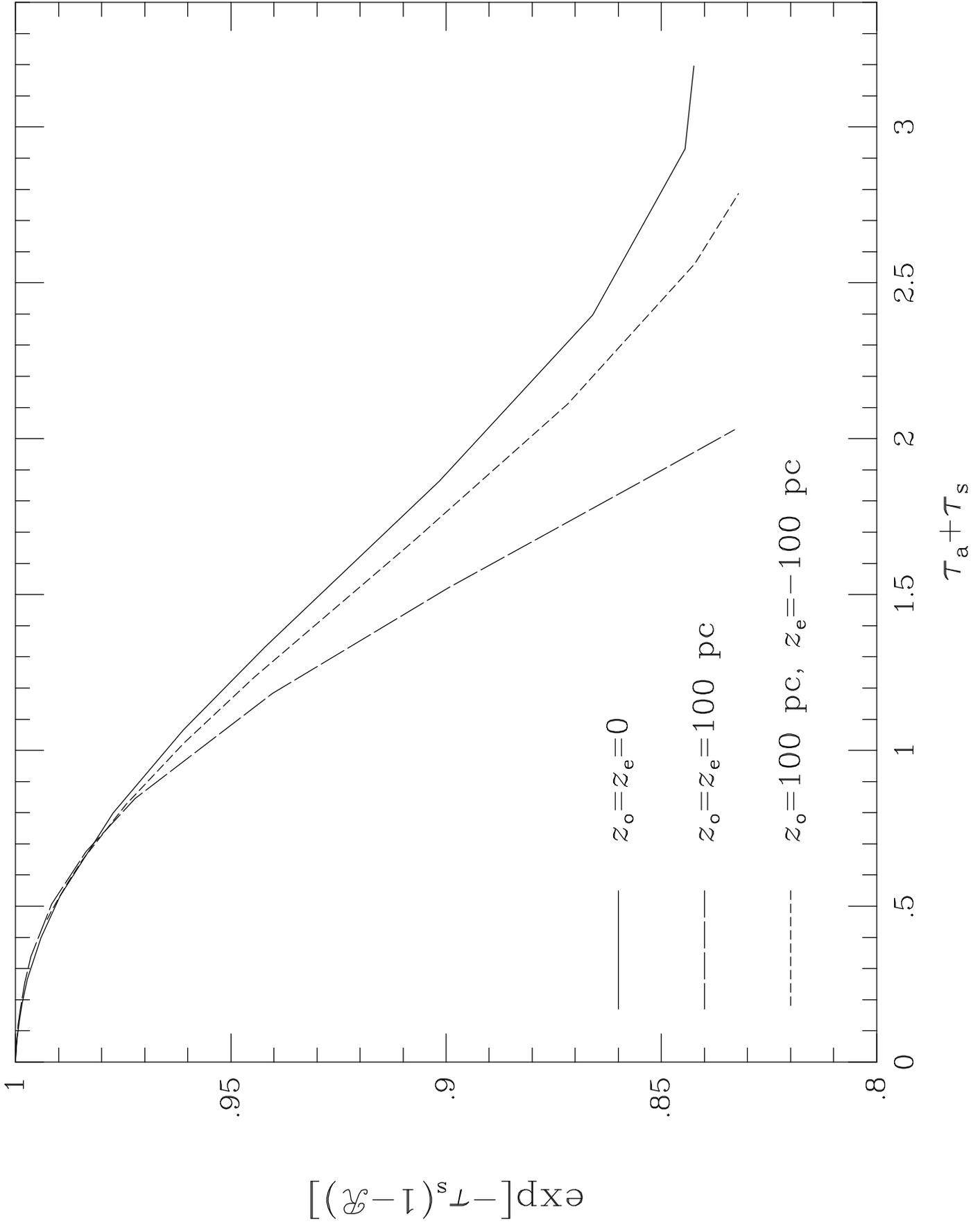
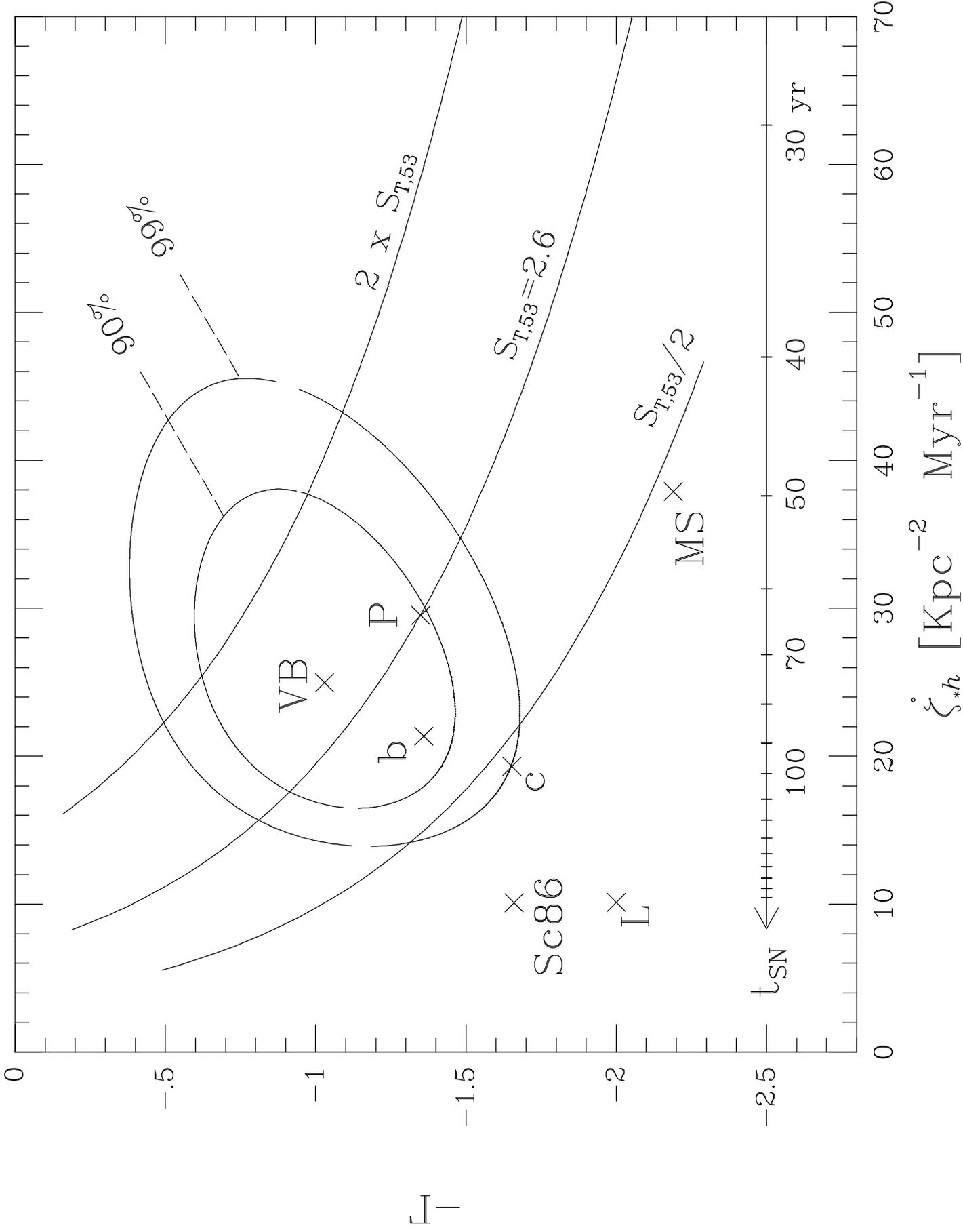
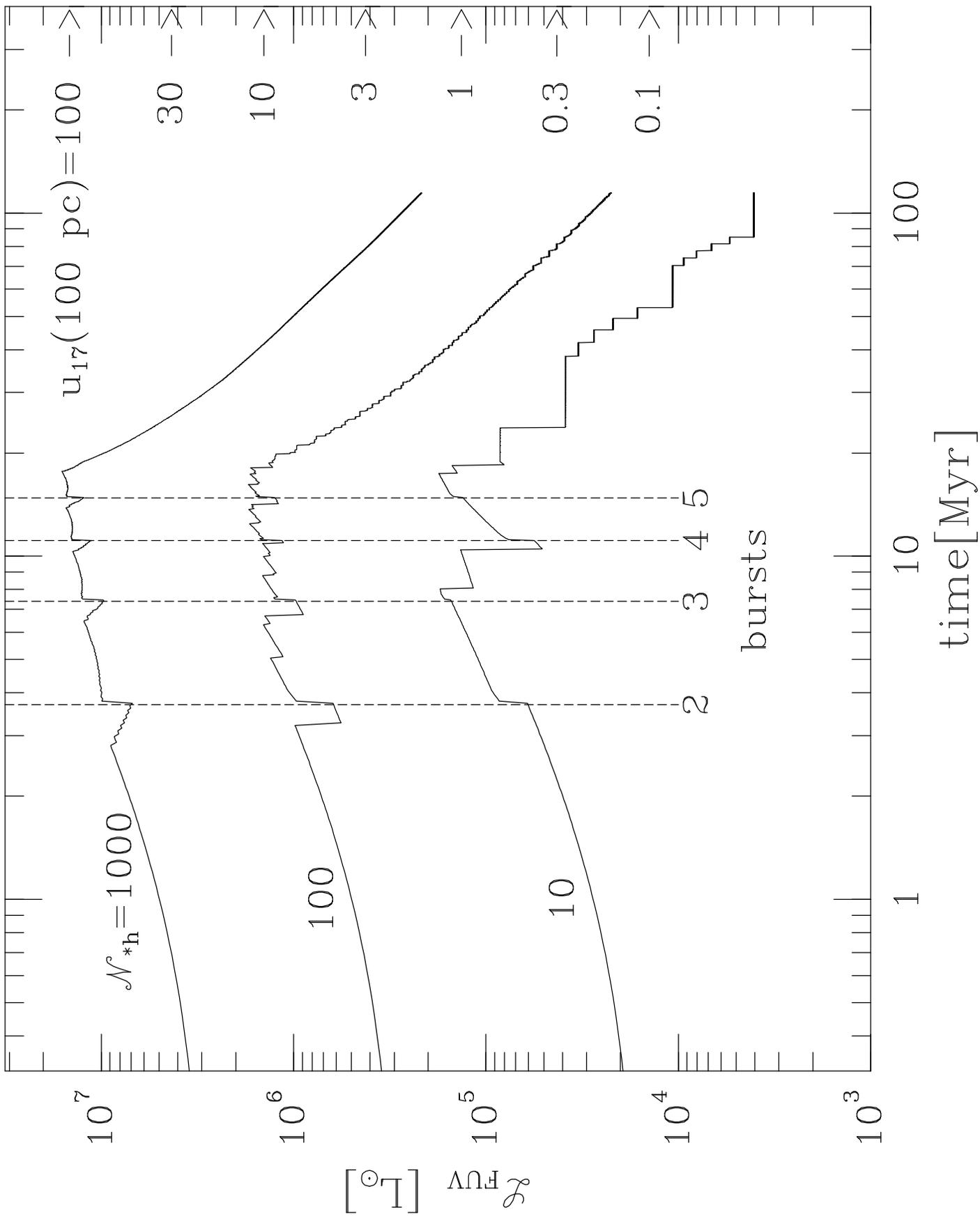
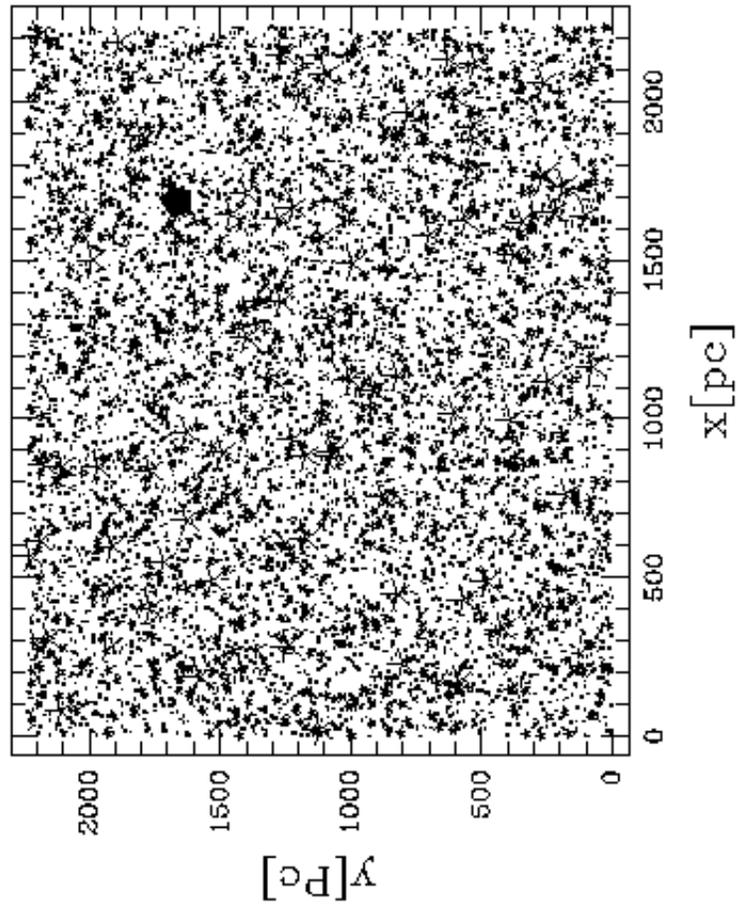
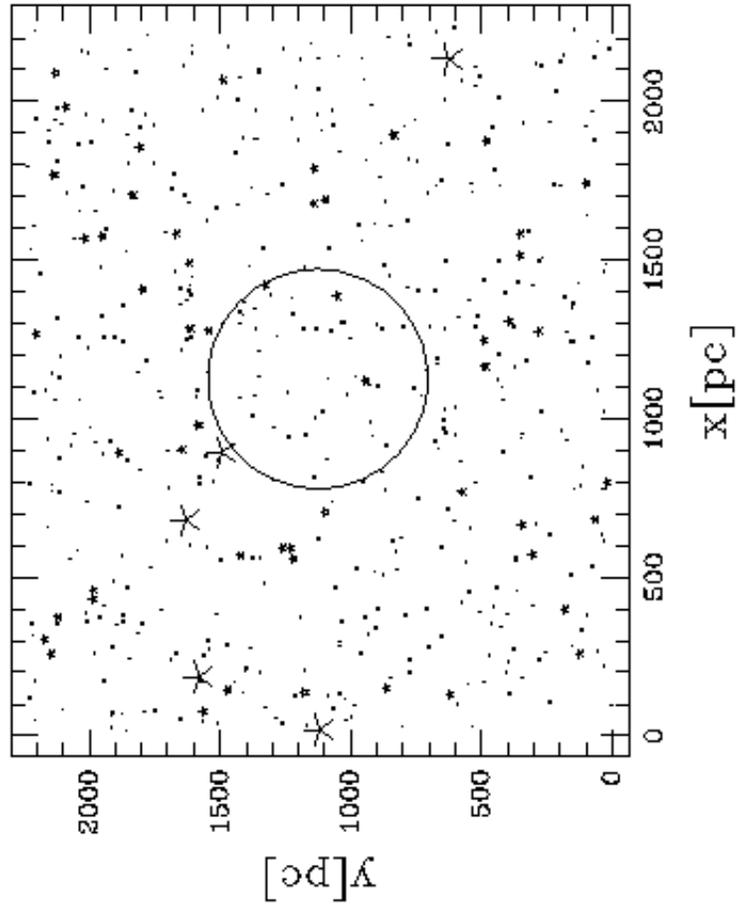
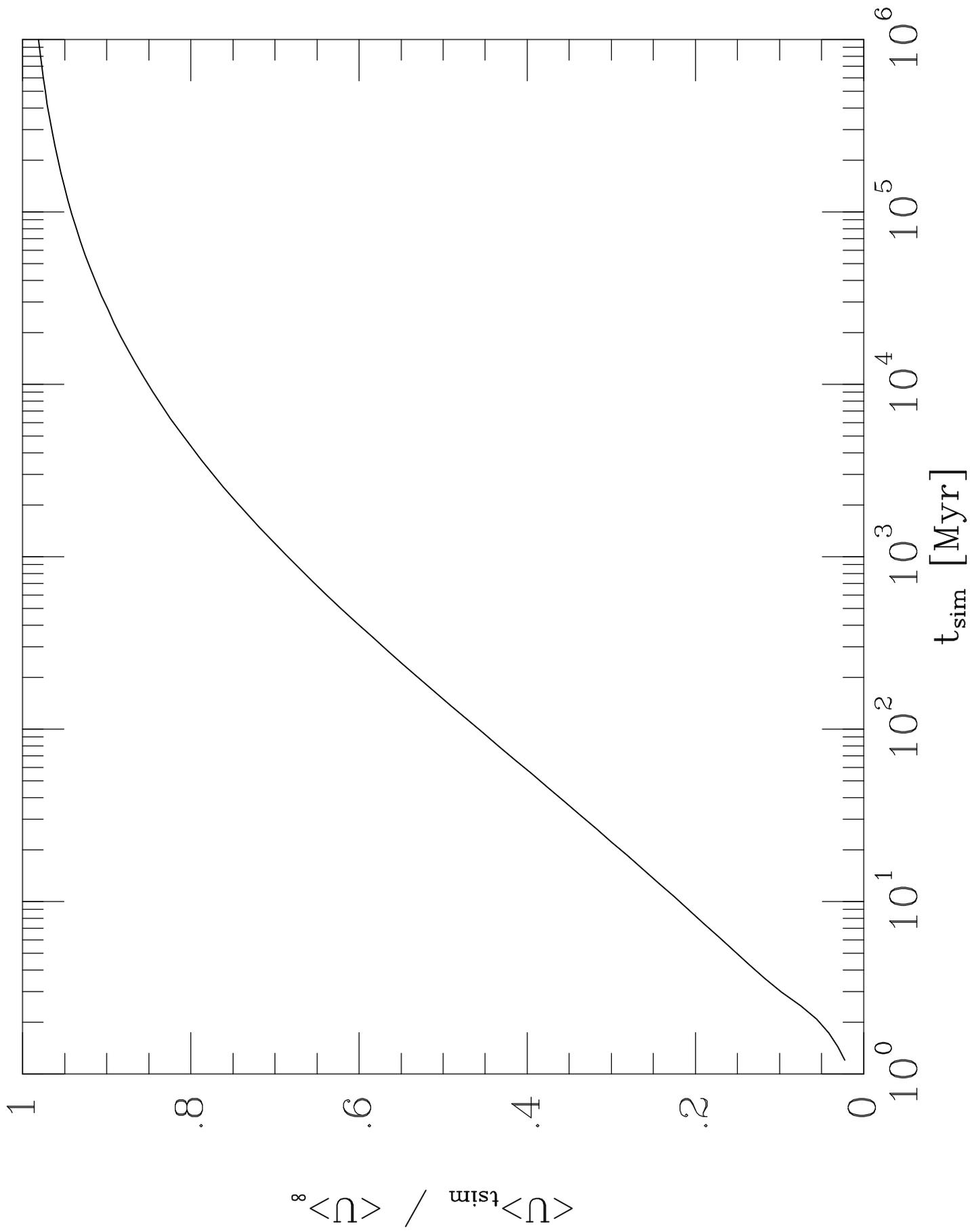
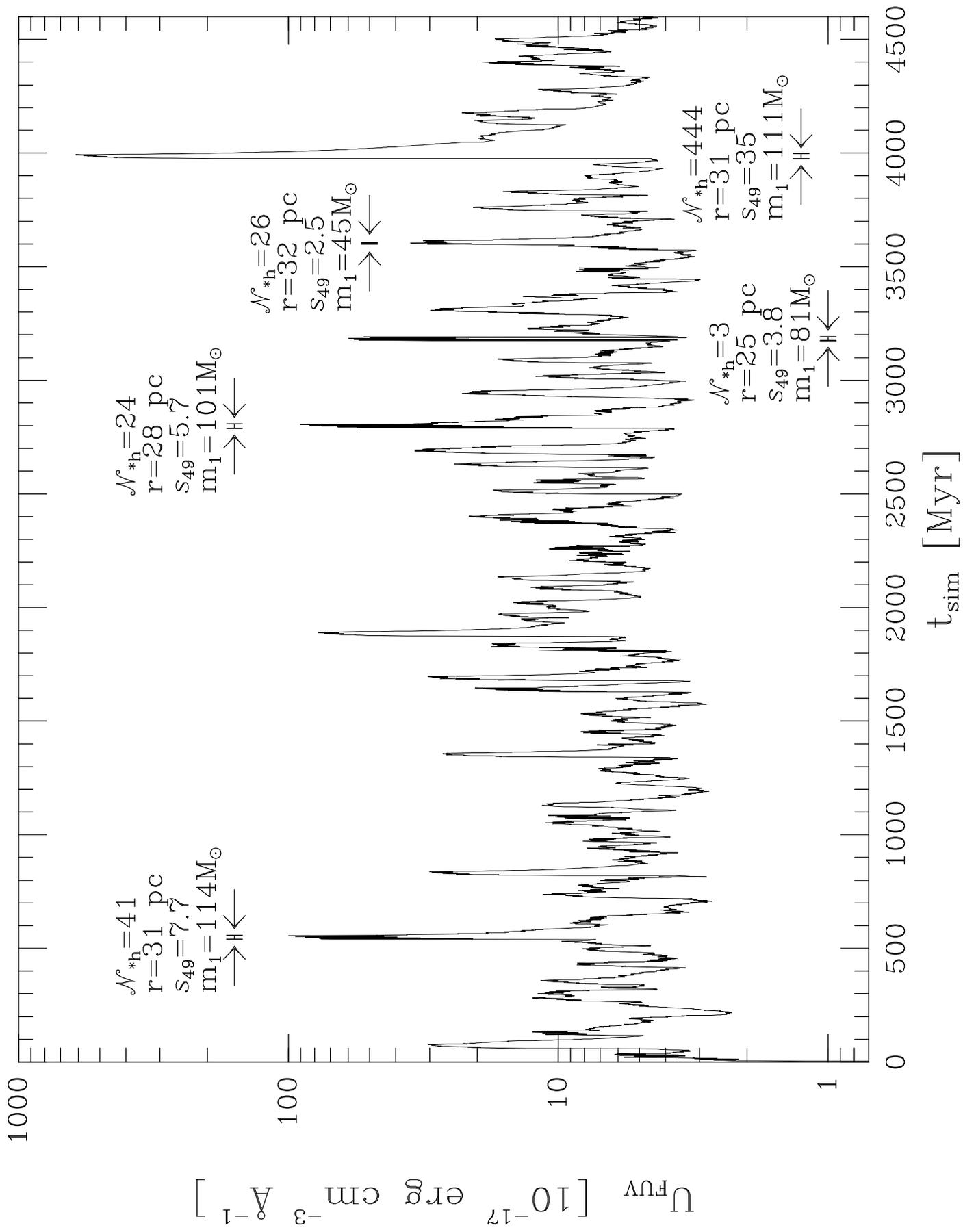
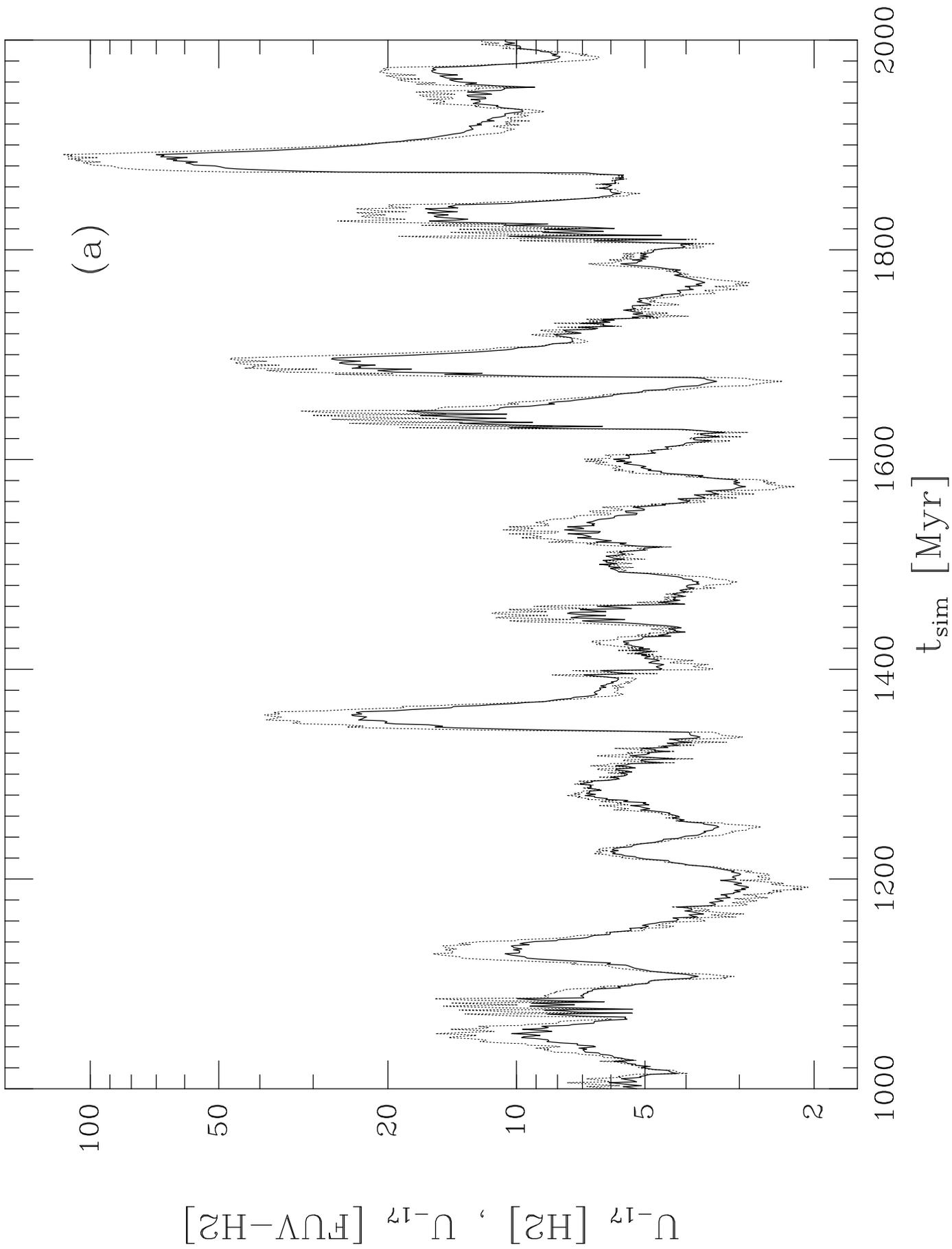
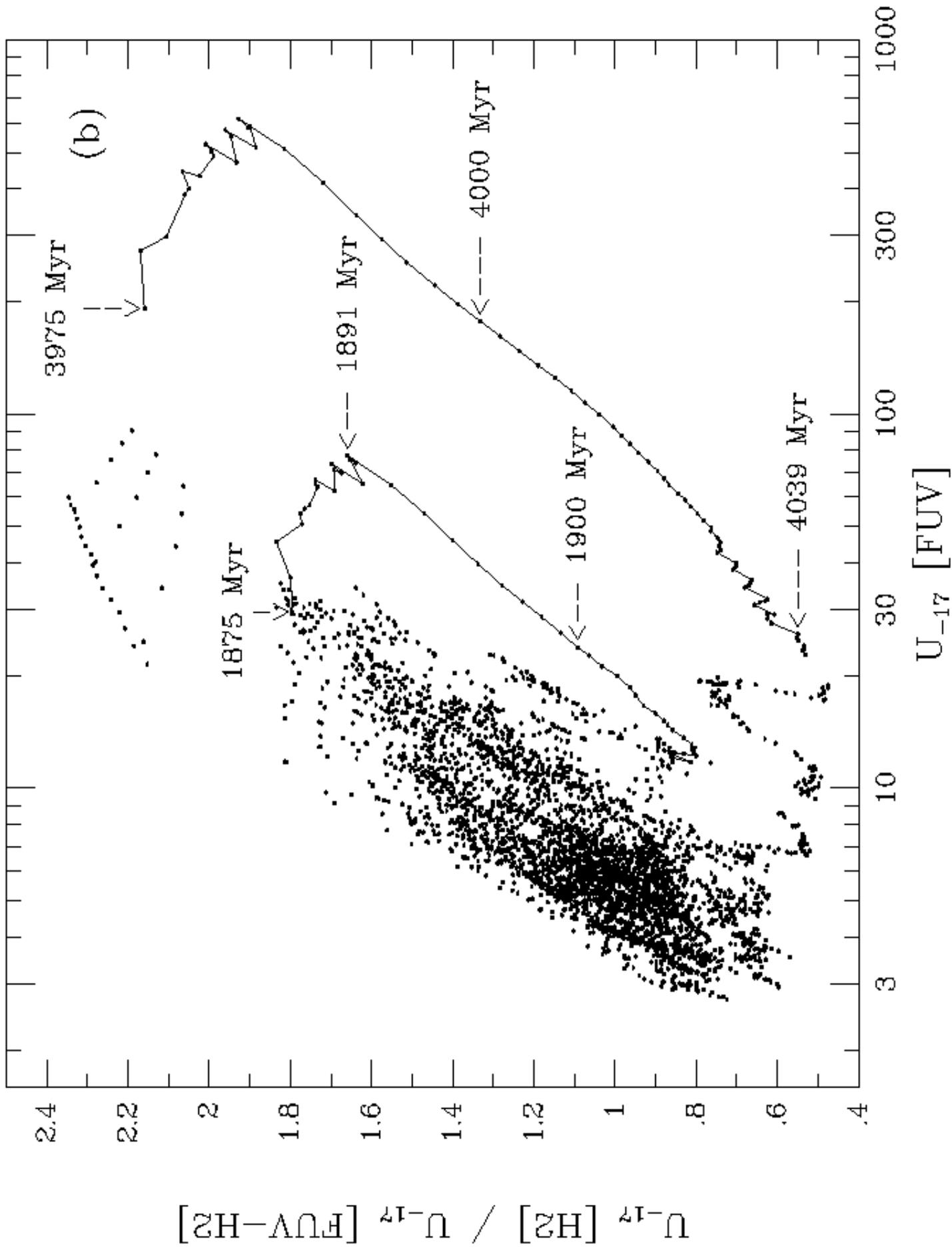
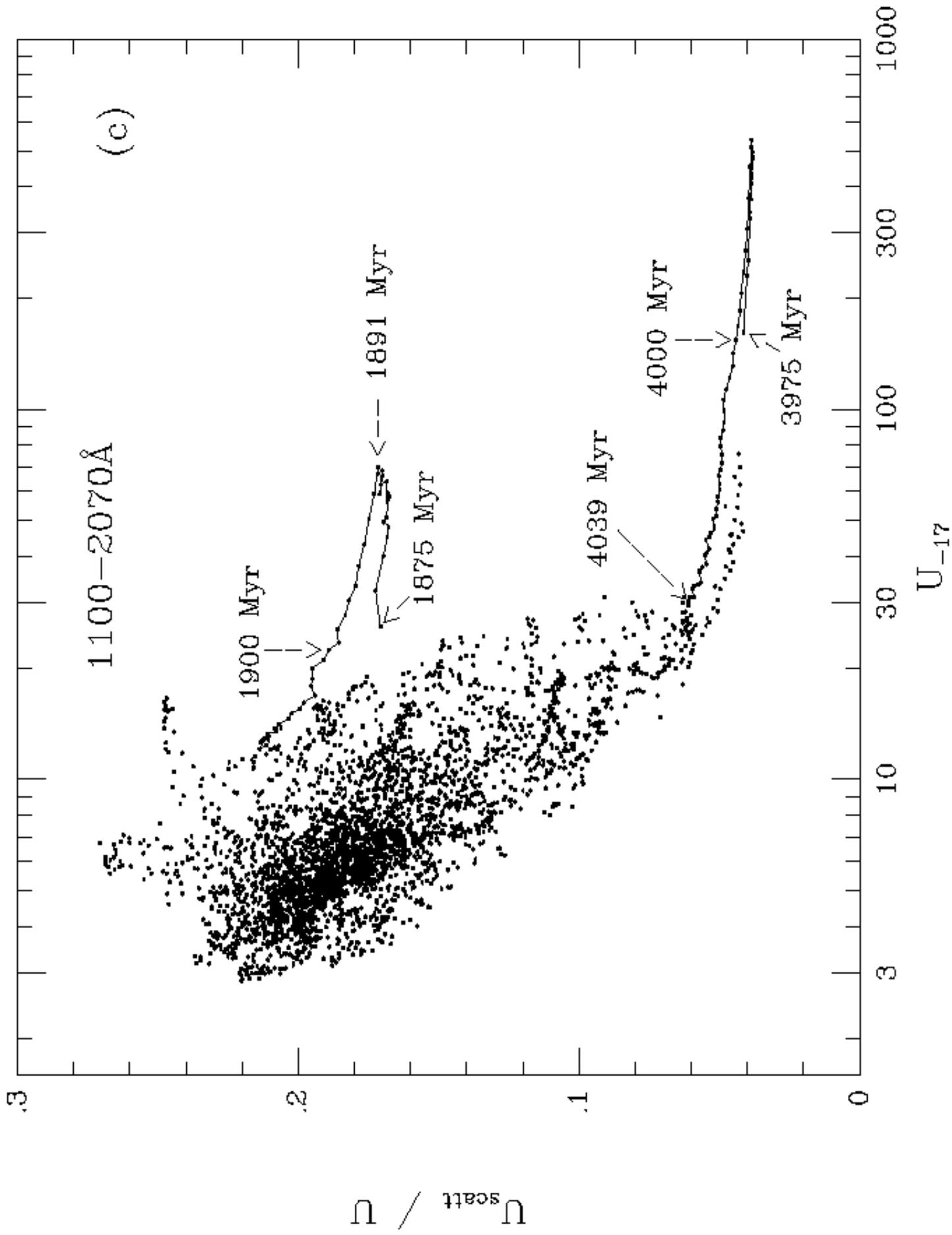
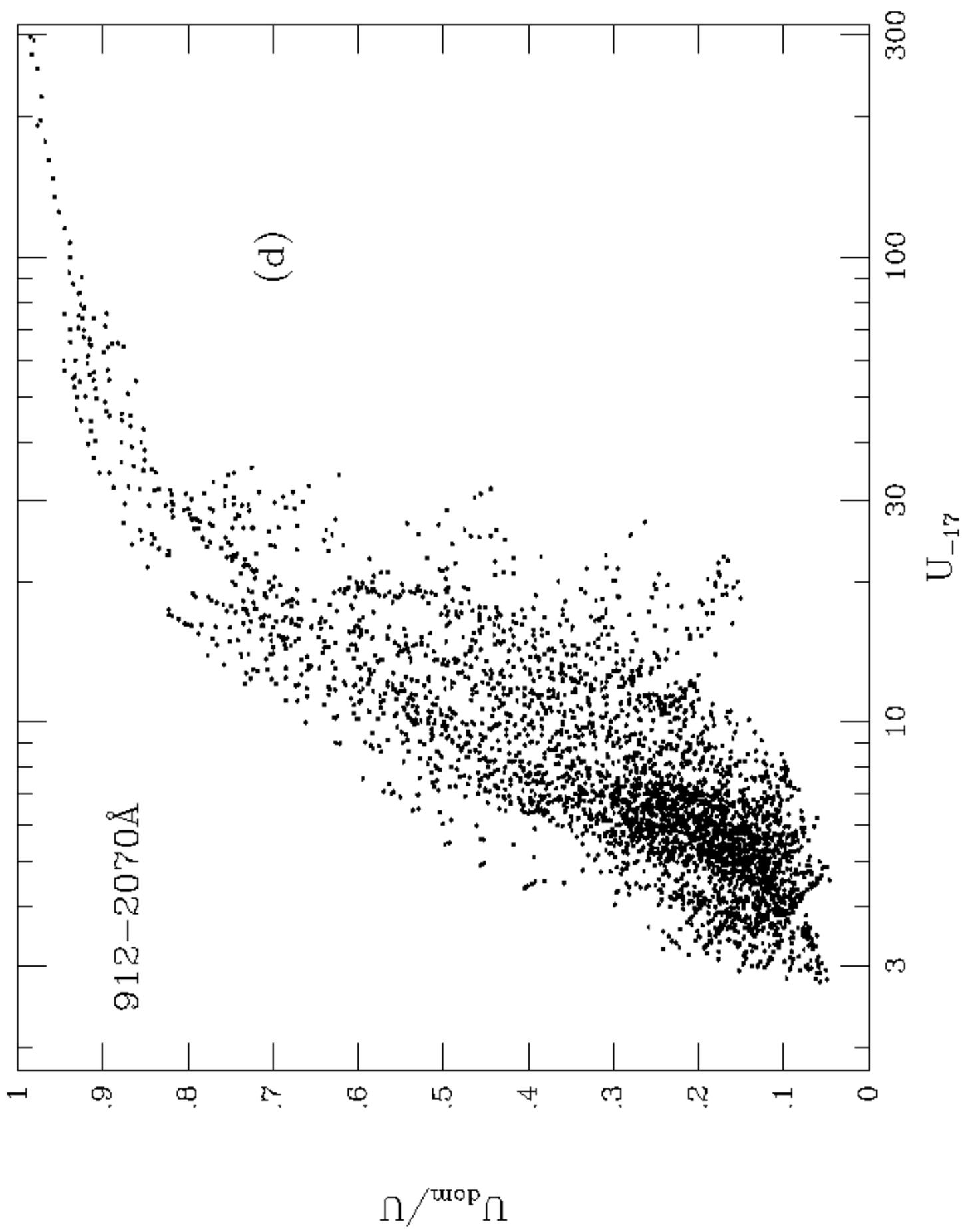
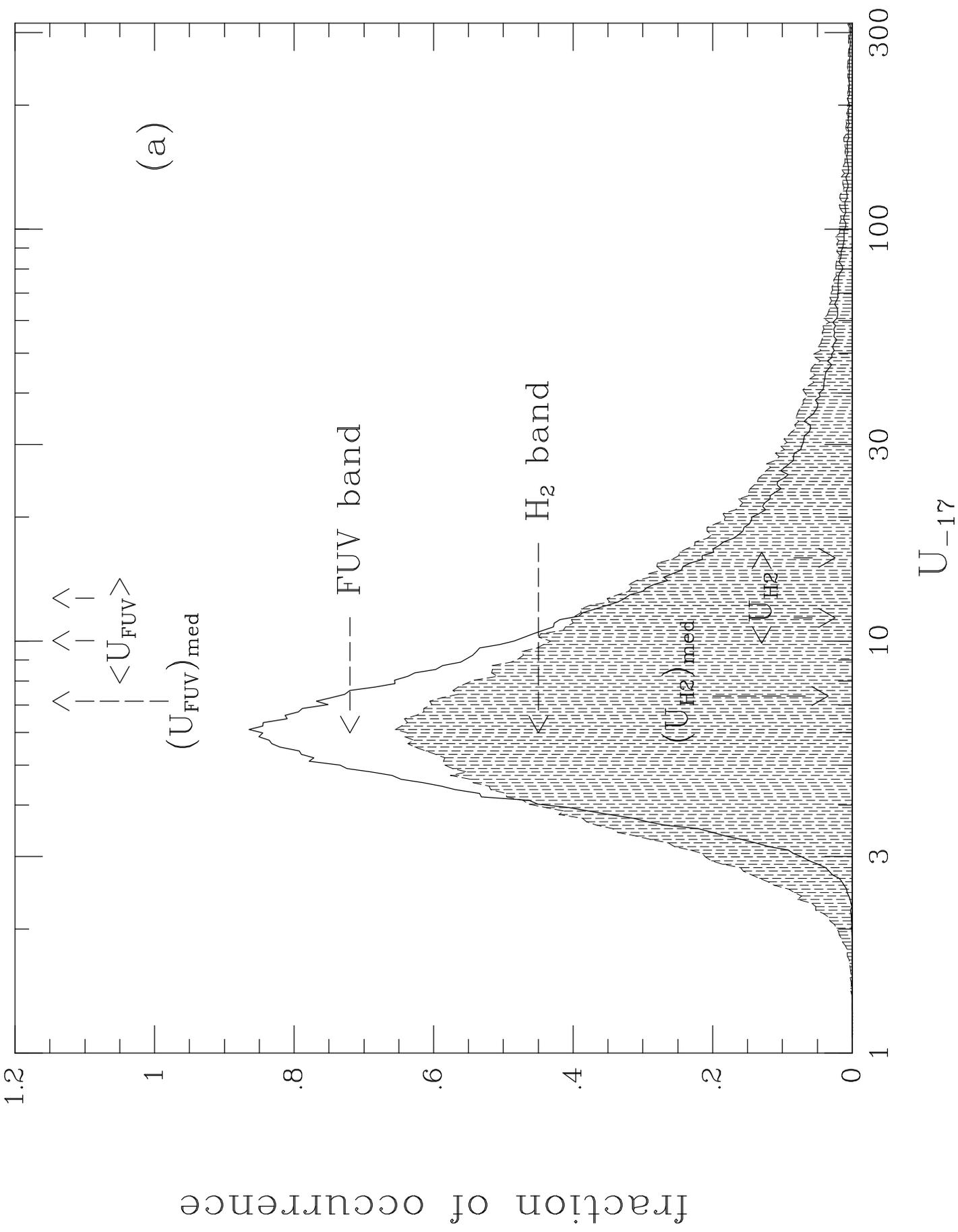
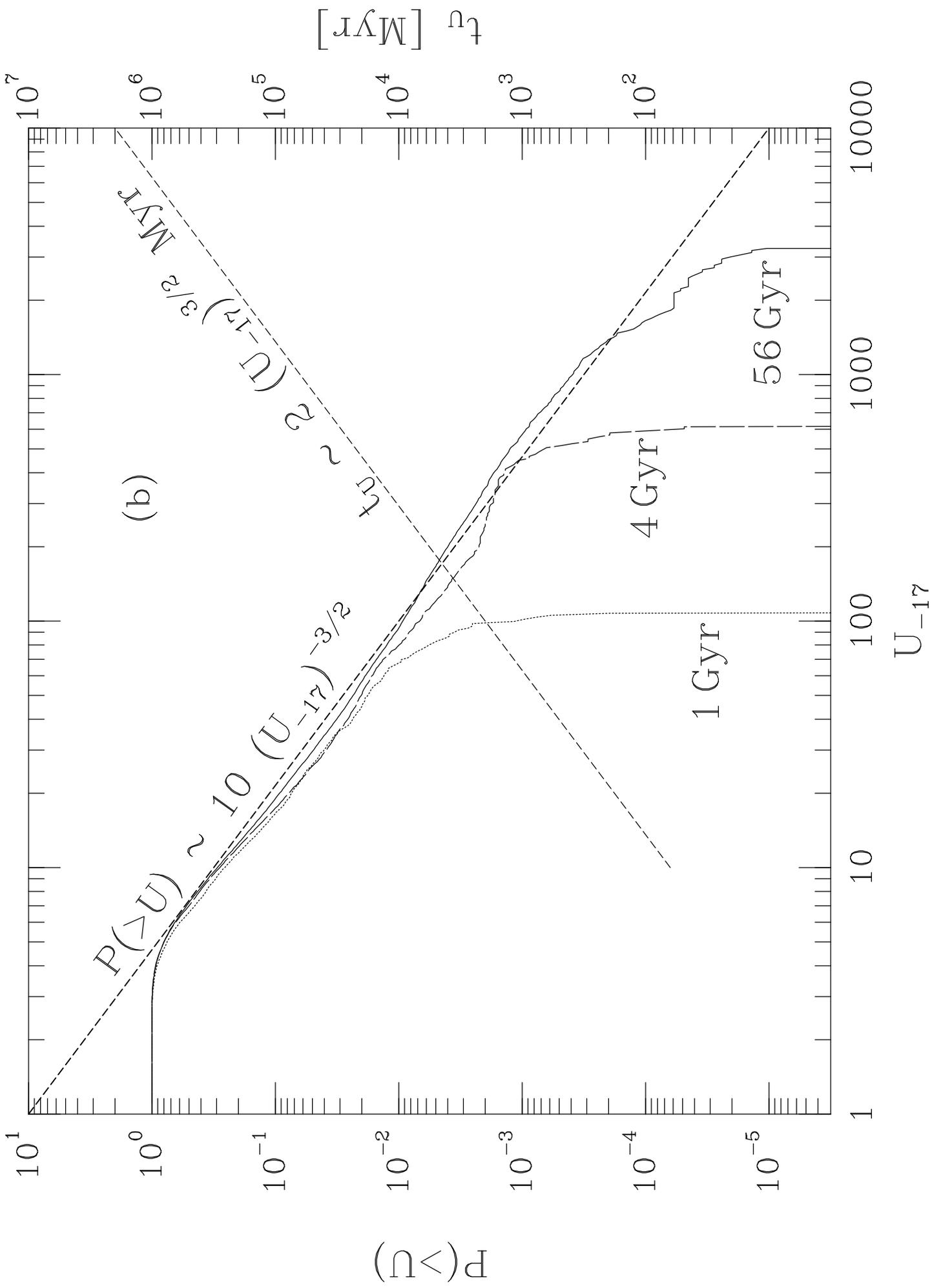
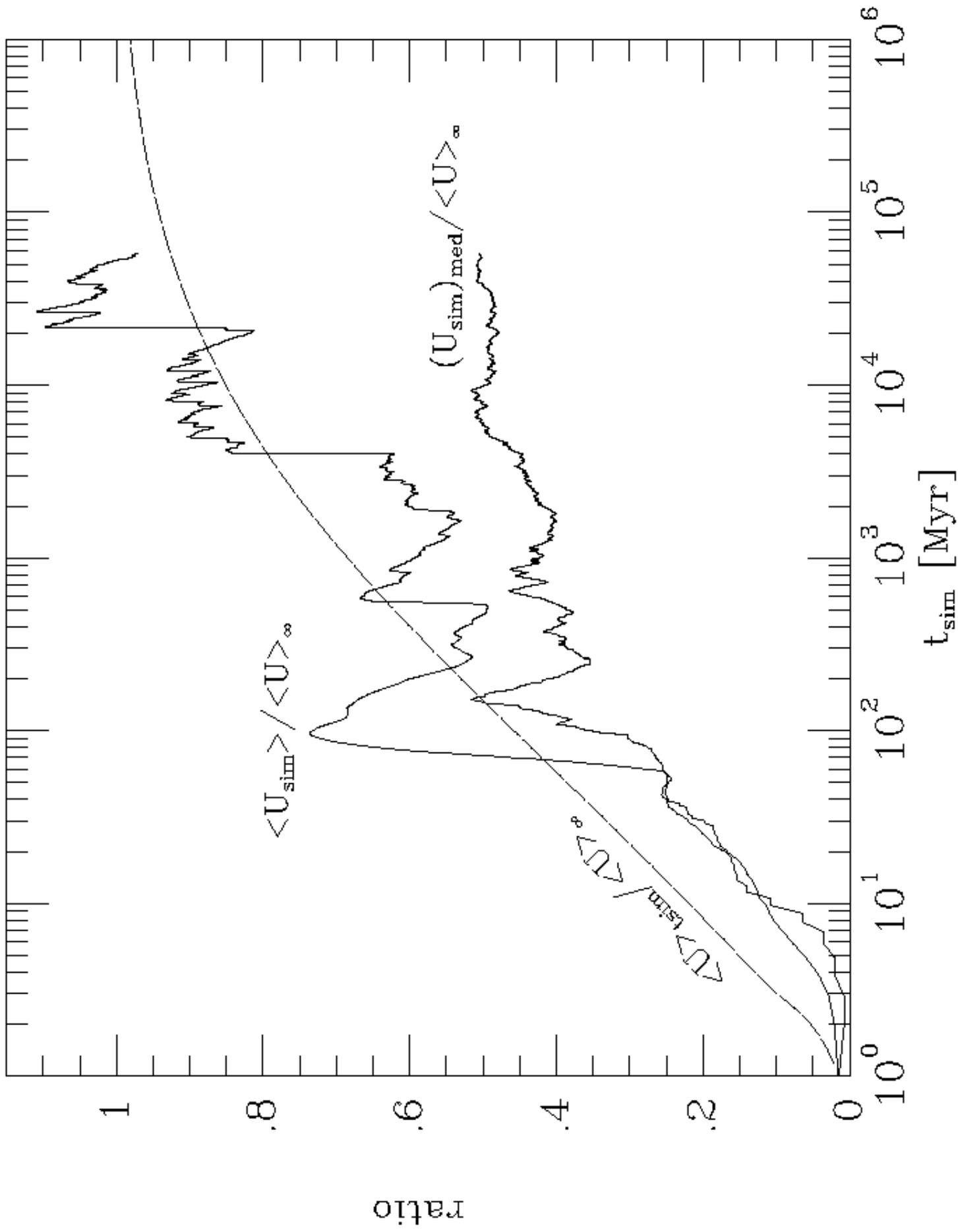
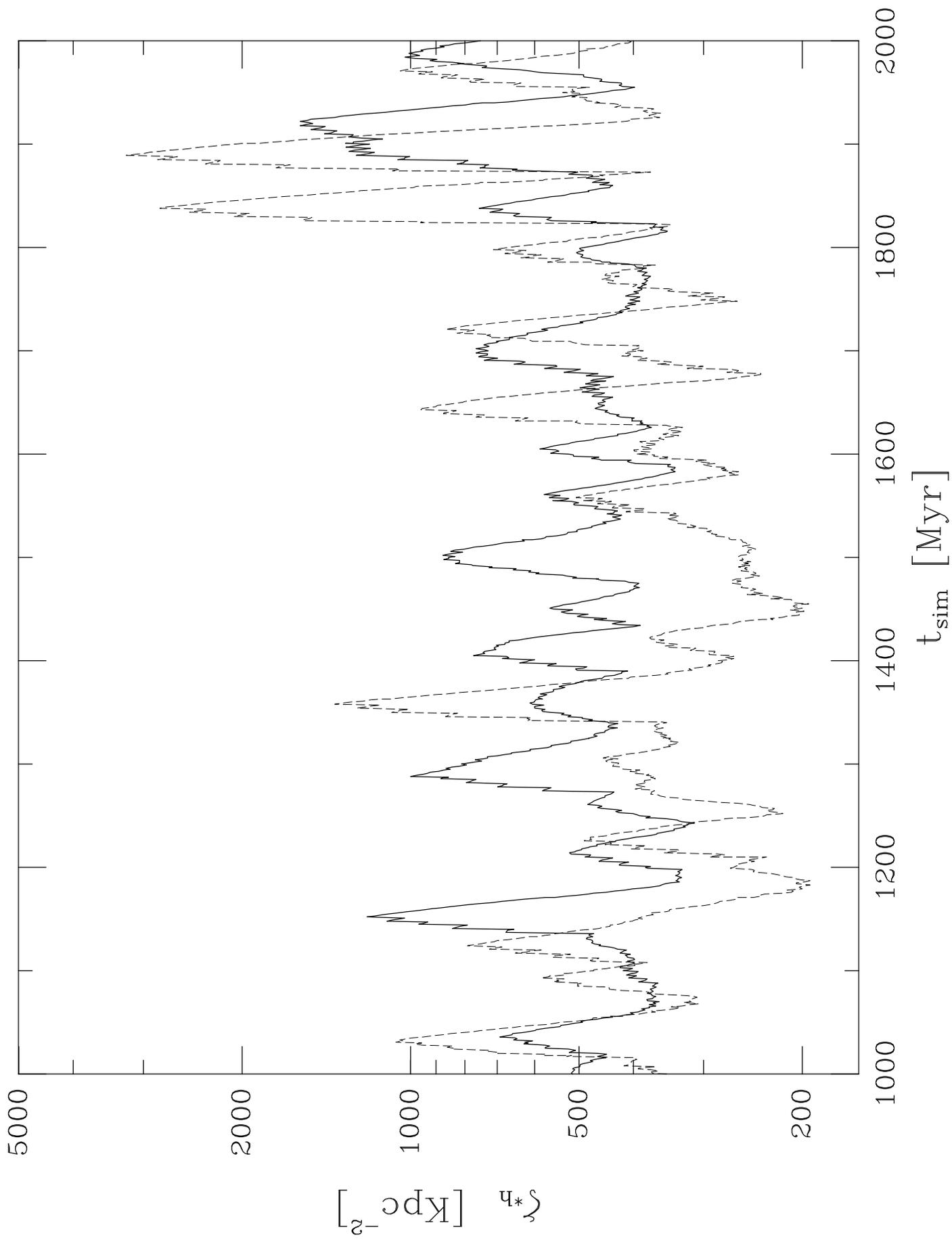
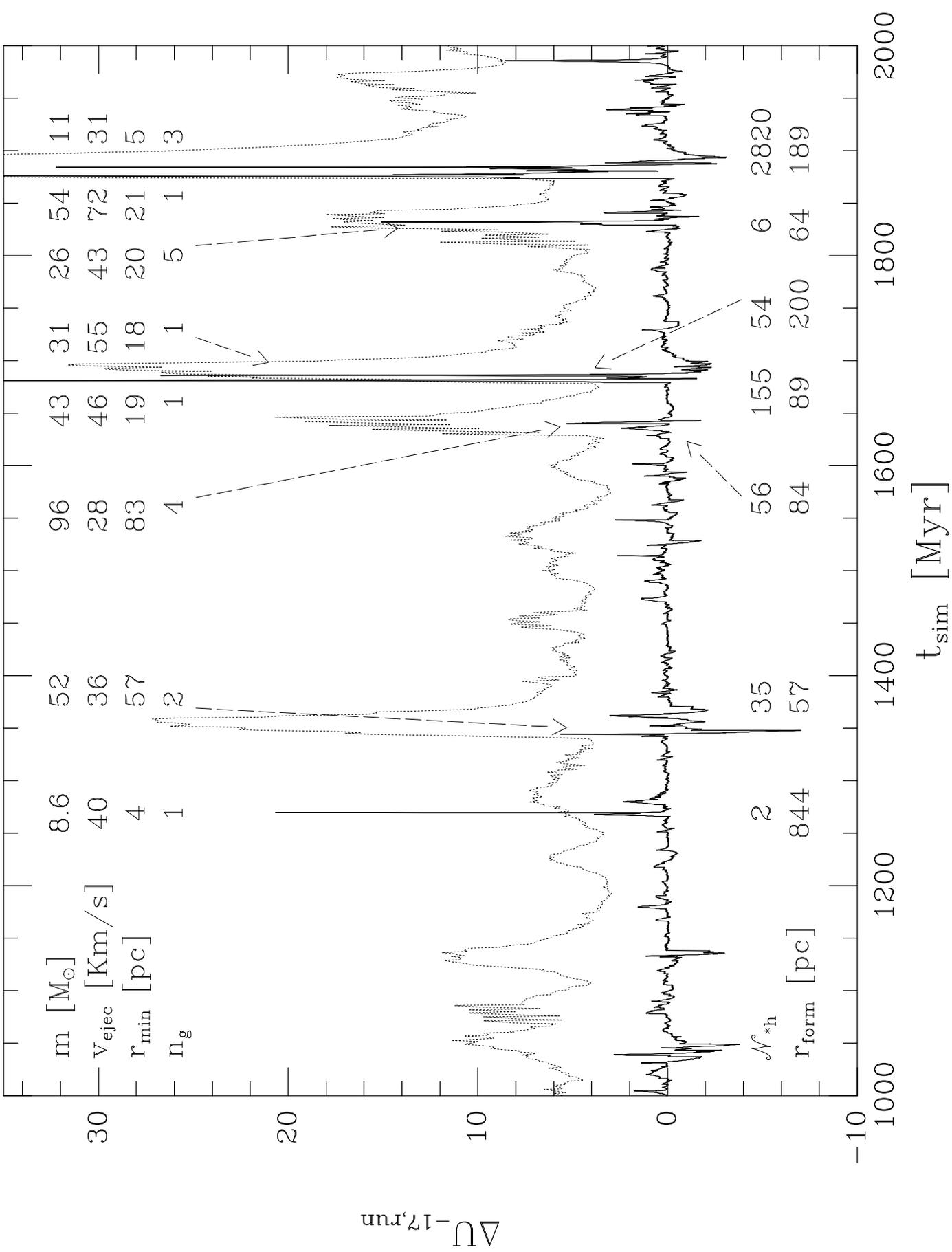
\end{document}